\pdfoutput=1

\documentclass[structabstract]{aa}
\usepackage[varg]{txfonts}

\usepackage{graphicx,amssymb}
\usepackage{epsf,psfig}
\usepackage{natbib}

\begin{document}

\title{Classifying orbits in galaxy models with a prolate or an oblate dark matter halo component}

\titlerunning{Classifying orbits in galaxy models with a prolate or an oblate dark matter halo component}
\authorrunning{Euaggelos E. Zotos}

\author{Euaggelos E. Zotos \inst{}}

\institute{Department of Physics, School of Science, Aristotle University of Thessaloniki, GR-541 24, Thessaloniki, Greece \\
\email{evzotos@physics.auth.gr}}

\date{Received 17 October 2013 / Accepted 3 January 2014}

\abstract{}
{The distinction between regular and chaotic motion in galaxies is undoubtedly an issue of paramount importance. We explore the nature of orbits of stars moving in the meridional plane $(R,z)$ of an axially symmetric galactic model with a disk, a spherical nucleus, and a flat biaxial dark matter halo component. In particular, we study the influence of all the involved parameters of the dynamical system, by computing both the percentage of chaotic orbits and the percentages of orbits of the main regular resonant families in each case.}
{To distinguish between ordered and chaotic motion, we use the smaller alignment index (SALI) method to extensive samples of orbits by numerically integrating the equations of motion as well as the variational equations. Moreover, a method based on the concept of spectral dynamics that utilizes the Fourier transform of the time series of each coordinate is used to identify the various families of regular orbits and also to recognize the secondary resonances that bifurcate from them. Two cases are studied for every parameter: (i) the case where the halo component is prolate and (ii) the case where an oblate dark halo is present.}
{Our numerical investigation indicates that all the dynamical quantities affect, more or less, the overall orbital structure. It was observed that the mass of the nucleus, the halo flattening parameter, the scale length of the halo, the angular momentum, and the orbital energy are the most influential quantities, while the effect of all the other parameters is much weaker. It was also found that all the parameters corresponding to the disk only have a minor influence on the nature of orbits. Furthermore, some other quantities, such as the minimum distance to the origin, the horizontal, and the vertical force, were tested as potential chaos detectors. Our analysis revealed that only general information can be obtained from these quantities. We also compared our results with early related work.}
{}


\maketitle

\section{Introduction}
\label{intro}

One of the most controversial topics in astronomy is that of dark matter, which was introduced by \citet{Z33} about a century ago. In his pioneer work, he discovered that the mass-to-light ratio $(M/L)$ of the Coma cluster had larger value than those known from luminous parts of nearby galaxies. Therefore, in an attempt to explain and justify this phenomenon, he claimed the existence of an invisible (dark) matter. About forty years later, \citet{OPY74} proposed that dark matter should be concentrated in extended galactic haloes. Hot X-ray emitting haloes have been used to estimate the total mass of galaxies \citep[e.g.,][]{ML99}. Computing the value of mass-to-light ratios in several galaxies shows that it always exceeds the maximum values of stellar populations and, therefore, only the presence of dark matter can explain the missing mass. Another solid indication for the presence of dark matter in galaxies is the flattened rotation curves at large galactocentric distances \citep{RFT80}. Analysis of rotation curves of spiral galaxies \citep[e.g.,][]{B81,RB85} denotes that their profiles cannot be explained without the presence of non–radiating (invisible) dark matter.

Furthermore, the presence of dark matter haloes in galaxies is indeed expected by the standard cold dark matter (CDM) cosmology models regarding formation of galaxies. The most well-known model for CDM haloes is the flattened cuspy Navarro Frenk White (NFW) model \citep{NFW96,NFW97}, which is simplified to be spherical. Most CDM models, however, take considerable deviations from the standard spherically symmetric dark matter halo distributions into account. For instance, the model of formation of dark matter haloes in a universe dominated by CDM by \citet{FWDE88} produced triaxial haloes with a preference for prolate configurations. In addition, numerical simulations of dark matter halo formation conducted by \citet{DC91} are consistent with haloes that are triaxial and flat. There are roughly equal numbers of dark halos with oblate and prolate forms.

Observational data indicate that disk galaxies are often surrounded by massive and extended dark matter haloes. The best tool to study dark matter haloes in galaxies, as explained earlier, is the galactic rotation curve derived from neutral hydrogen HI \citep[e.g.,][]{C85,PS95,HS97}. However, the determination of the exact shape of a dark matter halo is a challenging task. Numerical simulations suggest that dark matter galactic haloes are not only spherical, but may also be oblate, prolate, or even triaxial \citep[e.g.,][]{MF96,C00,KTT00,OM00,JS02,WBPKD02,KE05,AFP06,CLMV07,WMJ09,EB09,CZ09,CZ10,CZ11}. The variety of the shapes of galactic haloes points out that the structure of these objects plays an important role in the orbital behavior and, generally, in the dynamics of a galaxy.

Knowing the regular or chaotic nature of orbits in galaxies is an issue of great importance. This is true because this knowledge allow us to understand and interpret the formation and also predict the evolution of galaxies. In addition, families of regular orbits are often used as the basic tool in constructing a dynamical model for describing the main properties of galaxies. Over the last several decades, a huge amount of research work has been devoted to understanding the orbital structure in different types of galaxy models \citep[e.g.,][]{P84,CG89,SW93,P96,OP98,PMM04}. However, the vast majority of the existing literature deals only either with the distinction between regular and chaotic motion \citep[e.g.,][]{MA11,BMA12,MBS13} or the detection of periodic orbits and the analysis of their stability \citep[e.g.,][]{SPA02a,SPA02b,KP05}. We would like to note that all the above-mentioned references on the dynamics of galaxies are exemplary rather than exhaustive. In the present paper, on the other hand, we proceed one step further contributing to this active field by classifying ordered orbits into different regular families.

In \citet{CZ09}, a three-dimensional galactic model consisting of a disk, a spherical nucleus, and a logarithmic asymmetric dark matter halo component was used. For simplicity, we chose a nearly spherical dark matter halo with an internal, small deviation from spherical symmetry described by the term $-\lambda x^3$. The results of this work suggest that even small asymmetries in the galactic halo play a significant role in the nature of three-dimensional orbits stars, mainly by depopulating the box family; the box orbits become chaotic as the value of the internal perturbation increases. In the same vein, a similar three-dimensional composite galaxy model was utilized in \citet{CZ11}, however, in this case the dark mater halo was modeled by a mass Plummer potential. The mass of the halo was found to be an important physical quantity, acting as a chaos controller in galaxies. In particular, the percentage of chaotic orbits reduces rapidly as the mass of the spherical dark halo increases. Moreover, it was found that the amount of chaos is higher in asymmetric triaxial galaxies when they are surrounded by less concentrated spherical dark halo components.

In two earlier papers \citep{C97,PC06}, two-dimensional, axially symmetric or non-axially symmetric active galaxy models with an additional spherical halo component were studied. In both cases it was observed that the presence of a spherical halo resulted in reduced area in the phase space occupied by the chaotic orbits. Moreover, the behavior of orbits in an active galaxy with a biaxial (prolate or oblate) dark matter halo was investigated recently in \citet{CZ10} (hereafter Paper I). In this work, we studied how the regular or chaotic nature of orbits is influenced by some important quantities of the system, such as the flattening parameter of the halo, the scale length of the halo component, and the conserved component of the angular momentum. It was found that when a biaxial halo component is present there is a linear relationship between the chaotic percentage in the phase plane and the flattening parameter. In contrast, the relation between chaos and the scale length of the halo was found to be not linear but exponential. A similar model was used in \citet{H04} for numerical simulations of the evolution of a system like the Sagittarius dSph in a variety of galactic potentials varying the flattening parameter. In the current research, we use the results of Paper I as a starting point and we try to explore how all the involved parameters influence the overall orbital structure, not only distinguishing between order and chaos, but taking a step further by classifying and distributing all regular orbits into different families \citep[e.g.,][]{ZC13,CZ13,ZCar13,ZCar14}.

The present article is organized as follows. In Section \ref{GalMod}, we present in detail the structure and the properties of our gravitational galactic model. In Section \ref{CompMeth}, we describe the computational methods we used in order to determine the character of orbits. In the following section, we investigate how the involved parameters of the dynamical system influence the nature of the orbits when a prolate or an oblate dark halo component is present. In Section \ref{Anal}, we present some heuristic arguments in an attempt to support and explain the numerically obtained outcomes of the previous Section. Our paper ends with section \ref{Disc}, where the discussion and the conclusions of this research are presented.

\section{Presentation and properties of the galactic model}
\label{GalMod}

In the present work, we investigate the character of motion in the meridional plane of an axially symmetric galaxy model with a disk, a spherical nucleus, and a flat biaxial dark halo component. For this purpose, we use the usual cylindrical coordinates $(R, \phi, z)$, where $z$ is the axis of symmetry.

\begin{figure}
\includegraphics[width=\hsize]{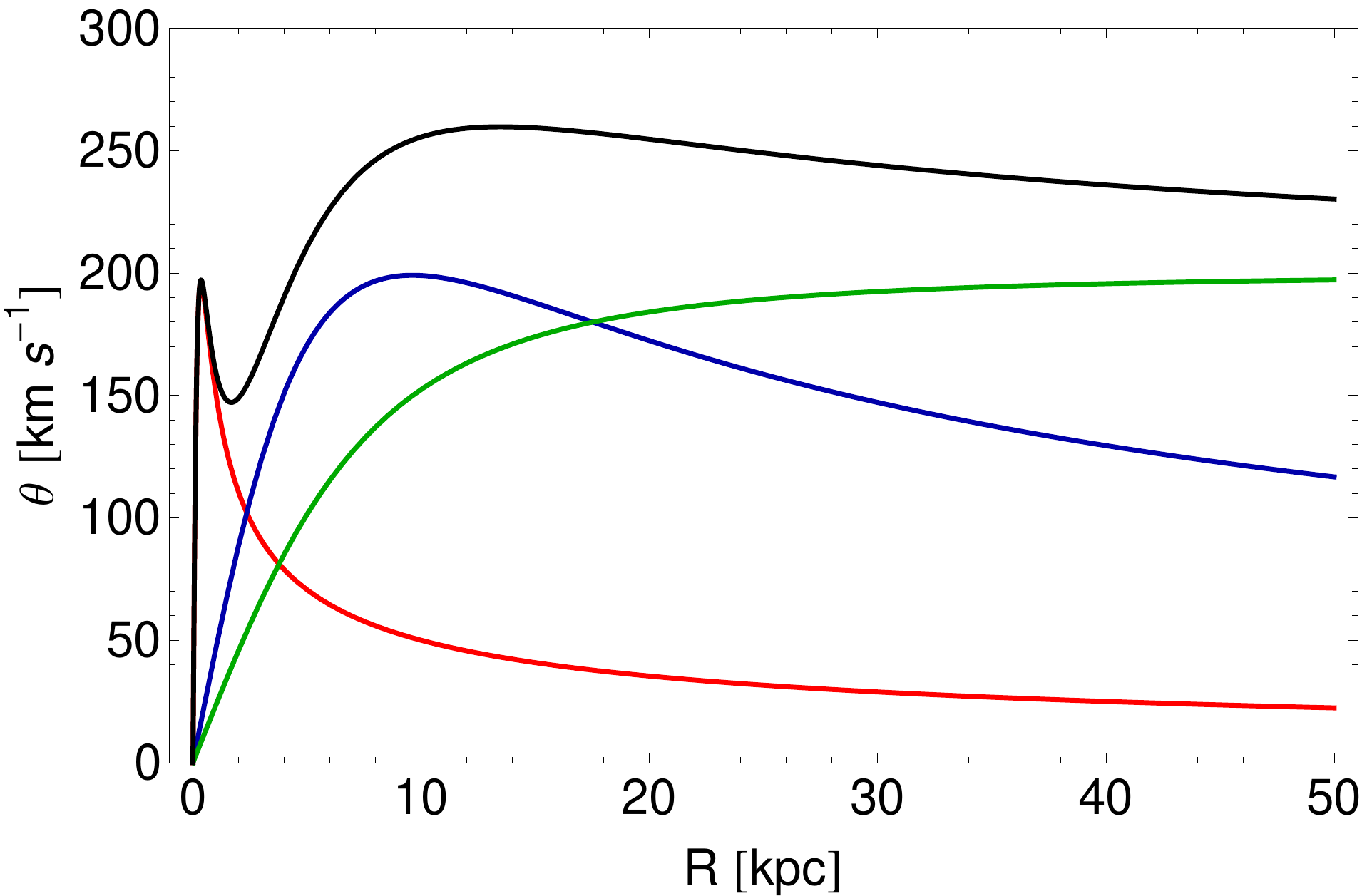}
\caption{A plot of the rotation curve in our galactic model. We distinguish the total circular velocity (black) and also the contributions from the spherical nucleus (red), the disk (blue) and that of the dark matter halo (green).}
\label{rotvel}
\end{figure}

The total potential $V(R,z)$ in our model consists of three components: (i) the disk potential $V_{\rm d}$, (ii) the central spherical component $V_{\rm n}$, and (iii) the dark matter halo component $V_{\rm h}$. The first one is represented by a generalization of the \citet{MN75} potential \citep[see also][]{CI87,CI91}
\begin{equation}
V_{\rm d}(R,z) = \frac{- G M_{\rm d}}{\sqrt{b^2 + R^2 + \left(\alpha + \sqrt{h^2 + z^2}\right)^2}}.
\label{Vd}
\end{equation}
Here $G$ is the gravitational constant, $M_{\rm d}$ is the mass of the disk, $b$ is the core radius of the disk-halo, $\alpha$ is the scale length of the disk, while $h$ corresponds to the disk's scale height. Here we should note that potential (\ref{Vd}) is very similar to the potential introduced by Satoh \citep{S80}. On the other hand, a new gravitational model, which is a combination of the standard logarithmic and the Miyamoto-Nagai model, was proposed by \citet{Z11} to describe the motion of stars both in elliptical and disk galaxies. For the description of the spherically symmetric nucleus, we use a Plummer potential \citep[e.g.,][]{BT08}
\begin{equation}
V_{\rm n}(R,z) = \frac{- G M_{\rm n}}{\sqrt{R^2 + z^2 + c_{\rm n}^2}},
\label{Vn}
\end{equation}
where $M_{\rm n}$ and $c_{\rm n}$ are the mass and the scale length of the nucleus, respectively. This potential has been used successfully in the past to model and, therefore, interpret the effects of the central mass component in a galaxy \citep[see, e.g.,][]{HN90,HPN93,Z12a}. At this point, we must clarify that we do not include any relativistic effects, because the nucleus represents a bulge rather than a black hole or any other compact object. The potential of the dark matter halo is modeled by the flattened axisymmetric logarithmic potential
\begin{equation}
V_{\rm h}(R,z) = \frac{\upsilon_0^2}{2}\ln \left(R^2 + \beta z^2 + c_{\rm h}^2 \right),
\label{Vh}
\end{equation}
where $\beta$ is the flattening parameter and $c_{\rm h}$ stands for the scale length of the dark halo component. The parameter $\upsilon_0$ is used for the consistency of the galactic units. The choice for the logarithmic potential was motivated for several reasons: (i) it can model a wide variety of shapes of galactic haloes by suitably choosing the parameter $\beta$. In particular, when $0.1 \leq \beta < 1$ the dark matter halo is prolate, when $\beta = 1$ is spherical, while when $1 < \beta < 2$ is oblate; (ii) it is appropriate for the description of motion in a dark matter halo as it produces a flat rotation curve at large radii (see Fig. \ref{rotvel}); (iii) it allows for the investigation of flattened configurations of the galactic halo at low computational costs; (iv) the relatively small number of input parameters of Eq. (\ref{Vh}) is an advantage concerning the performance and speed of the numerical model; and (v) the flattened logarithmic potential was utilized successfully in previous works to model a dark matter halo component \citep[see, e.g.][]{H04,RPT07}.

In this work, we use a system of galactic units where the unit of length is 1 kpc, the unit of velocity is 10 km s$^{-1}$, and $G = 1$. Thus, the unit of mass is $2.325 \times 10^7 {\rm M}_\odot$, that of time is $0.9778 \times 10^8$ yr, the unit of angular momentum (per unit mass) is 10 km kpc$^{-1}$ s$^{-1}$, and the unit of energy (per unit mass) is 100 km$^2$s$^{-2}$. Our main objective is to investigate the regular or chaotic nature of orbits in two different cases: that is when the dark matter halo component is (i) prolate (PH model) and (ii) oblate (OH model). Our models have the following standard values of the parameters: $M_{\rm d} = 7000$ (corresponding to $1.63\times 10^{11}$ M$_\odot$, i.e., a normal disk galaxy mass), $b = 6$, $\alpha = 3$, $h = 0.2$, $M_{\rm n} = 250$ (corresponding to $5.8\times 10^{9}$ M$_\odot$), $c_{\rm n} = 0.25$, $\upsilon_0 = 20$ and $c_{\rm h} = 8.5$, while $\beta = 0.5$ for the PH model and $\beta = 1.5$ for the OH model. The values of the disk and the nucleus were chosen with a Milky Way-type galaxy in mind \citep[e.g.,][]{AS91}. In the case of the prolate dark halo, the set of the values of the parameters define the standard prolate model (SPM), while when the dark halo is oblate we use the standard oblate model (SOM). The values of the parameters of the standard models secure positive density everywhere.

It is well known that in disk galaxies the circular velocity in the galactic plane $z=0$,
\begin{equation}
\theta(R) = \sqrt{R\left|\frac{\partial V(R,z)}{\partial R}\right|_{z = 0}},
\label{rcur}
\end{equation}
is a very important physical quantity. A plot of $\theta(R)$ for our galactic model\footnote{On the galactic plane applies $z=0$ so, the rotation curve is the same for both PH and OH models.} is presented in Fig. \ref{rotvel} as a black curve. Moreover, in the same plot, the red line shows the contribution from the spherical nucleus, the blue curve is the contribution from the disk, while the green line corresponds to the contribution form the dark halo. It is seen that each contribution prevails in different distances form the galactic center. In particular, at small distances when $R\leq 2$ kpc, the contribution from the spherical nucleus dominates, while at mediocre distances, $2 < R < 18$ kpc, the disk contribution is the dominant factor. On the other hand, at large galactocentric distances, $R > 18$ kpc, we see that the contribution from the dark halo prevails, thus forcing the rotation curve to remain flat with increasing distance from the center. We also observe the characteristic local minimum of the rotation curve due to the massive nucleus, which appears when fitting observed data to a galactic model \citep[e.g.,][]{GHBL10,IWTS13}.

\begin{figure*}
\centering
\resizebox{\hsize}{!}{\includegraphics{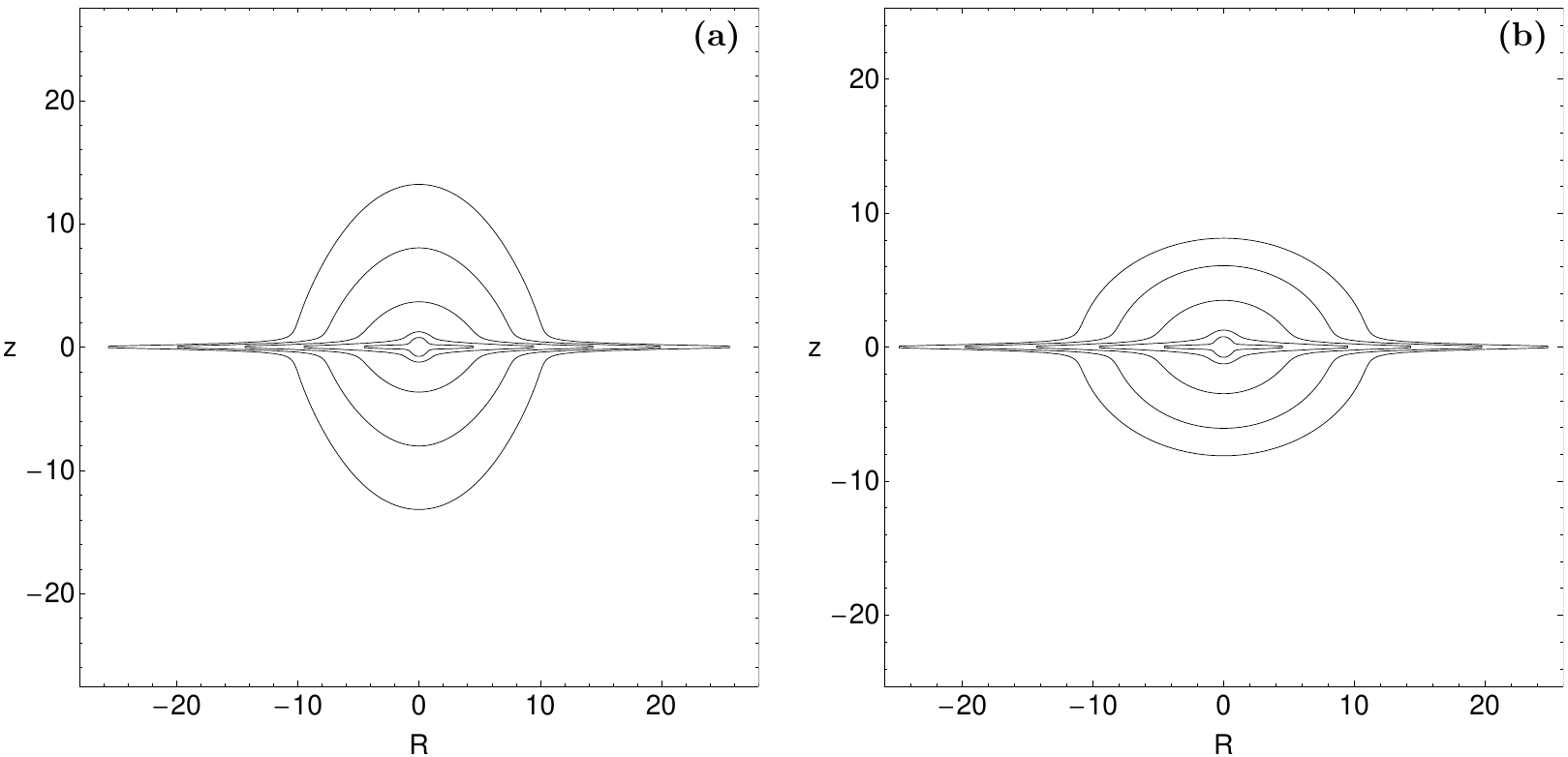}}
\caption{Iso-density contours for our galactic model when the dark matter halo is (a-left): prolate (PH model) and (b-right): oblate (OH model).}
\label{dencon}
\end{figure*}

\begin{figure}
\includegraphics[width=\hsize]{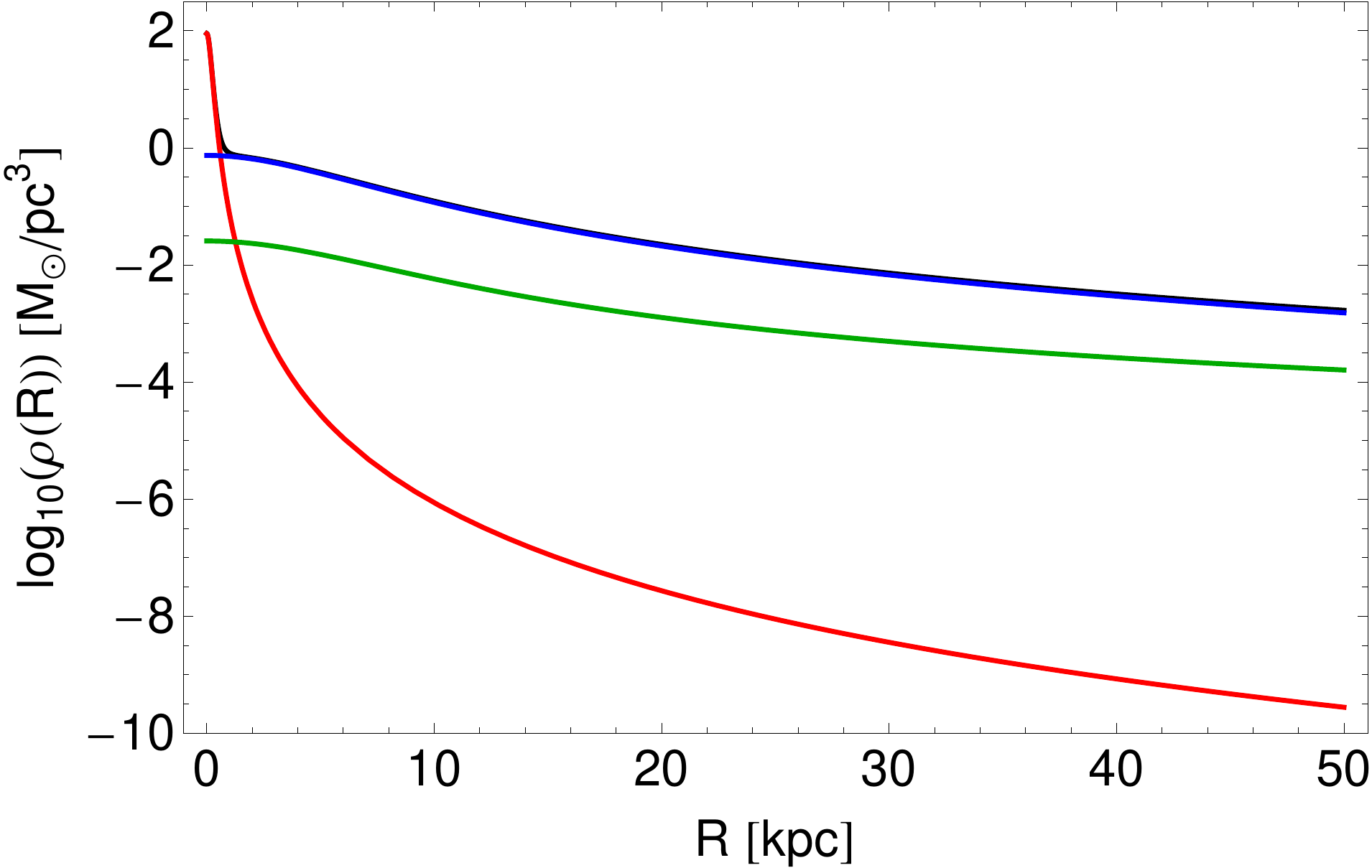}
\caption{Evolution of the mass density $\rho(R)$ in the galactic plane $(z=0)$, as a function of the distance $R$ from the center for SPM. The total mass density is shown in black, while we distinguish the three different contributions: the spherical nucleus (red), the disk (blue), and the dark matter halo (green).}
\label{denevol}
\end{figure}

It is very useful to compute the mass density $\rho(R,z)$ derived from the total potential $V(R,z)$ using the Poisson's equation
\begin{equation}
\begin{split}
&\rho(R,z) = k \frac{1}{4 \pi G} \nabla^2 V(R,z) = \\
&= k \frac{1}{4 \pi G} \left(\frac{\partial^2}{\partial R^2} + \frac{1}{R}\frac{\partial}{\partial R}
+ \frac{1}{R^2} \frac{\partial^2}{\partial \phi^2} + \frac{\partial^2}{\partial z^2}\right) V(R,z).
\end{split}
\label{dens}
\end{equation}
Recall, that due to the axial symmetry the third term in Eq. (\ref{dens}) is zero. In the same equation, we observe the presence of an additional parameter $k = 2.325 \times 10^{-2}$, which is simply a numerical coefficient dictated by the current system of galactic units to obtain the density in units of ${\rm M}_\odot/pc^3$. Fig. \ref{dencon}(a-b) shows the iso-density contours $\rho(R,z) = const$, for the PH and OH galaxy models, respectively. In Fig. \ref{denevol}, the evolution of the total mass density $\rho(R,z=0)$ in the galactic plane for SPM, as a function of the radius $R$ from the galactic center is shown as the black curve. In the same diagram, the red line shows the contribution from the spherical nucleus, the blue curve is the contribution from the disk, while the green line corresponds to the contribution form the dark matter halo. It is evident that the density of the nucleus decreases rapidly obtaining very low values, while on the other hand, the density of both the disk and the halo continues to hold significantly larger values. We also see that when $R > 1$ kpc, the total mass density coincides with the density of the disk. Therefore, at large galactocentric distances, the mass density should vary like $1/R^3$ (to be more precise, from the nonlinear fit, we derived that the exact power of the $1/R^n$ decrease law of the total mass density is $n = 2.847$). This means that the total mass $M(R)$, enclosed in a sphere of radius $R$, increases with the distance. This fact explains why the circular velocity profile shown in Fig. \ref{rotvel} remains flat. Things are quite similar for the SOM. Here, we must point out that our gravitational potential is truncated ar $R_{max} = 50$ kpc, otherwise the total mass of the galaxy modeled by this potential would be infinite, which is obviously not physical.

Taking into account that the total potential $V(R,z)$ is axisymmetric, the $z$-component of the angular momentum $L_z$ is conserved. With this restriction, orbits can be described by means of the effective potential \citep[e.g.,][]{BT08}
\begin{equation}
V_{\rm eff}(R,z) = V(R,z) + \frac{L_z^2}{2R^2}.
\label{veff}
\end{equation}

The equations of motion on the meridional plane are
\begin{equation}
\ddot{R} = - \frac{\partial V_{\rm eff}}{\partial R}, \ \ \ \ddot{z} = - \frac{\partial V_{\rm eff}}{\partial z},
\label{eqmot}
\end{equation}
while the equations governing the evolution of a deviation vector $\vec{w} = (\delta R, \delta z, \delta \dot{R}, \delta \dot{z})$, which joins the corresponding phase space points of two initially nearby orbits, needed for the
calculation of the standard indicators of chaos (the SALI in our case), are given by the variational equations
\begin{eqnarray}
\dot{(\delta R)} &=& \delta \dot{R}, \ \ \ \dot{(\delta z)} = \delta \dot{z}, \nonumber \\
(\dot{\delta \dot{R}}) &=&
- \frac{\partial^2 V_{\rm eff}}{\partial R^2} \delta R
- \frac{\partial^2 V_{\rm eff}}{\partial R \partial z}\delta z,
\nonumber \\
(\dot{\delta \dot{z}}) &=&
- \frac{\partial^2 V_{\rm eff}}{\partial z \partial R} \delta R
- \frac{\partial^2 V_{\rm eff}}{\partial z^2}\delta z.
\label{vareq}
\end{eqnarray}

Consequently, the corresponding Hamiltonian to the effective potential given in Eq. (\ref{veff}) can be written as
\begin{equation}
H = \frac{1}{2} \left(\dot{R}^2 + \dot{z}^2 \right) + V_{\rm eff}(R,z) = E,
\label{ham}
\end{equation}
where $\dot{R}$ and $\dot{z}$ are momenta per unit mass, and conjugate to $R$ and $z$, respectively, while $E$ is the numerical value of the Hamiltonian, which is conserved. Therefore, an orbit is restricted to the area in the meridional plane satisfying $E \geq V_{\rm eff}$.

\section{Computational methods}
\label{CompMeth}

In our study, we seek to determine whether an orbit is regular or chaotic. Several indicators of chaos are available in the literature; we chose the SALI indicator \citep{S01}. We use the usual method in which we check after a certain and predefined time interval of numerical integration whether the value of SALI is less than a very small threshold value, in order to decide whether an orbit is ordered or chaotic. In our current research, we define this value to be equal to $10^{-8}$. However, depending on the particular location of each orbit, this threshold value can be reached more or less quickly, as there are phenomena that can hold off the final classification of the orbit (i.e., there are special orbits called ``sticky" orbits, which behave regularly for long time periods before they finally drift away from the regular regions and start to wander in the chaotic domain, revealing their true chaotic nature fully.

To determine the chaoticity of our models, we chose, for each set of values of the parameters of the potential, a dense grid of initial conditions in the $(R,\dot{R})$ plane, regularly distributed in the area allowed by the value of the energy $E$. In cases, $z_0 = 0$, while $\dot{z_0}$ is found from the energy integral (Eq. \ref{ham}). The points of the grid were separated 0.1 units in $R$ and 0.5 units in $\dot{R}$ direction. For each initial condition, we integrated the equations of motion (\ref{eqmot}) as well as the variational equations (\ref{vareq}) with a double precision Bulirsch-Stoer algorithm \citep[e.g.,][]{PTVF92} with a small time step of the order of $10^{-2}$, which is sufficient enough for the desired accuracy of our computations (i.e., our results practically do not change by halving the time step). In all cases, the energy integral (Eq. \ref{ham}) was conserved better than one part in $10^{-10}$, although for most orbits it was better than one part in $10^{-11}$.

Each orbit was integrated numerically for a time interval of $10^4$ time units ($10^{12}$ yr), which corresponds to a time span of the order of hundreds of orbital periods. The particular choice of the total integration time is an element of great importance, especially in the case of the sticky orbits. A sticky orbit could be easily misclassified as regular by any chaos indicator\footnote{Generally, dynamical methods are broadly split into two types: (i) those based on the evolution of sets of deviation vectors to characterize an orbit and (ii) those based on the frequencies of the orbits that extract information about the nature of motion only through the basic orbital elements without the use of deviation vectors.}, if the total integration interval is too small, so that the orbit does not have enough time to reveal its true chaotic character. Thus, all the initial conditions of the orbits of a given grid were integrated, as we already said, for $10^4$ time units, thus avoiding sticky orbits with a stickiness at least of the order of $10^2$ Hubble time. All the sticky orbits that do not show any signs of chaoticity for $10^4$ time units are counted as regular orbits since such vast sticky periods are completely out of the scope of our research.

A first step toward the understanding of the overall behavior of our system is knowing whether the orbits in the galactic model are regular or chaotic. Also of particular interest is the distribution of regular orbits into different families. Therefore, once the orbits have been characterized as regular or chaotic, we then further classified the regular orbits into different families by using a frequency analysis method \citep{CA98,MCW05}. Initially, \citet{BS82,BS84} proposed a technique, dubbed spectral dynamics, for this particular purpose. Later on, this method has been extended and improved by \citet{CA98} and \citet{SN96}. In a recent work, \citep{ZC13} the algorithm was refined even further so it can be used to classify orbits in the meridional plane. In general terms, this method computes the Fourier transform of the coordinates of an orbit, identifies its peaks, extracts the corresponding frequencies, and searches for the fundamental frequencies and their possible resonances. Thus, we can easily identify the various families of regular orbits and also recognize the secondary resonances that bifurcate from them.

\begin{figure*}
\centering
\resizebox{\hsize}{!}{\includegraphics{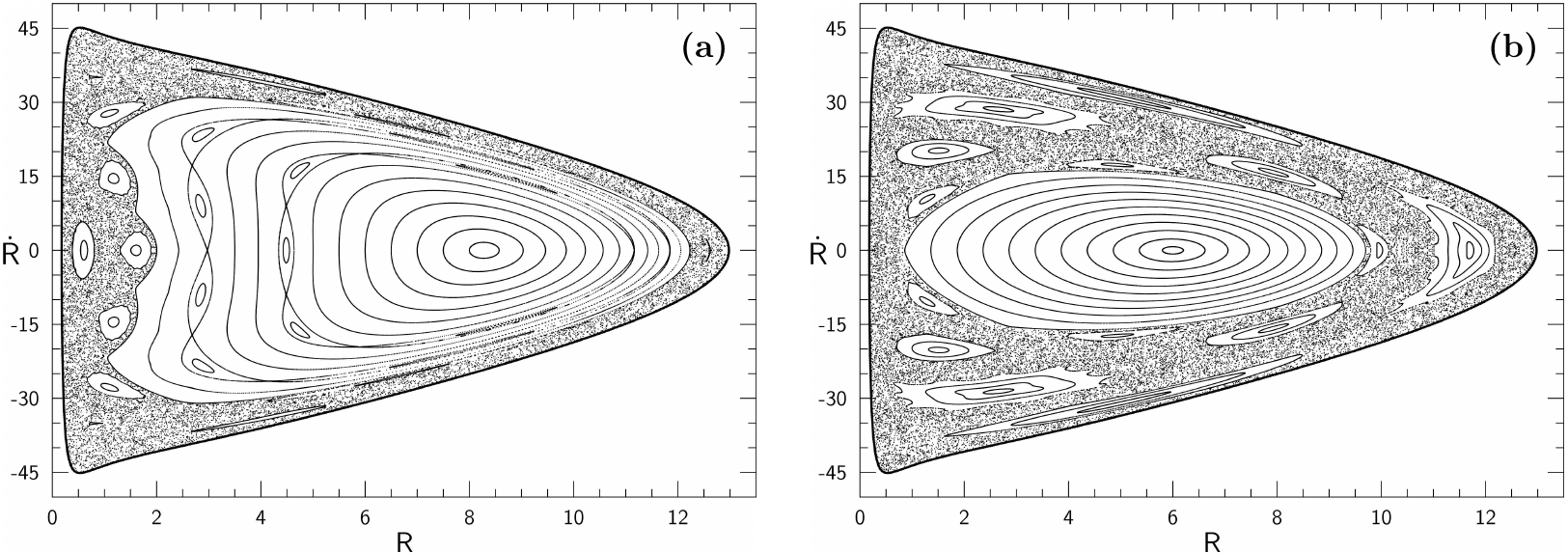}}
\caption{The $(R,\dot{R})$ $z = 0$, $\dot{z} > 0$ Poincar\'{e} surface of section (PSS )for the (a-left): standard prolate model (SPM) and (b-right): standard oblate model (SOM).}
\label{PSSs}
\end{figure*}

\begin{figure*}
\centering
\resizebox{\hsize}{!}{\includegraphics{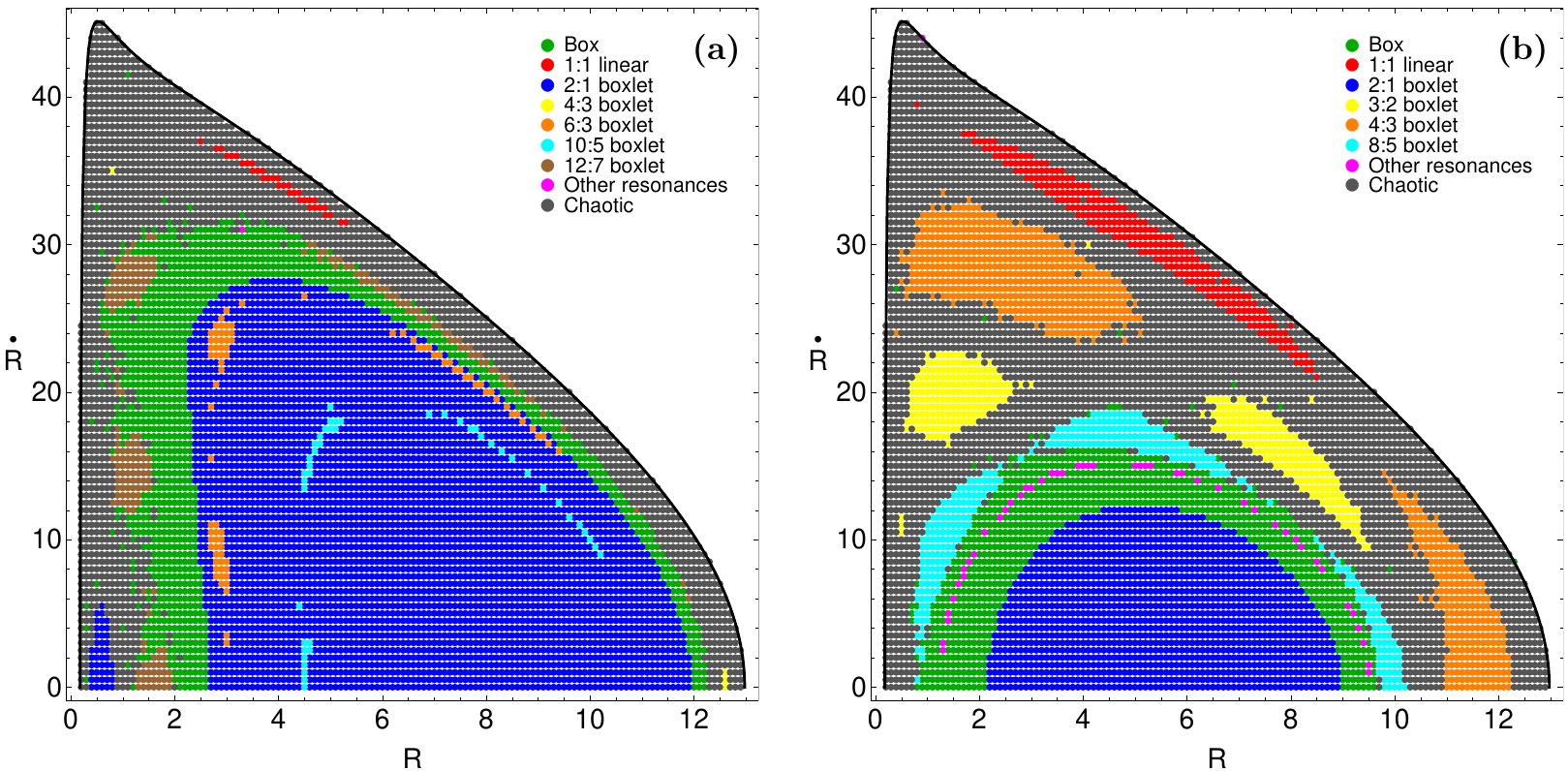}}
\caption{Grid of initial conditions revealing the orbital structure of the $(R,\dot{R})$ phase plane for the (a-left): PH model and (b-right): OH model.}
\label{Clas}
\end{figure*}

Before closing this section, we would like to make a clarification about the nomenclature of orbits. All the orbits of an axisymmetric potential are in fact three-dimensional (3D) loop orbits, i.e., orbits that always rotate around the axis of symmetry in the same direction. However, in dealing with the meridional plane, the rotational motion is lost, so the path that the orbit follows onto this plane can take any shape, depending on the nature of the orbit. Following the same approach of the previous papers of this series, we characterize an orbit according to its behavior in the meridional plane. If, for example, an orbit is a rosette lying in the equatorial plane of the axisymmetric potential, it will be a linear orbit in the meridional plane, a tube orbit it will be a 2:1 orbit, etc. We should emphasize that we use the term ``box orbit" for an orbit that conserves circulation, but this refers \textbf{only} to the meridional plane. Because of the their boxlike shape in the meridional plane, such orbits were originally called ``boxes" \citep[e.g.,][]{O62}, even though their three-dimensional shapes are more similar to doughnuts \citep[see the review of][]{M99}. Nevertheless, we kept this formalism to maintain continuity with all the previous papers of this series.

\section{Numerical results - Orbit Classification}
\label{NumRes}

In this section, we will numerically integrate several sets of orbits to distinguish the regular or chaotic nature of motion. We use the initial conditions mentioned in the previous section to construct the respective grids, always adopting values inside the zero velocity curve (ZVC) defined by
\begin{equation}
\frac{1}{2} \dot{R}^2 + V_{\rm eff}(R,0) = E.
\label{zvc}
\end{equation}
In all cases, the energy was set to $600$ and the angular momentum of the orbits is $L_z = 10$. We chose, for both PH and OH models, the particular energy level, which yields $R_{\rm max} \simeq 15$ kpc, where $R_{\rm max}$ is the maximum possible value of $R$ on the $(R,\dot{R})$ phase plane. Once the values of the parameters were chosen, we computed a set of initial conditions as described in Sec. \ref{CompMeth} and integrated the corresponding orbits calculating the value of SALI and then classifying the regular orbits into different families. Each grid contains roughly a total of 20000 initial conditions $(R_0,\dot{R_0})$ of orbits with, $z_0 = 0$, while $\dot{z_0}$ is always obtained from the energy integral (Eq. \ref{ham}). In every case, we let one quantity vary in a predefined range, while fixing the values of all the other parameters, according to SPM and SOM. Color-coded grids of initial conditions $(R_0,\dot{R_0})$, equivalent to surfaces of section, that allow us to determine what types of orbits occupy specific areas in the phase plane are presented in Appendix \ref{os}.

Fig. \ref{PSSs}a depicts the $(R,\dot{R}$), $z = 0$, $\dot{z} > 0$ Poincar\'e surface of section (PSS) of the PH model. We observe that majority of the phase plane is covered by different types of regular orbits, while there is also a unified chaotic domain separating the areas of regularity. The outermost black thick curve is the ZVC. In Fig. \ref{PSSs}b, we present the phase plane of the OH model. It is evident that when the dark matter halo has an oblate shape, the observed amount of chaos is considerably higher than that in the case where we have a prolate dark halo component.

To identify all the different regular families belonging to each of the islands seen in the PSSs, we present, in Figs. \ref{Clas}a and \ref{Clas}b, grids of orbits that we have classified on the PSSs of Figs. \ref{PSSs}a and \ref{PSSs}b, respectively. In Fig. \ref{Clas}a, we distinguish eight main families of regular orbits: (i) 2:1 banana-type orbits correspond mainly to the invariant curves surrounding the central periodic point in the PSS; (ii) box orbits are situated mainly outside of the 2:1 resonant orbits; (iii) 1:1 open linear orbits form the double set of elongated islands in the PSS; (iv) 4:3 resonant orbits form the outer triple set of small islands of the PSS; (v) 6:3 resonant orbits correspond to the middle triple set of islands in the PSS; (vi) 10:5 resonant orbits form the chain of five islands of the PSS inside the 2:1 area; (vii) 12:7 resonant orbits produce the set of seven islands of invariant curves of the PSS; and (viii) other types of resonances produce extremely small islands, embedded in the chaotic layers. The outermost black thick curve is the ZVC. In the OH model shown in Fig. \ref{Clas}b the box orbits, 2:1 resonant orbits, 1:1 linear open orbits, 4:3 resonant orbits. Other types of orbits are also observed. However, two new families of orbits are introduced: (i) 3:2 resonant orbits correspond to the middle double set of islands in the PSS; and (ii) 8:5 resonant orbits form the chain of five islands of the PSS outside the box orbits. Note that every resonance $n:m$ is expressed in such a way that $m$ is equal to the total number of islands of invariant curves produced in the $(R,\dot{R})$ phase plane by the corresponding orbit.

\begin{figure*}
\centering
\resizebox{\hsize}{!}{\includegraphics{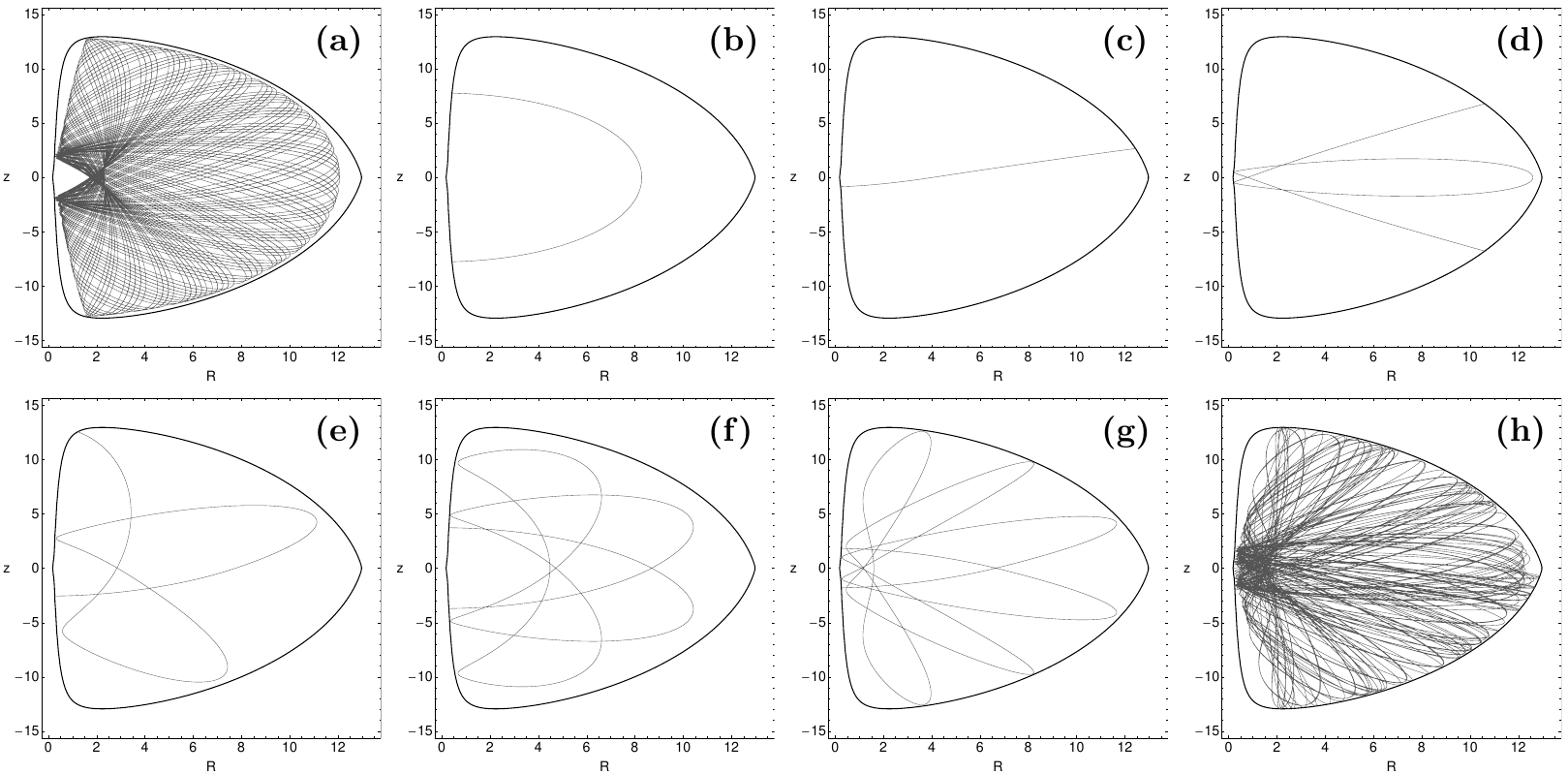}}
\caption{Orbit collection of the eight basic types in the PH galactic model: (a) box orbit; (b) 2:1 banana-type orbit; (c) 1:1 linear orbit; (d) 4:3 boxlet orbit; (e) 6:3 boxlet orbit; (f) 10:5 boxlet orbit; (g) 12:7 boxlet orbit; (h) chaotic orbit.}
\label{orbP}
\end{figure*}

\begin{figure*}
\centering
\resizebox{\hsize}{!}{\includegraphics{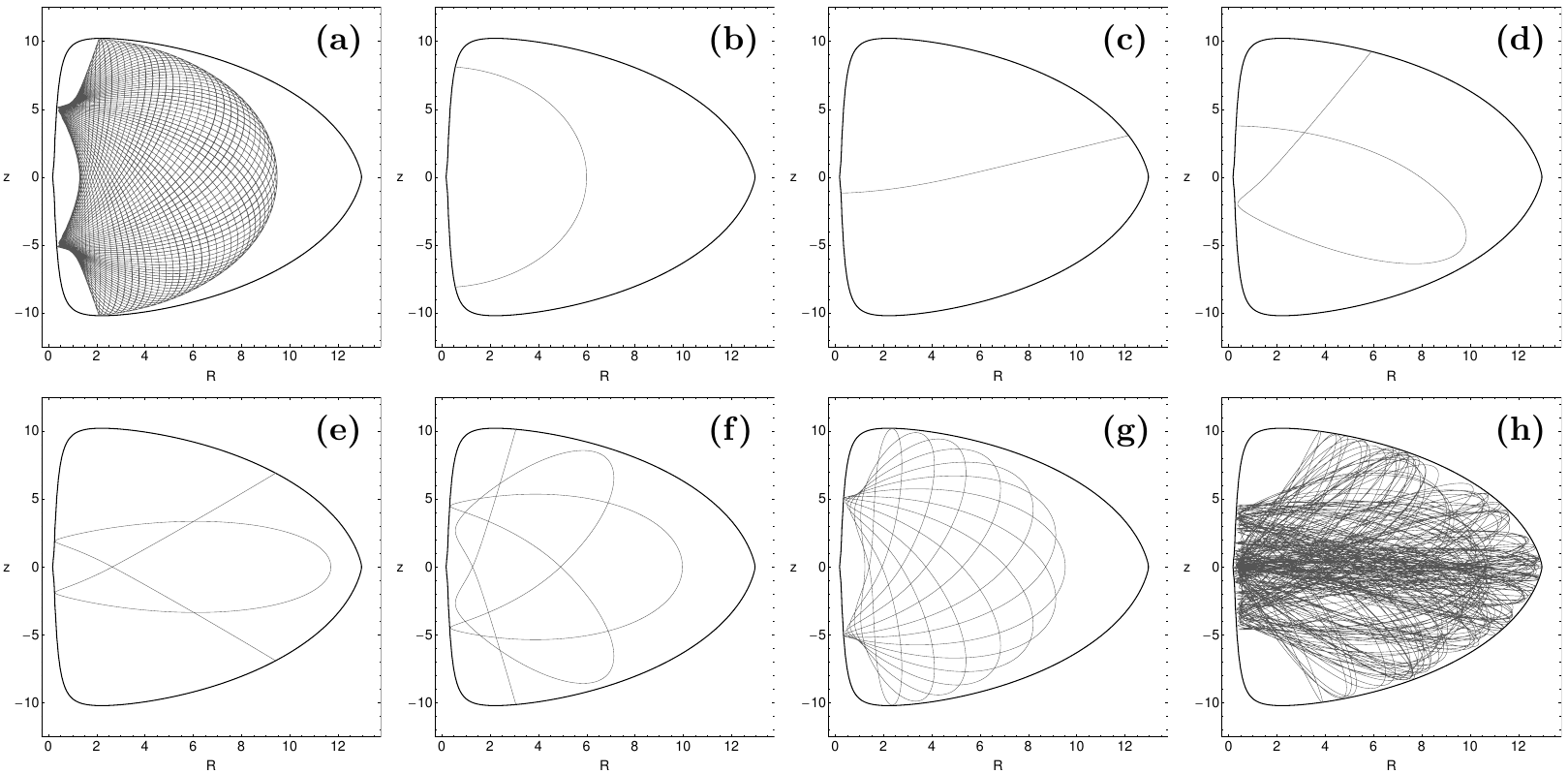}}
\caption{Orbit collection of the eight basic types in the OH galactic model: (a) box orbit; (b) 2:1 banana-type orbit; (c) 1:1 linear orbit; (d) 3:2 boxlet orbit; (e) 4:3 boxlet orbit; (f) 8:5 boxlet orbit; (g) 13:8 boxlet orbit, one of our ``orbits with higher resonance"; (h) chaotic orbit.}
\label{orbO}
\end{figure*}

The basic types of regular orbits, plus an example of a chaotic one, for both PH and OH models are shown in Fig. \ref{orbP}(a-h) and Fig. \ref{orbO}(a-h), respectively. The box and chaotic orbits were computed up to $t = 100$ time units, while all the rest of the parent\footnote{For every orbital family there is a parent (or mother) periodic orbit, that is, an orbit that describes a closed figure. Perturbing the initial conditions which define the exact position of a periodic orbit we generate quasi-periodic orbits that belong to the same orbital family and librate around their closed parent periodic orbit.} periodic orbits were computed until one period was completed. The curve circumscribing each orbit is the limiting curve in the meridional plane defined as $V_{\rm eff}(R,z) = E$. Table \ref{table1} shows the type and the initial conditions for each of the depicted orbits; for the resonant cases, the initial conditions and the period $T_{\rm per}$ correspond to the parent periodic orbit. We observe that the orbits in the PH model obtain larger values of the $z$ coordinate ($z_{max} \simeq 13$ kpc), while the orbits of the OH model shown in Fig. \ref{orbO}(a-h) stay closer to the galactic plane ($z_{max} \simeq 10$ kpc). Here we note that in the PH model two of the basic types of orbits (the 6:3 orbits and the 10:5 orbits), are in fact subfamilies of the main 2:1 family. An orbit with an improper ratio of frequencies (i.e., a ratio that is a reducible fraction) is a member of a subfamily of the orbits with a ratio of frequencies that is the irreducible corresponding fraction. For instance, a 6:3 orbit torus always surrounds a 2:1 torus; a 6:3 entire subfamily (as opposed to a single, invisible 6:3 orbit) appears when the parent 6:3 is stable and so it spawns its own subfamily. The 6:3 family has a parent closed periodic orbit the torus of which surrounds the 2:1 parent torus. All of this is beautifully seen in the PSS plot shown in Fig. \ref{PSSs}a and better viewed in the corresponding grid of Fig. \ref{Clas}a. So, in one sense, 6:3 orbits are the same as the 2:1 ones (they are members of the same family), but in another sense they are not the same (6:3 is a separated subfamily). In our research, we consider that both 6:3 and 10:5 orbits form separate families of orbits.

\begin{table}
\begin{center}
   \caption{Types and initial conditions $(R_0,\dot{R_0})$ of the orbits shown in Figs. \ref{orbP}(a-h) and \ref{orbO}(a-h). In all cases, $z_0 = 0$ and $\dot z_0$ are found from the energy integral, Eq. (\ref{ham}). $T_{\rm per}$ is the period of the parent periodic orbits.}
   \label{table1}
   \setlength{\tabcolsep}{3.0pt}
   \begin{tabular}{@{}llccc}
      \hline
      Figure & Type & $R_0$ & $\dot{R_0}$ & $T_{\rm per}$  \\
      \hline
      \ref{orbP}a &  box        & 2.34000000 & 0.00000000 &          - \\
      \ref{orbP}b &  2:1 banana & 8.28929529 & 0.00000000 & 2.19869972 \\
      \ref{orbP}c &  1:1 linear & 4.05253686 & 34.1059591 & 1.31155743 \\
      \ref{orbP}d &  4:3 boxlet & 12.6071432 & 0.00000000 & 5.29392834 \\
      \ref{orbP}e &  6:3 boxlet & 2.84576284 & 9.03961105 & 14.1859568 \\
      \ref{orbP}f & 10:5 boxlet & 4.48765950 & 0.00000000 & 11.4173302 \\
      \ref{orbP}g & 12:7 boxlet & 1.60478185 & 0.00000000 & 16.9198138 \\
      \ref{orbP}h & chaotic     & 0.18000000 & 0.00000000 &          - \\
      \ref{orbO}a &  box        & 1.30000000 & 0.00000000 &          - \\
      \ref{orbO}b &  2:1 banana & 6.00338292 & 0.00000000 & 1.93777021 \\
      \ref{orbO}c &  1:1 linear & 5.04686633 & 30.9252396 & 1.30240567 \\
      \ref{orbO}d &  3:2 boxlet & 1.50953110 & 20.2079017 & 3.75035732 \\
      \ref{orbO}e &  4:3 boxlet & 11.7079492 & 0.00000000 & 5.12892115 \\
      \ref{orbO}f &  8:5 boxlet & 9.97146568 & 0.00000000 & 9.79669202 \\
      \ref{orbO}g & 13:8 boxlet & 9.53161067 & 0.00000000 & 15.8235994 \\
      \ref{orbO}h & chaotic     & 0.18000000 & 0.00000000 &          - \\
      \hline
   \end{tabular}
\end{center}
\end{table}

Here we should note that the 1:1 resonance is usually the hallmark of the loop orbits and both coordinates oscillate with the same frequency in their main motion. Their mother orbit is a closed loop orbit. Moreover, when the oscillations are in phase, the 1:1 orbit degenerates into a linear orbit (the same as in Lissajous figures made with two oscillators). In our meridional plane, however, 1:1 orbits do not have the shape of a loop. In fact, their mother orbit is linear (as in Figs. \ref{orbP}c and \ref{orbO}c), and thus they do not have a hollow (in the meridional plane), but fill a region around the linear mother, always oscillating along the $R$ and $z$ directions with the same frequency. We designate these orbits ``1:1 linear open orbits" to differentiate them from true meridional plane loop orbits, which have a hollow and also always rotate in the same direction.

\subsection{Influence of the mass of the nucleus}

\begin{figure*}
\centering
\resizebox{\hsize}{!}{\includegraphics{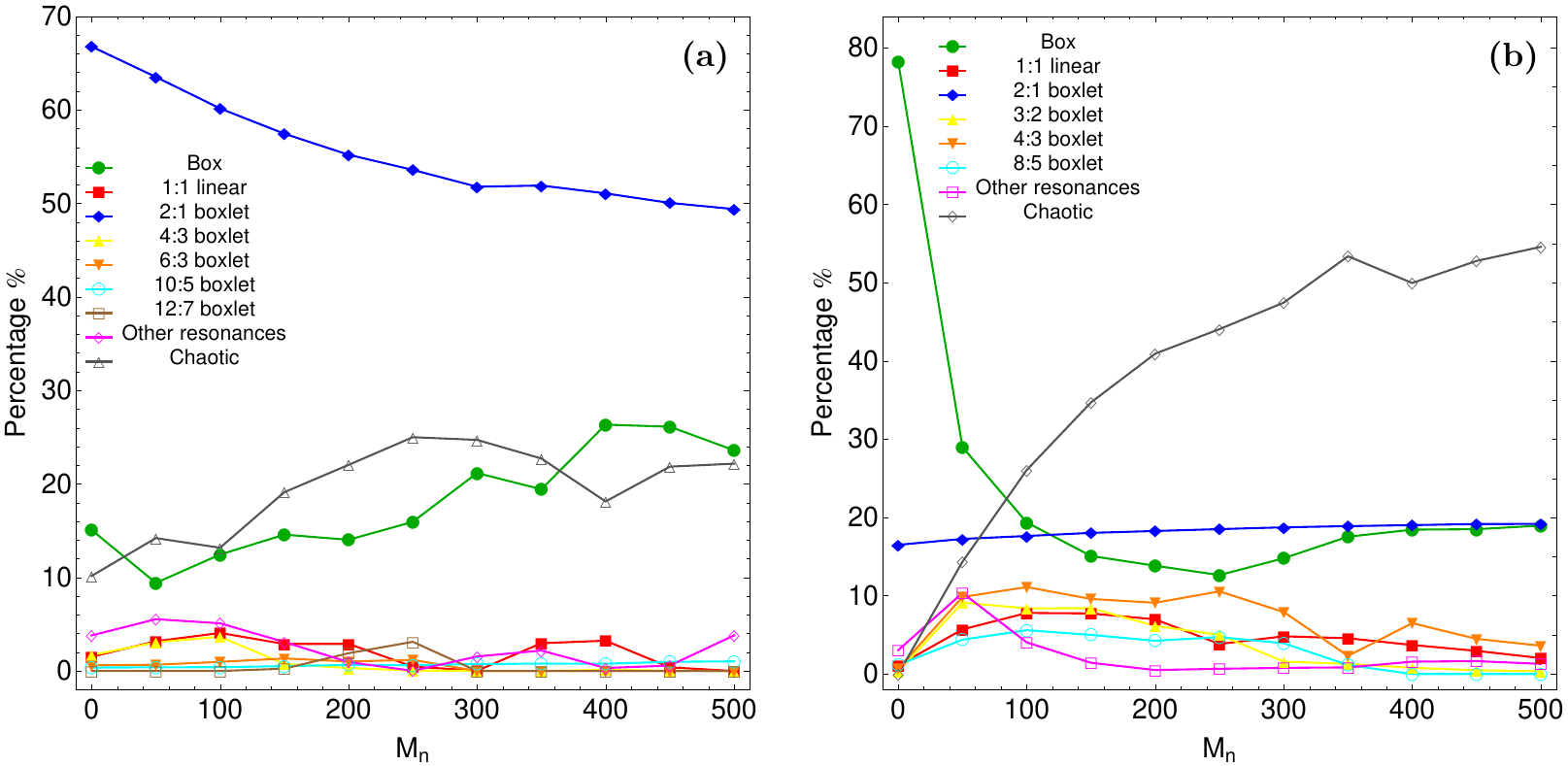}}
\caption{Evolution of the different kinds of orbits, varying $M_{\rm n}$ for (a-left): PH models and (b-right): OH models.}
\label{percMn}
\end{figure*}

To study how the mass of the nucleus $M_{\rm n}$ influences the level of chaos, we let it vary while fixing all the other parameters of our model, and integrate orbits in the meridional plane for the set $M_{\rm n} = \{0,50,100,...,500\}$. In all cases, the energy was set to $600$ and the angular momentum of the orbits $L_z = 10$. Once the values of the parameters were chosen, we computed a set of initial conditions as described in Sec. \ref{CompMeth}, and integrated the corresponding orbits computing the SALI of the orbits and then classified regular orbits into different families.

The evolution of the resulting percentages of the chaotic orbits and of the mean families of regular orbits as $M_{\rm n}$ varies for the PH model is shown in Fig. \ref{percMn}a. It can be seen that when the nucleus is absent, the amount of chaos is low and the majority of orbits is 2:1 banana-type orbits. In fact, the 2:1 orbits are reduced as the nucleus gains mass, but the 2:1 family always remains the dominant one. Moreover, the evolution of box and chaotic orbits is very similar, while all the rest of the resonant families continue to have low percentages (less that 10\%) with increasing $M_{\rm n}$. Thus, we could argue that in galaxy models with prolate dark halo, the mass of the nucleus influences mainly the box, the 2:1 banana-type, and the chaotic orbits. In Fig. \ref{percMn}b, we present a similar figure regarding the OH models. We observe that when the central nucleus is not present there is no chaos whatsoever and almost all orbits are box orbits. However, even a small nucleus is enough to trigger chaotic phenomena, whereas the box orbits are gradually depleted. This trend continues, although at a lower rate, as the nucleus grows in mass, i.e., the percentage of box orbits is reduced, and at the same time that of chaotic orbits is increased. The remaining orbits change very little; the meridional bananas, in fact, are almost unperturbed by the shifting of the mass of the nucleus. From this figure, one may conclude that $M_{\rm n}$ affects mostly the box and chaotic orbits in oblate dark halo galactic models.

\subsection{Influence of the scale length of the nucleus}

\begin{figure*}
\centering
\resizebox{\hsize}{!}{\includegraphics{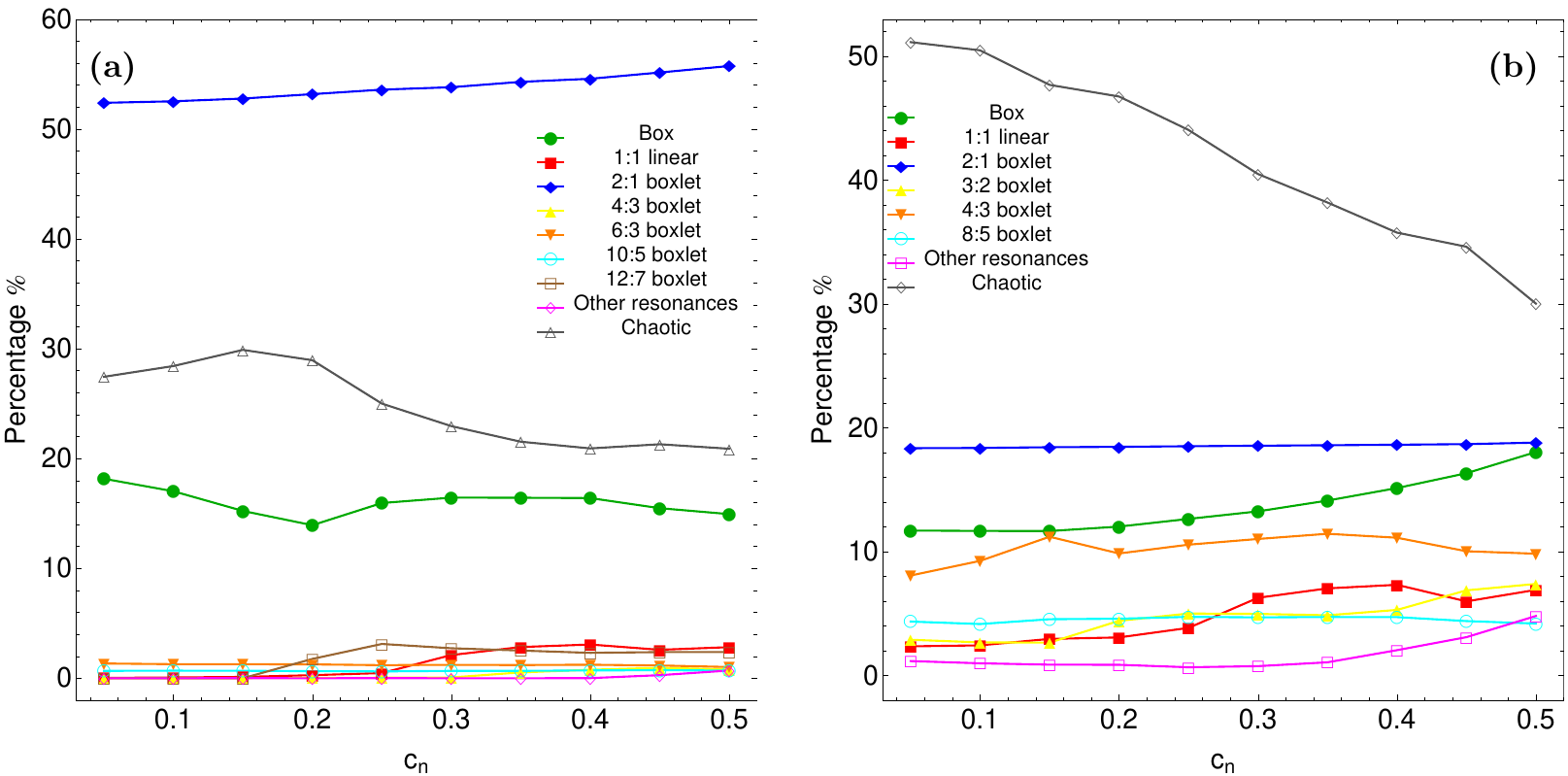}}
\caption{Evolution of the different kinds of orbits, varying $c_{\rm n}$ for (a-left): PH models and (b-right): OH models.}
\label{perccn}
\end{figure*}

Now we proceed to investigate how the scale length of the nucleus $c_{\rm n}$ influences the amount of chaos in our PH and OH models. Again, we let it vary while fixing all the other parameters of our galactic model and integrating orbits in the meridional plane for the set $c_{\rm n} = \{0.05,0.10,0.15,...,0.50\}$.

The resulting percentages of chaotic and regular orbits for the PH models as the scale length of the nucleus $c_{\rm n}$ varies are shown in Fig. \ref{perccn}a. It is evident that the 2:1 banana-type orbits exhibit an almost monotone evolution as the nucleus become less dense, but nevertheless this family always prevails over all other regular families. In contrast, box and chaotic orbits are the most affected types of orbits, especially when the nucleus is fairly concentrated ($c_{\rm n} < 0.25$). All the rest of the resonant families raise their percentages with increasing $c_{\rm n}$, but with a significant smaller rate, being always less than 5\%. Therefore, we could say that in prolate dark halo models, the box and chaotic orbits are most affected by the scale length of the nucleus. In Fig. \ref{perccn}b, we present the evolution of the percentages of orbits for the OH galaxy models. We observe that there is a strong correlation between the percentage of chaotic orbits and the value of $c_{\rm n}$. However, chaotic orbits ate always the dominant type of orbits. At the same time, as the nucleus become less concentrated, there is a gradual increase in the percentage of almost all of the regular families, most noticeably the box and the high resonant boxlets. Once again, the meridional 2:1 bananas and the 8:5 resonant orbits are immune to changes of the value of the scale length of the nucleus. Thus, decreasing the scale length of the nucleus (in other words, the nucleus becomes more concentrated and dense) in oblate dark halo models turns box and high resonant orbits into chaotic orbits, while those with low resonances are less affected.

\subsection{Influence of the mass of the disk}

\begin{figure*}
\centering
\resizebox{\hsize}{!}{\includegraphics{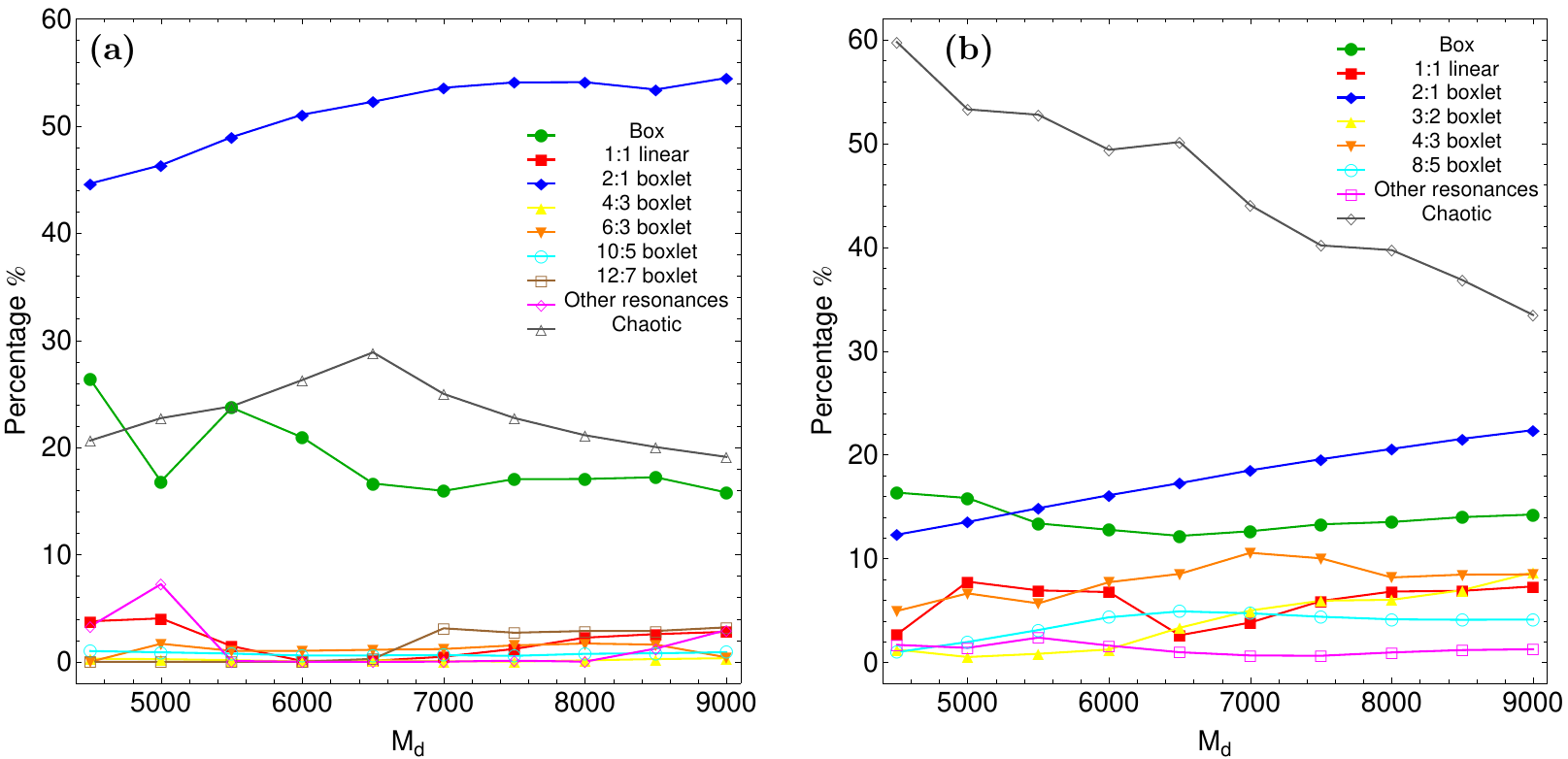}}
\caption{Evolution of the different kinds of orbits, varying $M_{\rm d}$ for (a-left): PH models and (b-right): OH models.}
\label{percMd}
\end{figure*}

Our next step is to reveal how the overall orbital structure in our PH and OH models is affected by the mass of the disk $M_{\rm d}$. As usual, we let this quantity vary while fixing the values of all the other parameters of our galactic model and integrating orbits in the meridional plane for the set $M_{\rm d} = \{4500,5000,5500,...,9000\}$.

In Fig. \ref{percMd}(a-b), we present the resulting percentages of chaotic and regular orbits for both PH, and OH models as the mass of the disk $M_{\rm d}$ varies. In Fig. \ref{percMd}a, we observe that in galaxy models with prolate dark matter haloes the 2:1 meridional bananas are always the dominant type of orbits. As the disk gains mass there is a continuous and similar variation at the percentage of box and chaotic orbits. At the higher value of the mass of the disk studied, the percentages of chaotic and box orbits tend to a common value (around 20\%), thus sharing four-tenths of the entire phase plane. Moreover, it is evident that all the other types of regular orbits remain almost unperturbed and with very low percentages throughout (less than 5\%). Therefore, one may reasonably conclude that in prolate dark matter halo models the chaotic, box and the 2:1 orbits are mostly affected by the mass of the disk. The evolution of the resulting percentages of the types of orbits in the OH models is shown in Fig. \ref{percMd}b. In the case of OH models the motion of stars is highly chaotic throughout the range of the values of $M_{\rm d}$. However, our numerical experiments suggest that the mass of disk plays an important role on the amount of chaos. In particular, the percentage of chaos decreases following an almost linear trend with increasing $M_{\rm d}$. At the same time, the 2:1 banana-type orbits exhibit a perfect linear increase, while all the other types of regular orbits change very little when $M_{\rm d}$ varies. Thus, we may say that the change in the mass of the disk in oblate dark matter halo galaxy models affects mostly chaotic and 2:1 resonant orbits.

\subsection{Influence of the core radius of the disk-halo}

\begin{figure*}
\centering
\resizebox{\hsize}{!}{\includegraphics{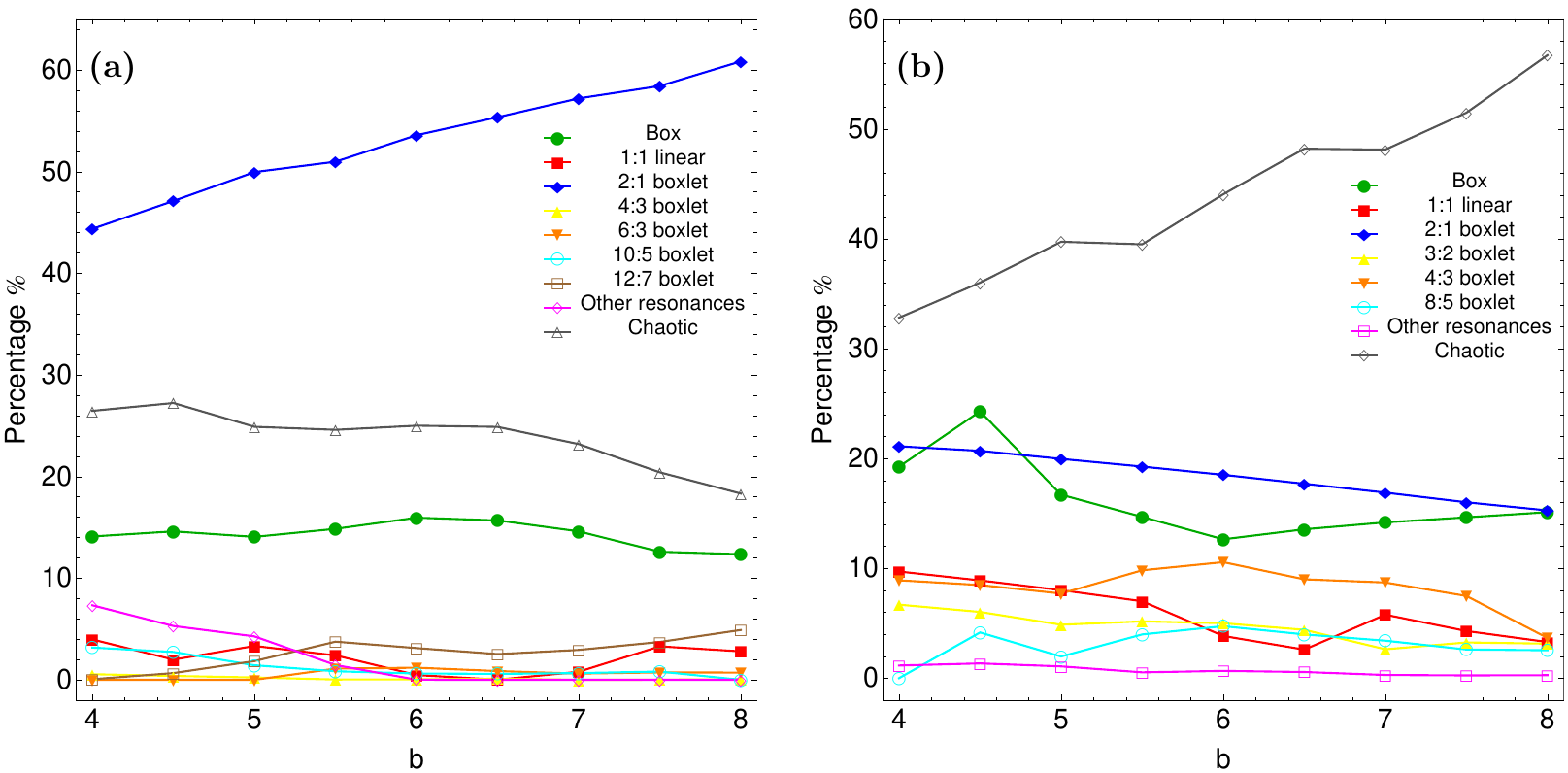}}
\caption{Evolution of the different kinds of orbits, varying $b$ for (a-left): PH models and (b-right): OH models.}
\label{percb}
\end{figure*}

The next parameter under investigation is the core radius of the disk-halo $b$. We will try to understand how the overall orbital structure in our PH and OH galaxy models is influenced by $b$. Again, we let this quantity vary while fixing the values of all the other parameters of our galactic model and integrating orbits in the meridional plane for the set $b = \{4,4.5,5,...,8\}$.

Fig. \ref{percb}(a-b) shows the resulting percentages of chaotic and regular orbits for both PH and OH galaxy models as the core radius of the disk-halo $b$ varies. Looking at Fig. \ref{percb}a, it becomes evident that once more, the meridional 2:1 banana-type orbit is the dominant type of star orbits when the dark matter halo has a prolate shape. We also observe that the core radius parameter has no influence to the box orbits whose rate always remains at about 15\%. On the other hand, we see that the chaotic orbits are affected by $b$ and their percentage exhibits a decrease as we proceed in models with larger values of the core radius parameter. All the other resonant families hold small percentages (less than 10\%) and varying the value of $b$ only shuffles the orbital content among them. Hence, we argue that in galaxy models with prolate dark matter haloes the core radius of the disk halo $b$ mainly influences the portion of chaotic and 2:1 banana-type orbits. Interestingly, the same parameter $(b)$ affects the families of orbits in the case of oblate dark matter haloes completely differently, as we can see in Fig. \ref{percb}b. Here, chaotic motion prevails throughout the range of $b$. In fact, the percentage of chaotic orbits increases constantly as the value of $b$ grows, thus suppressing the rates of almost all the regular families. It seems that only box orbits are able to sustain their percentage, which hovers around 20\%. At the same time, however, the percentages of the 2:1, 1:1, 3:2, and 4:3 resonant orbits are being reduced. Therefore, increasing the core radius of the disk-halo in OH galaxy models turns the majority of different kinds of low resonant orbits into chaotic orbits, while those with high resonances are considerably less affected.

\subsection{Influence of the scale length of the disk}

\begin{figure*}
\centering
\resizebox{\hsize}{!}{\includegraphics{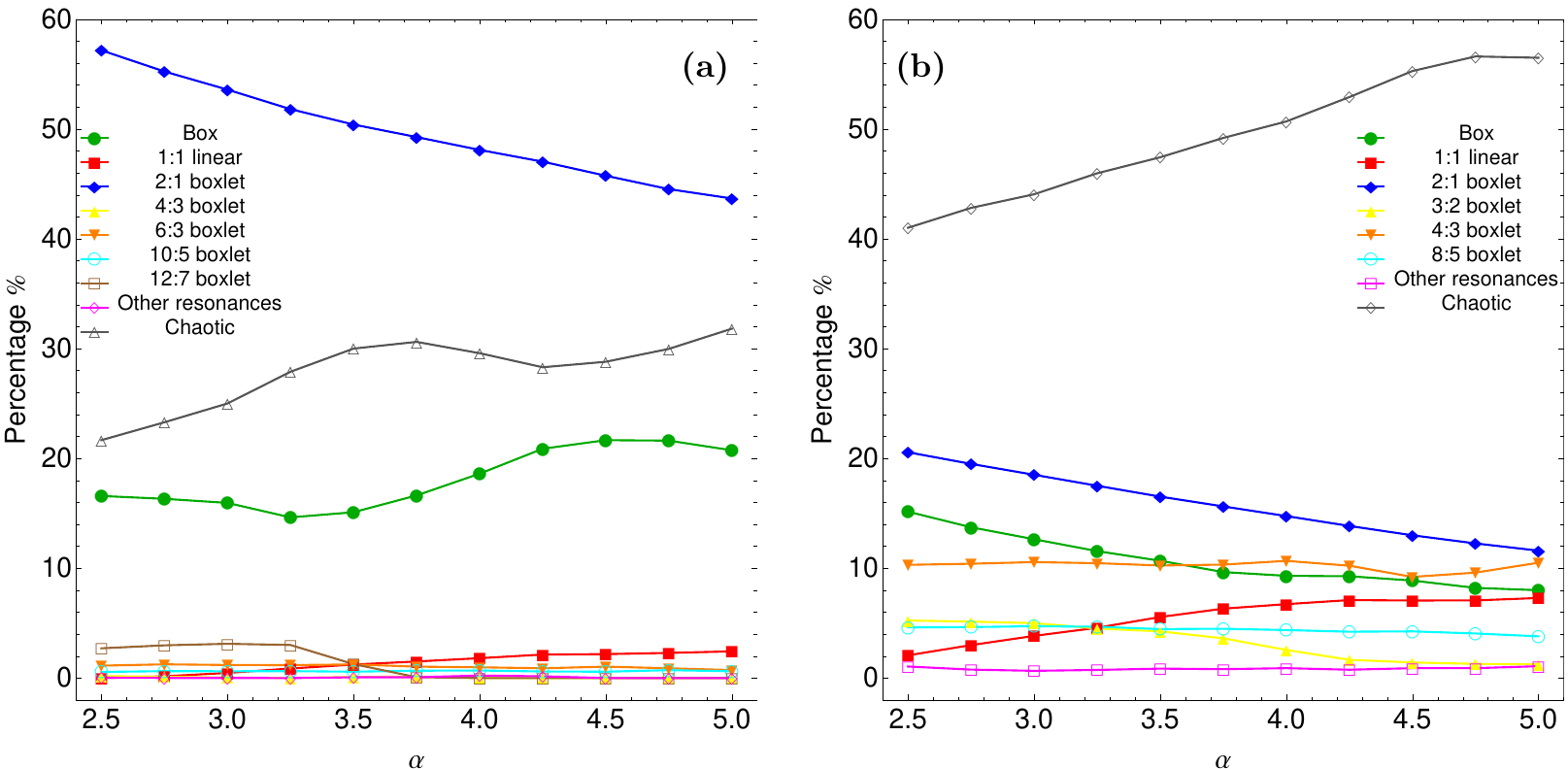}}
\caption{Evolution of the different kinds of orbits, varying $\alpha$ for (a-left): PH models and (b-right): OH models.}
\label{perca}
\end{figure*}

We continue our quest of trying to understand how the scale length of galaxy $\alpha$ influences the the overall orbital structure in our PH and OH galaxy models. As usual, we let this parameter vary while fixing the values of all the other quantities of our galactic model and integrating orbits in the meridional plane for the set $\alpha = \{2.5,2.75,3,...,5\}$.

The evolution of the resulting percentages of both the chaotic orbits and the different families of regular orbits for both the PH and OH galaxy models, as the scale length of the disk $\alpha$ varies is shown in Fig. \ref{perca}(a-b). As Fig. \ref{perca}a clearly indicates, in galaxy models with prolate dark matter haloes the motion of stars is highly regular since the 2:1 meridional banana-type orbits account for the largest proportion throughout. However, we observe that as we proceed to larger values of $\alpha$, the percentage of the 2:1 resonant orbits declines linearly, although such resonances still dominate the distribution of orbits. At the same time, the percentages of box and chaotic orbits are elevated, while all the other resonant families are only marginally affected by the change in the value of $\alpha$. Furthermore, with a much closer look at the diagram, we see that the 12:7 family disappears when $\alpha > 3.75$. Therefore, taking all the above into account, we may conclude that in galaxy models with prolate dark matter haloes the scale length of the disk halo $\alpha$ affects mainly the chaotic, box, and 2:1 banana-type orbits. In galaxy models with an oblate dark matter halo on the other hand, the vast majority of stars move in chaotic orbits. We see in Fig. \ref{perca}b that the percentages of the chaotic and the 1:1 resonant orbits increase almost linearly at the expense of box, 2:1, 3:2 orbits. Only higher resonant orbits (i.e., 4:3, 8.5) are able to maintain their rates. Thus, increasing the scale length of the disk in OH galaxy models turns different kinds of low resonant orbits either into chaotic orbits or 1:1 resonant orbits, while high resonant families are practically unaffected.

\subsection{Influence of the scale height of the disk}

\begin{figure*}
\centering
\resizebox{\hsize}{!}{\includegraphics{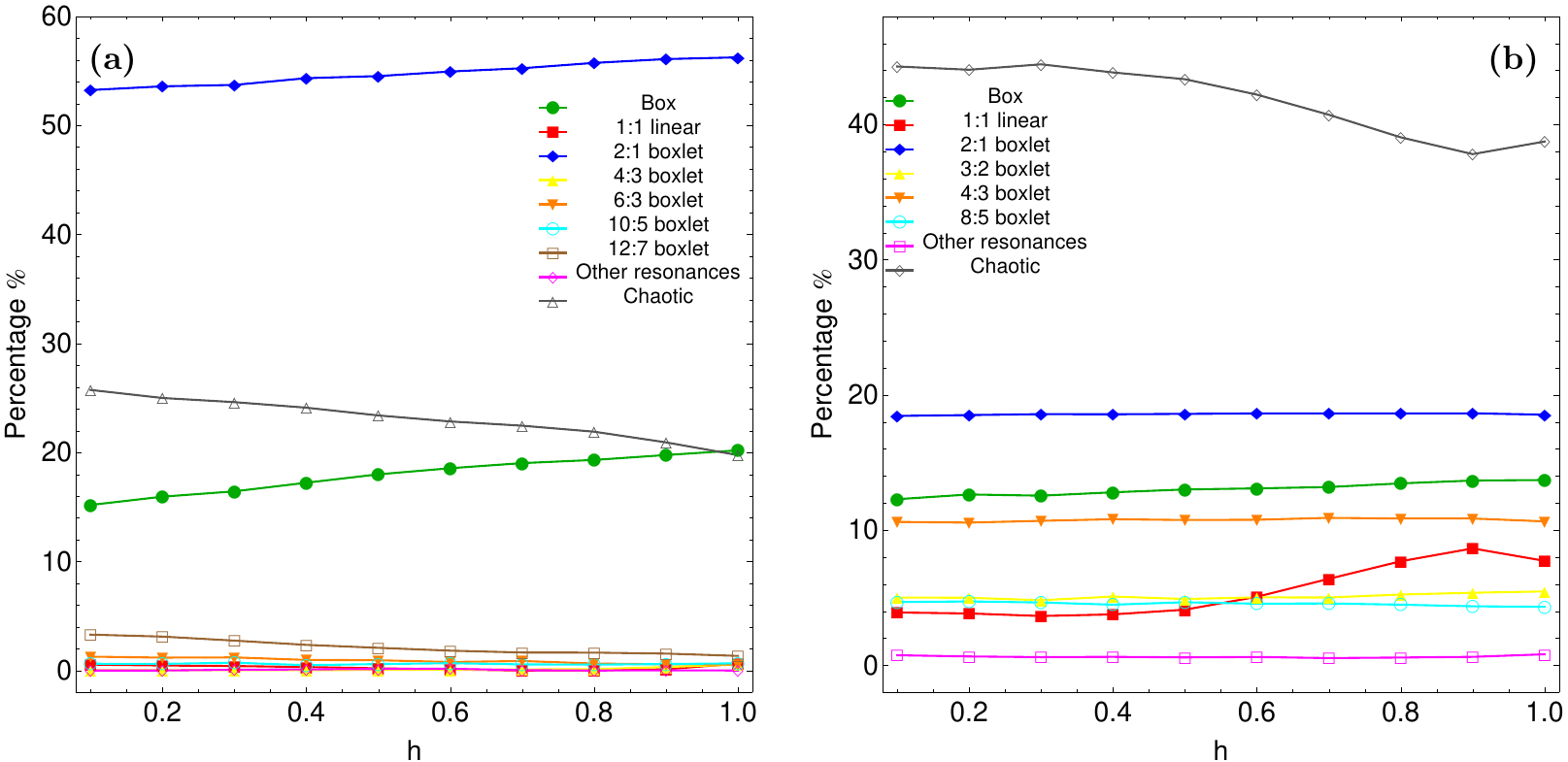}}
\caption{Evolution of the different kinds of orbits, varying $h$ for (a-left): PH models and (b-right): OH models.}
\label{perch}
\end{figure*}

The next stop of our investigation is to determine how the overall orbital structure of our PH and OH galaxy models is affected by the scale height of disk $h$. Following the usual procedure, we let this parameter vary while fixing the values of all the other quantities of our galactic models and integrating orbits in the meridional plane for the set $h = \{0.1,0.2,0.3,...,1\}$.

Fig. \ref{perch}(a-b) demonstrates the influence of the scale height of disk $h$ to the percentages both of the chaotic orbits and the different families of regular orbits for both the PH and OH galaxy models. Based on both diagrams, we observe that $h$ is the least influential parameter we encountered so far. In galaxy models with prolate dark matter haloes, only the chaotic and the box orbits are affected by the variation of the value of $h$, while the percentages of all the other families of orbits present a monotone evolution as $h$ varies. In particular, we see that as the scale height of the disk is amplified, the rates of chaotic and box orbits are reduced and increased, respectively, following a linear trend. Specifically, at the higher value of $h$ studied, the percentages of the chaotic and box orbits tend to a common value of around 20\%, thus sharing two fifths of the entire phase plane. Fig. \ref{perch}b shows the evolution of the resulting percentages of the chaotic orbits, as well as the different families of regular orbits for the OH galaxy models, as the scale height of the disk $h$ varies. Again, except for the chaotic and 1:1 resonant orbits, all the other families of orbits are completely unperturbed by $h$, exhibiting a monotone evolution. Specifically, the percentage of the 1:1 resonant orbits starts to arise as soon as the percentage of the chaotic orbits begins to deteriorate. Therefore, we conclude that in galaxy models with oblate dark matter haloes, the scale height of the disk $h$ affects only the chaotic and 1:1 resonant orbits.

\subsection{Influence of the halo flattening parameter}

\begin{figure*}
\centering
\resizebox{\hsize}{!}{\includegraphics{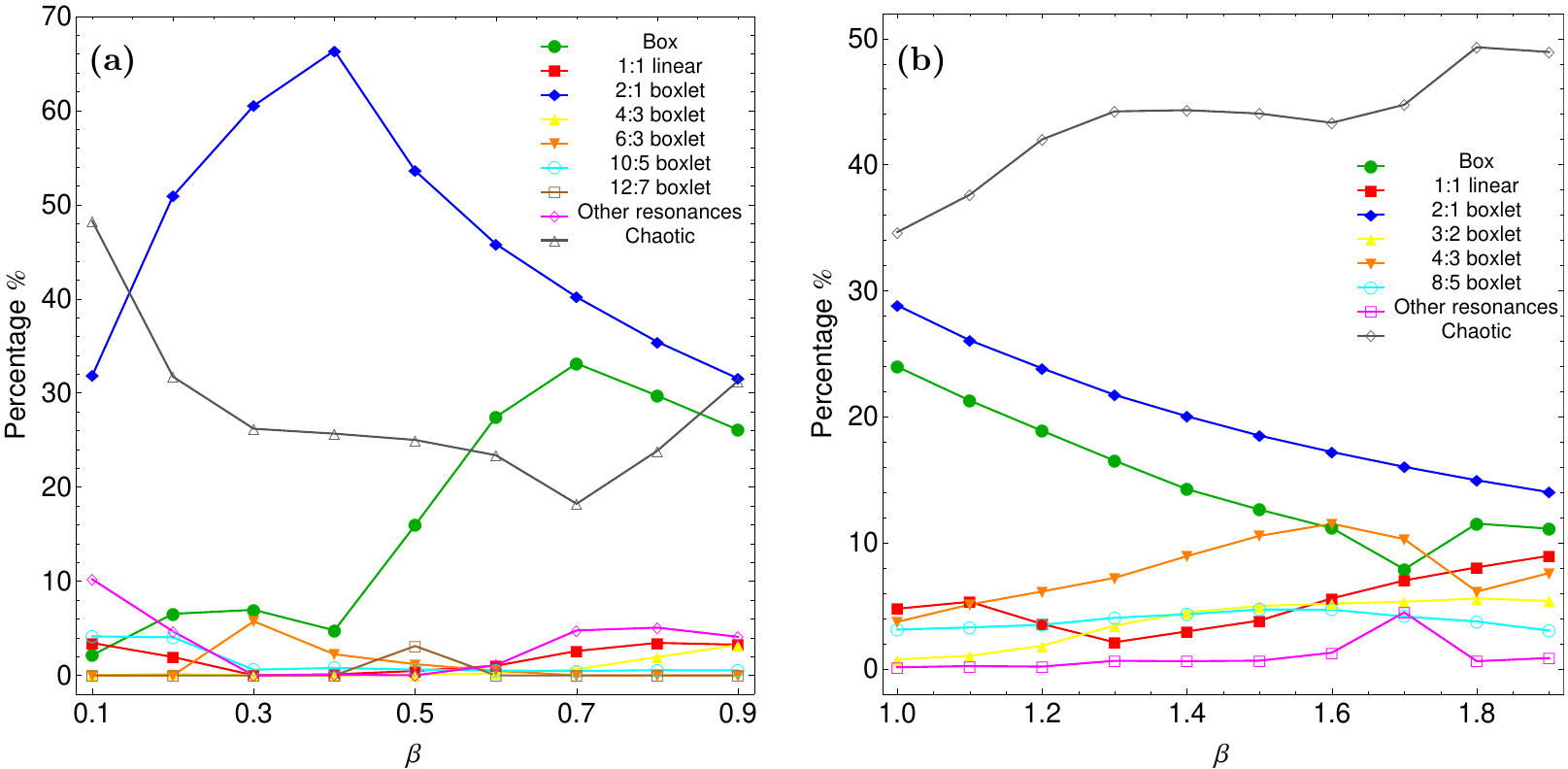}}
\caption{Evolution of the different kinds of orbits, varying $\beta$ for (a-left): PH models and (b-right): OH models.}
\label{percbeta}
\end{figure*}

The exact shape (prolate, spherical, or oblate) of the dark matter halo is determined by the flattening parameter $\beta$. So, it would be of particular interest to define how this parameter influences the overall orbital structure of our PH and OH galaxy models. Once more, we let this parameter vary while fixing the values of all the other parameters of our galactic models and integrating orbits in the meridional plane for the set $\beta = \{0.1,0.2,0.3,...,1.9\}$.

We see that the flattening parameter $\beta$ significantly influences the orbital structure in the phase plane. In Fig. \ref{percbeta}(a-b), we observe exactly how the flattening parameter affects both the percentages of the chaotic orbits and the different families of regular orbits for both PH and OH galaxy models. Here we should mention that if all the types of orbits were the same for both PH and OH models, we could merge these two plots. We demonstrate in Fig. \ref{percbeta}a that when $0.1 \leq \beta \leq 0.9$, which is a prolate dark matter halo, the majority of stars perform regular 2:1 banana-type orbits. The percentage of the 2:1 resonant orbits increases sharply for small values of $\beta$, while this tendency is reversed at higher values of the flattening parameter $(\beta > 0.4)$. The rate of the chaotic orbits is reduced to $\beta = 0.7$, while for higher values it increases. On the other hand, the percentage of box orbits grows steadily when $0.4 < \beta < 0.7$ and then begins to diminish. When $\beta = 0.9$ the rates of chaotic and 2:1 resonant orbits tend to a common value (around 30\%), thus sharing about the two thirds of the entire phase plane. The rest of the families of orbits change little, always having small percentages (less than 10\%). To summarize, in prolate dark matter halo models, the box, 2:1, and chaotic orbits are mostly affected by the flattening parameter. In Fig. \ref{percbeta}b, we see that the downward trend of the resonant 2:1 and box orbits continues to exist in the case of an oblate halo. At the same time, the percentage of the chaotic orbits keeps growing throughout the range of $\beta$. At extreme flattened oblate halo models $(\beta > 1.8)$ almost half of the phase plane is occupied by chaotic orbits. Moreover, the rates of 1:1, 3:2, and 4:3 resonant orbits rise with increasing $\beta$, while the rates of all the other regular families remain unperturbed. However, when $\beta = 1.7$, we observe a sudden increase at the percentage of higher resonances (in this case, the 13:8 resonance) accompanied by another sudden drop of the box orbits. Once more, the box, 2:1, and chaotic orbits are the types of orbits that are influenced greatly by the flattening parameter in oblate dark halo models. Here we should point out that our initial estimate suggested that the greater amount of chaos should exist at highly flattened halo models (prolate or oblate), while the lower amount should be observed when the dark matter halo is spherically symmetric $(\beta = 1)$. However, our numerical experiments only confirm the first part of our assumptions regarding the most chaotic model. Indeed, in Fig. \ref{percbeta}(a-b), the greater rate of chaos (around 50\%) occurs at the two opposite, extreme values of the flattening parameter (0.1 and 1.9). In contrast, and surprisingly enough, the smallest portion of chaotic orbits is observed when $\beta = 0.7$ thus dispelling our initial assumption.

\subsection{Influence of the scale length of the halo}

\begin{figure*}
\centering
\resizebox{\hsize}{!}{\includegraphics{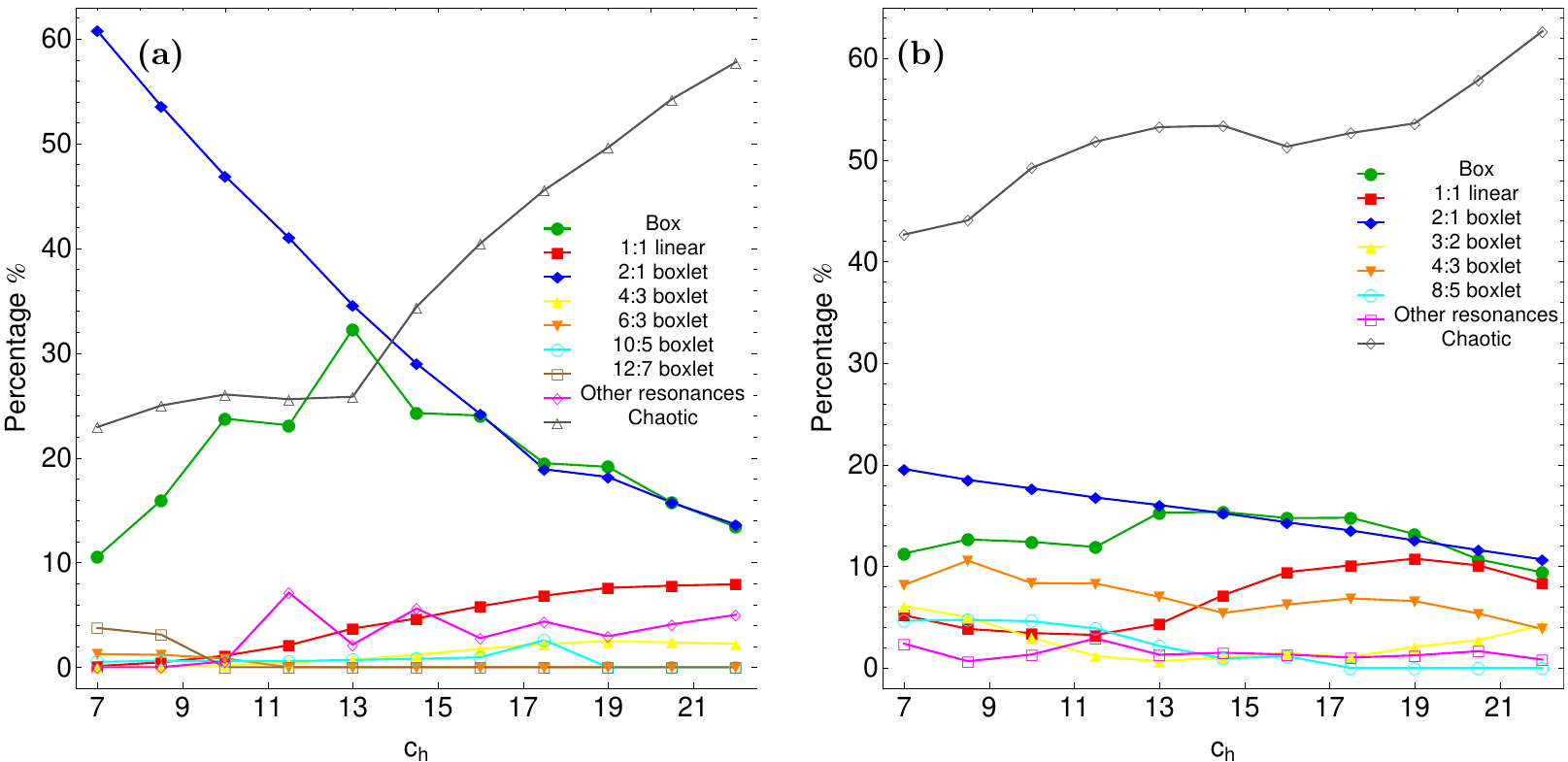}}
\caption{Evolution of the different kinds of orbits, varying $c_{\rm h}$ for (a-left): PH models and (b-right): OH models.}
\label{percch}
\end{figure*}

The concentration of the logarithmic, flattened dark matter halo is controlled by its scale length $c_{\rm h}$. In the following, we try to reveal how the scale length of the dark halo influences the overall orbital structure of our PH and OH galaxy models. As usual, we let this parameter vary while fixing the values of all the other parameters of our galactic models and integrating orbits in the meridional plane for the set $c_{\rm h} = \{7,8.5,10,...,22\}$.

The evolution of both the resulting percentages of the chaotic orbits and the different families of regular orbits for both PH and OH galaxy models, as the scale length of the halo $c_{\rm h}$ varies, is presented in Fig. \ref{percch}(a-b). In prolate dark matter halo galaxy models, Fig. \ref{percch}a shows that, as $c_{\rm h}$ increases, the percentage of the 2:1 meridional banana-type orbits decreases following almost a linear drop, while that of the chaotic orbits grows steadily when $c_{\rm h} > 13$. In fact, when $c_{\rm h} > 14$ chaotic orbits are the all-dominant types of orbits. Furthermore, the rate of box orbits increases until $c_{\rm h} = 13$, while for higher values of the scale length of the halo it decreases. On the other hand, the percentage of the 1:1 resonant family exhibits a small, but constant growth. The remaining families of orbits change insignificantly. Thus, taking all of the above into account, we conclude that in galaxy models with prolate dark matter haloes, the scale length of the halo affects mostly the box, the 2:1 banana-type, 1:1 resonant and chaotic orbits. As Fig. \ref{percch}b indicates, things are very different when the halo has an oblate shape. Here, the bulk of the stars move in chaotic orbits. Specifically, as we move on to less concentrated oblate halo models (larger value of $c_{\rm h}$) the percentage of the chaotic orbits increases, and at the higher value of the scale length of the halo studied $(c_{\rm h} = 22)$, around two thirds of the total orbits are chaotic. The meridional 2:1 banana-type orbits exhibit a linear decrease, while the percentage of the 1:1 resonant orbits grows when $c_{\rm h} > 13$. The rates of the remaining families of orbits perform small fluctuations and change little. Therefore, our calculations suggest that $c_{\rm h}$ influences mostly the chaotic, 2:1, and 1:1 resonant orbits in oblate dark halo galaxy models. In Figs. \ref{percch}(a-b), we also observe that when $c_{\rm h} = 22$ (the maximum studied value of the scale length of the dark halo), the percentages of box, 2:1, and 1:1 resonant orbits tend to a common value (around 10\%), thus sharing three tens of the entire phase plane, while the rate of the chaotic orbits is about 60\%. This evidence supports the conclusion derived from the above grids (see Figs. \ref{Gridsch}b and \ref{Gridsch}d), where we claimed that the overall orbital structure in high values of $c_{\rm h}$ is totally independent of the particular shape of the dark halo.

\subsection{Influence of the angular momentum}

\begin{figure*}
\centering
\resizebox{\hsize}{!}{\includegraphics{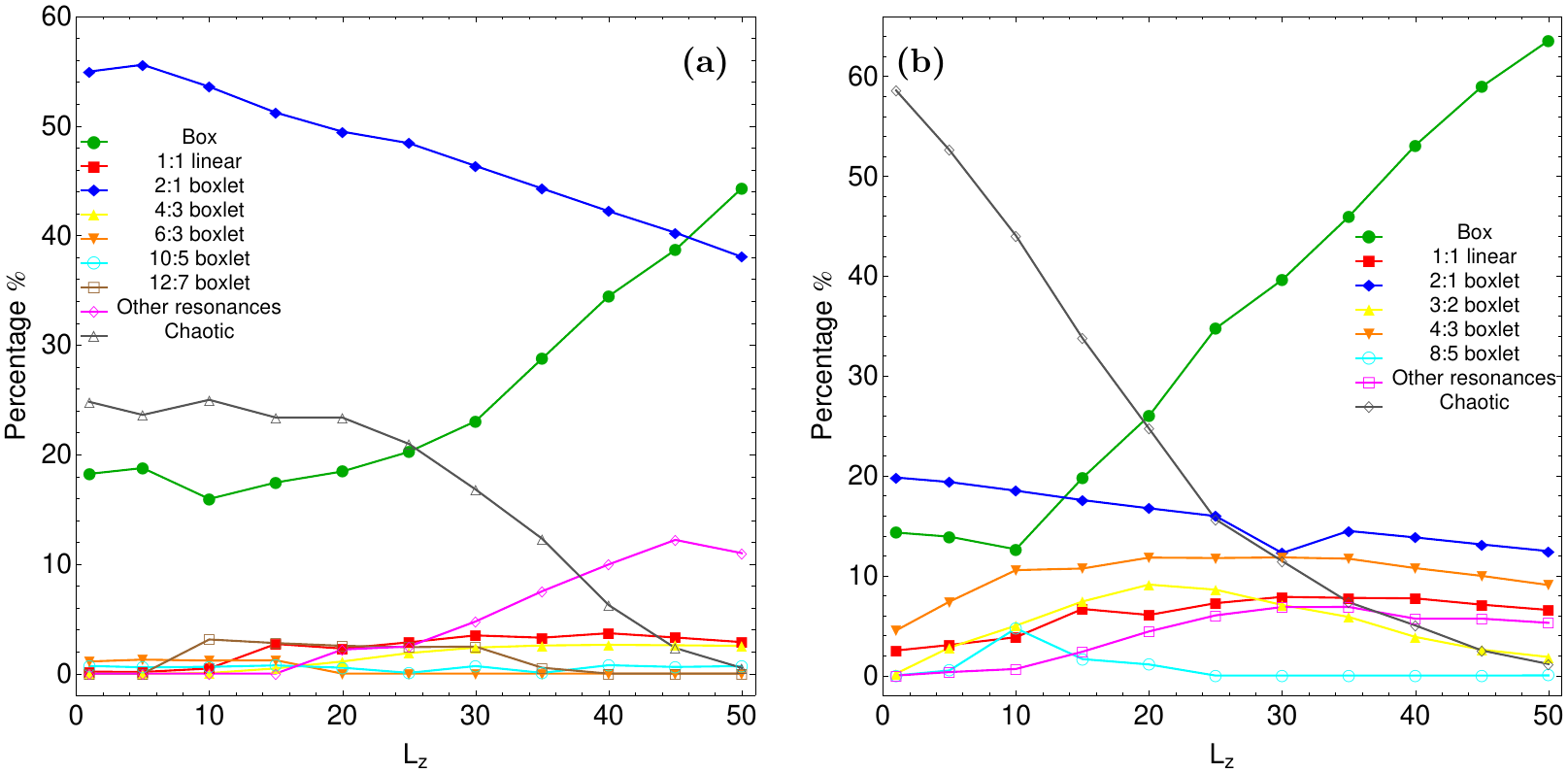}}
\caption{Evolution of the different kinds of orbits, varying $L_{\rm z}$ for (a-left): PH models and (b-right): OH models.}
\label{percLz}
\end{figure*}

One of the most important quantities, which plays a vital role in the nature of star orbits in the meridional plane $(R,z)$, is the angular momentum $L_{\rm z}$. Therefore, it is of paramount significance to investigate how the angular momentum affects the overall orbital structure of our PH and OH galaxy models. Using the same technique as in all previous cases, we let this quantity vary while fixing the values of all the other parameters of our galactic models and integrating orbits in the meridional plane for the set $L_{\rm z} = \{1,5,10,...,50\}$.

The following Fig. \ref{percLz}(a-b) shows the evolution of the resulting percentages of both the chaotic orbits and the different families of regular orbits for both PH and OH galaxy models as the value of the angular momentum $L_{\rm z}$ varies. We see in Fig. \ref{percLz}a that when the dark matter halo is prolate, the vast majority of low angular momentum stars move in regular orbits; 2:1 banana-type orbits to be exact. We note that the percentages of the 2:1 resonant and chaotic orbits is reduced as the value of the angular momentum increases. The rate of the box orbits, on the other hand, exhibits a rapid increase when $L_{\rm z} > 25$ and, in high angular momentum models, is the dominant type of orbits. We also observe that the percentage of the high resonant orbits grows significantly when $L_{\rm z} > 25$ reaches just over 10\% at the height studied value of the angular momentum $(L_{\rm z} = 50)$. The percentages of the remaining families of orbits are practically unperturbed by the shifting of the $L_{\rm z}$. To summarize, the angular momentum in prolate dark halo models affects mostly the chaotic, box, 2:1, and higher resonant orbits. Fig. \ref{percLz}b presents the evolution of the percentages of orbits, as $L_{\rm z}$ varies when the dark halo is oblate. One may observe that the percentage of chaotic orbits decreases almost linearly, while that of box orbits raises steadily when $L_{\rm z} > 10$. In particular, when $L_{\rm z} > 20$, box orbits are the dominant types of orbits. The rest of the families of orbits change less. In fact, the percentage of the meridional 2:1 banana-type orbits is minimally affected by the increase of the angular momentum, unlike in the previous case. We also note that when $L_{\rm z} = 50$ the percentages of 1:1, 4:3, and other types of resonant orbits tend to a common value of around 8\%. Thus, one may conclude that the angular momentum influences mostly box and chaotic orbits in galaxy models with an oblate dark matter halo.

\subsection{Influence of the orbital energy}

\begin{figure*}
\centering
\resizebox{\hsize}{!}{\includegraphics{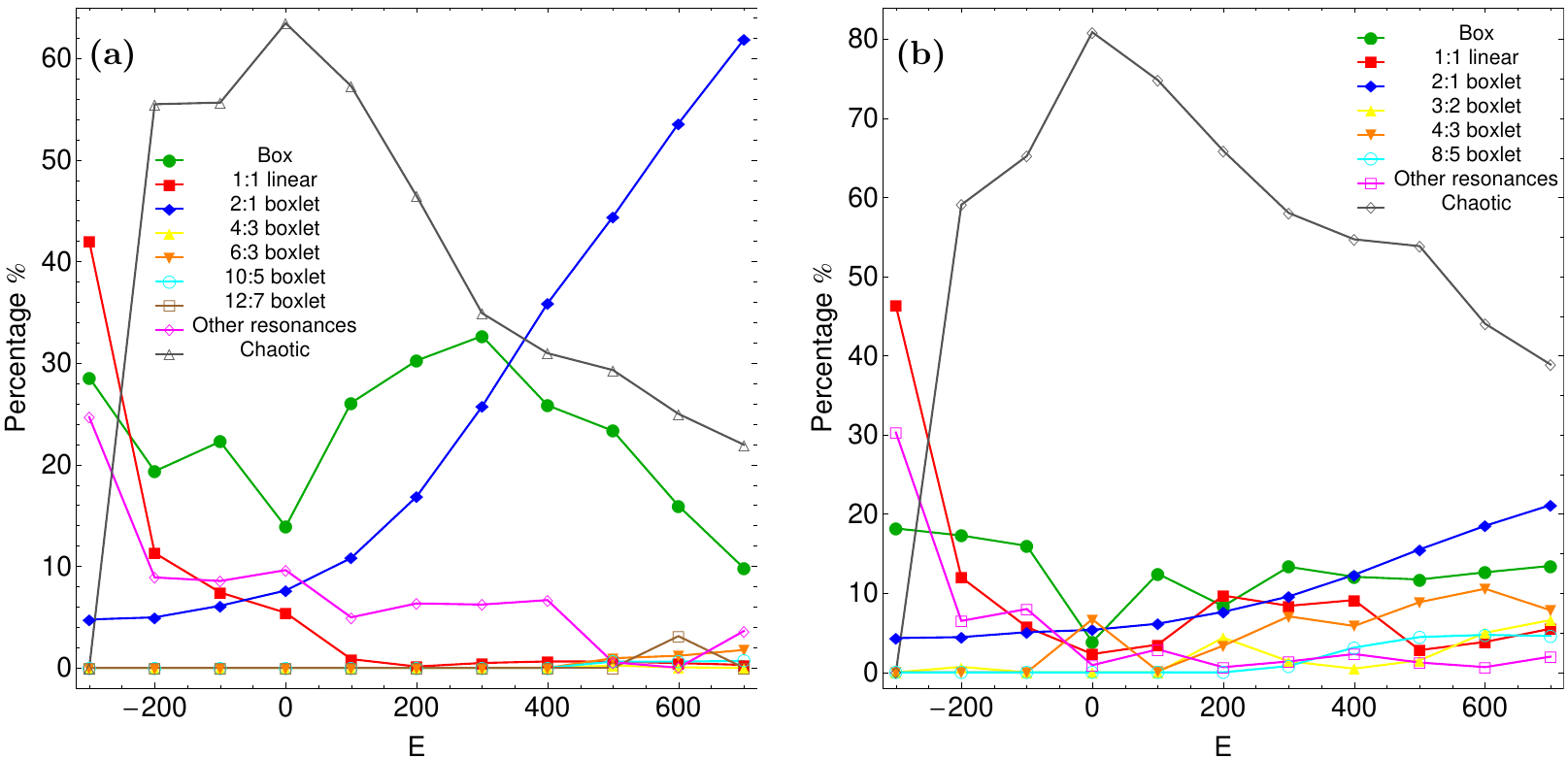}}
\caption{Evolution of the different kinds of orbits, varying $E$ for (a-left): PH models and (b-right): OH models.}
\label{percEn}
\end{figure*}

The last parameter under investigation is the total orbital energy $E$. To explore how the energy level affects the overall orbital structure of our PH and OH galaxy models, we use the normal procedure according to which we let the energy vary while fixing the values of all the other parameters of our galactic models and integrating orbits in the meridional plane for the set $E = \{-300,-200,-100,0, 100,...,700\}$. At this point we should point out that the particular value of the energy determines the maximum possible value of the $R$ coordinate $(R_{\rm max})$ on the $(R,\dot{R})$ phase plane. The energy values in the above interval result in $1.4 \lesssim R_{\rm max} \lesssim 15$.

The evolution of the resulting percentages of both the chaotic orbits and the different families of regular orbits for both PH and OH galaxy models, as the value of the orbital energy $E$ varies, is presented in Fig. \ref{percEn}(a-b). In Fig. \ref{percEn}a, we show that in low energy prolate dark halo models the motion is entirely regular. However, when $E > -300$ the percentage of chaotic orbits increases sharply and reaches its maximum value, of around 65\% when $E = 0$, while for all positive energy values it decreases almost linearly. The rate of the box orbits, on the other hand, fluctuates when $-300 < E < 300$, while for lager energies it decreases. Furthermore, we see that for low energy models, 1:1 and higher resonant orbits possess high rates, which, however, decrease when $E > 0$. At the same time, the percentage of the meridional 2:1 banana-type orbits grows rapidly for positive energy levels and when $E > 400$, the 2:1 resonant orbit is the most populated family. All the other resonant families remain completely unperturbed having infinitesimal rates throughout. Thus, one may reasonably conclude that in prolate dark halo models the energy mostly affects the chaotic, box, 1:1, 2:1, and higher resonant orbits. Fig. \ref{percEn}b shows the resulting percentages of chaotic and regular orbits for the oblate halo models when $E$ varies. We can identify many similarities regarding the evolution of the percentages between prolate and oblate dark halo models. To begin with, the rate of chaotic orbits follows a similar pattern, increasing rapidly for negative energies and maximizing its value when $E = 0$, while dropping for negative energy levels, although chaotic orbits remain by far the dominant type of orbits. Furthermore, the 1:1, 2:1, and higher resonant orbits evolve similarly. In contrast the percentages of box and 4:3 resonant orbits seem to change little, fluctuating around 15\% and 10\% respectively, while that of the 3:2 and 8:5 resonant families are very low (less than 10\%). Therefore, our numerical experiments reveal that in oblate dark halo models, varying the value of the energy mainly shuffles the orbital content among the families of regular orbits and only chaotic orbits suffer the most.

\section{Analysis of the results}
\label{Anal}

\begin{figure*}
\centering
\resizebox{\hsize}{!}{\includegraphics{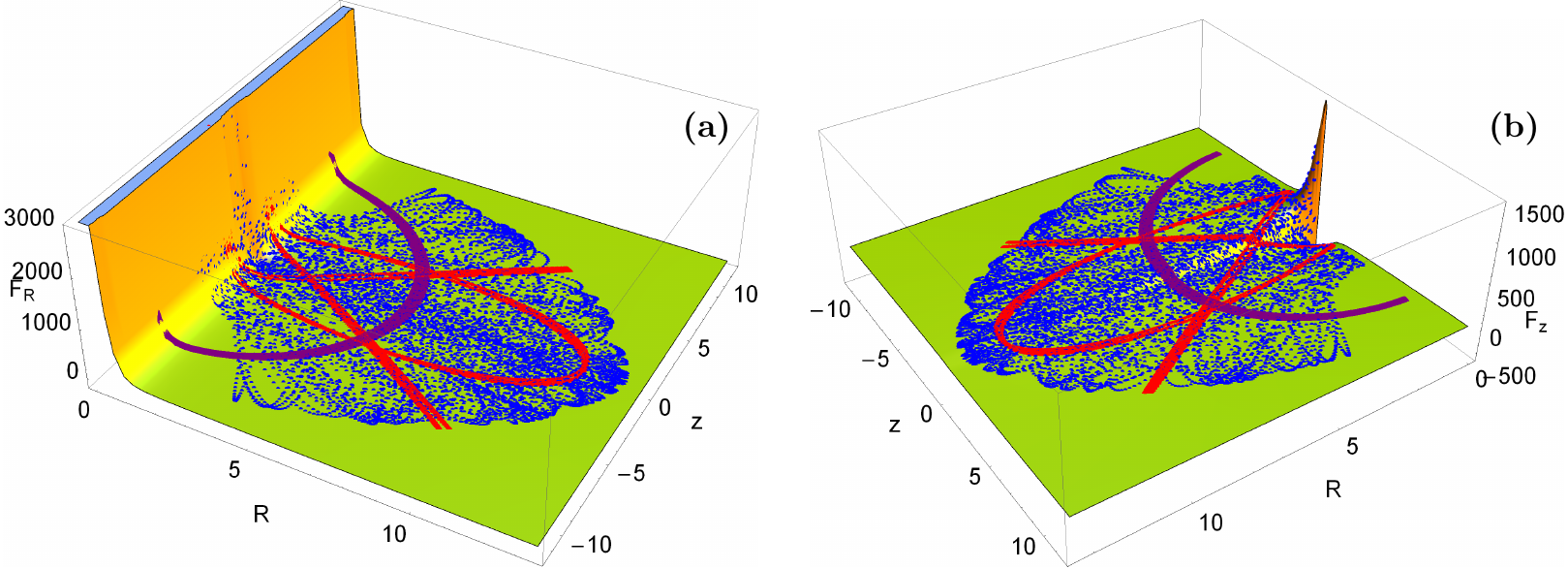}}
\caption{Forces acting at every time step of the numerical integration along the path of a regular 2:1 banana-type orbit (purple color), a 4:3 resonant orbit (red color), and a chaotic orbit (blue color) in the $M_{\rm n} = 250$ OH galaxy model. (a-left): the horizontal force $F_R$ and (b-right): the vertical force $F_z$.}
\label{FRz3D}
\end{figure*}

\begin{figure*}
\centering
\resizebox{\hsize}{!}{\includegraphics{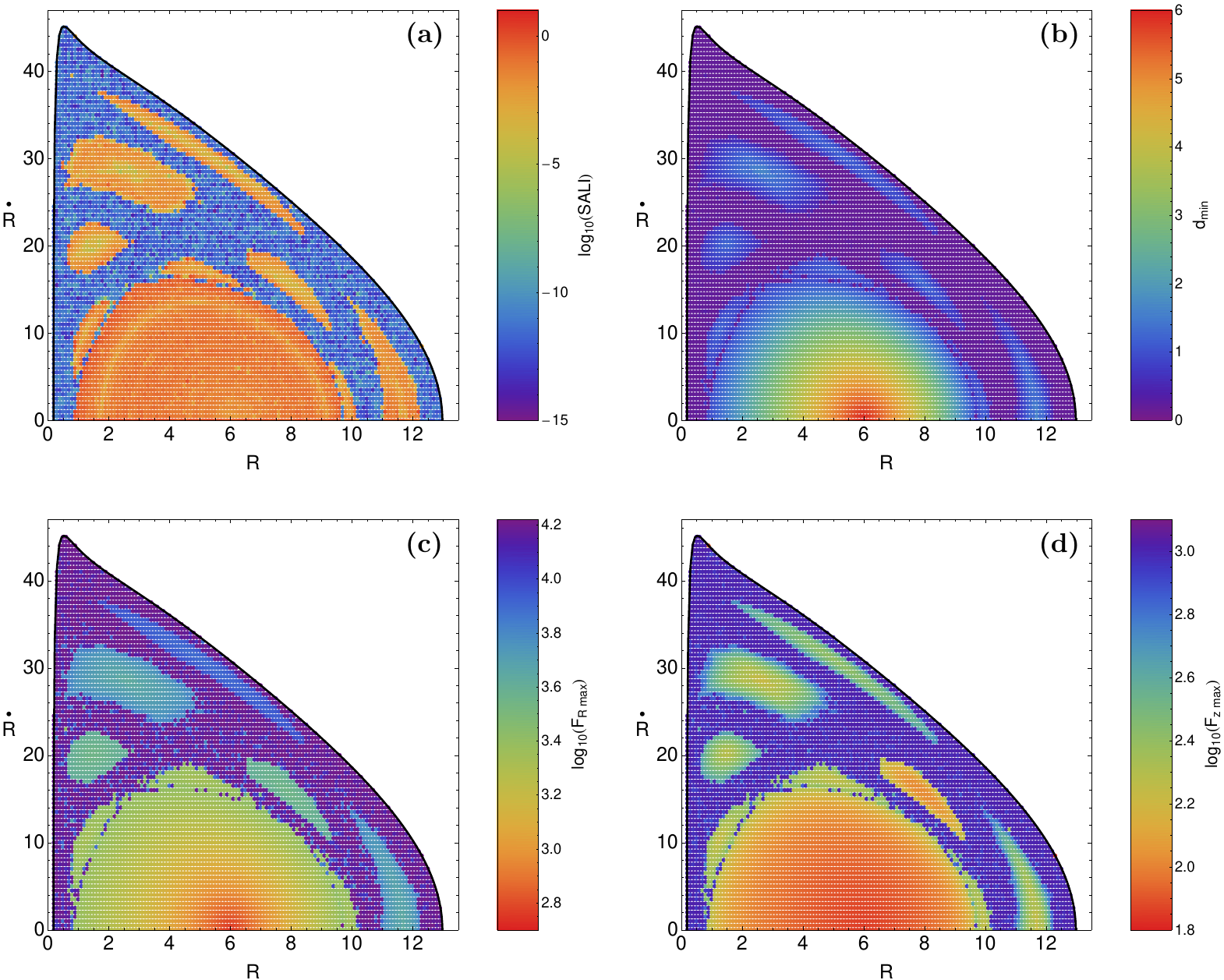}}
\caption{Grids of initial conditions $(R_0,\dot{R_0})$ at the $M_{\rm n} = 250$ OH galaxy model. Each point is colored according to its (a-upper left): SALI value, (b-upper right): minimum distance to the origin $d_{min}$, (c-lower left): maximum force along the $R$ direction $F_{Rmax}$, and (d-lower right): maximum force along the $z$ direction $F_{zmax}$.}
\label{miscGR}
\end{figure*}

\begin{figure*}
\centering
\resizebox{\hsize}{!}{\includegraphics{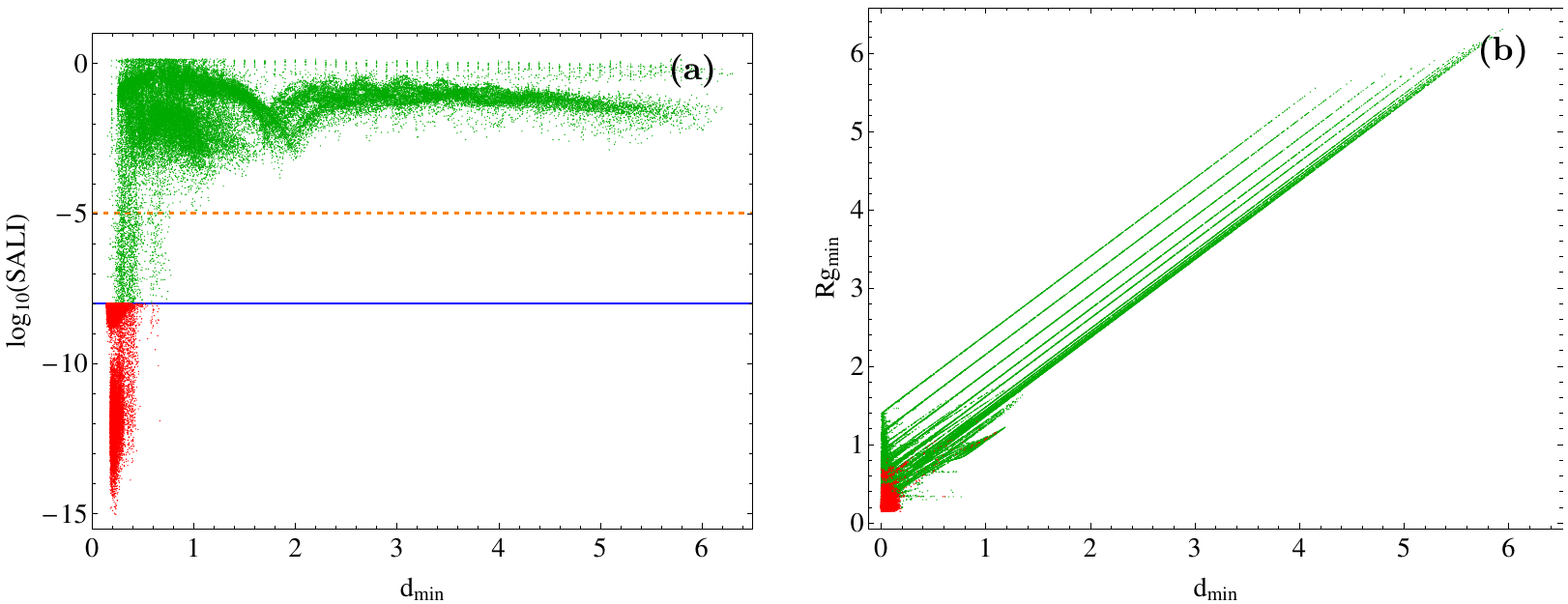}}
\caption{(a-left): Minimum distance of orbits to the origin versus SALI. The horizontal, blue line shows the limit separating ordered from chaotic orbits. All the orbits between the two horizontal lines (orange and blue) are probably sticky orbits, which require more than $10^4$ time units of integration time, so as to reveal their true chaotic nature. (b-right): Minimum distances of orbits to the origin versus minimum distances to the minimum of the effective potential, located at $(R_g,0)$. Green color corresponds to regular motion, while red corresponds to chaotic.}
\label{RminRS}
\end{figure*}

\begin{figure*}
\centering
\resizebox{\hsize}{!}{\includegraphics{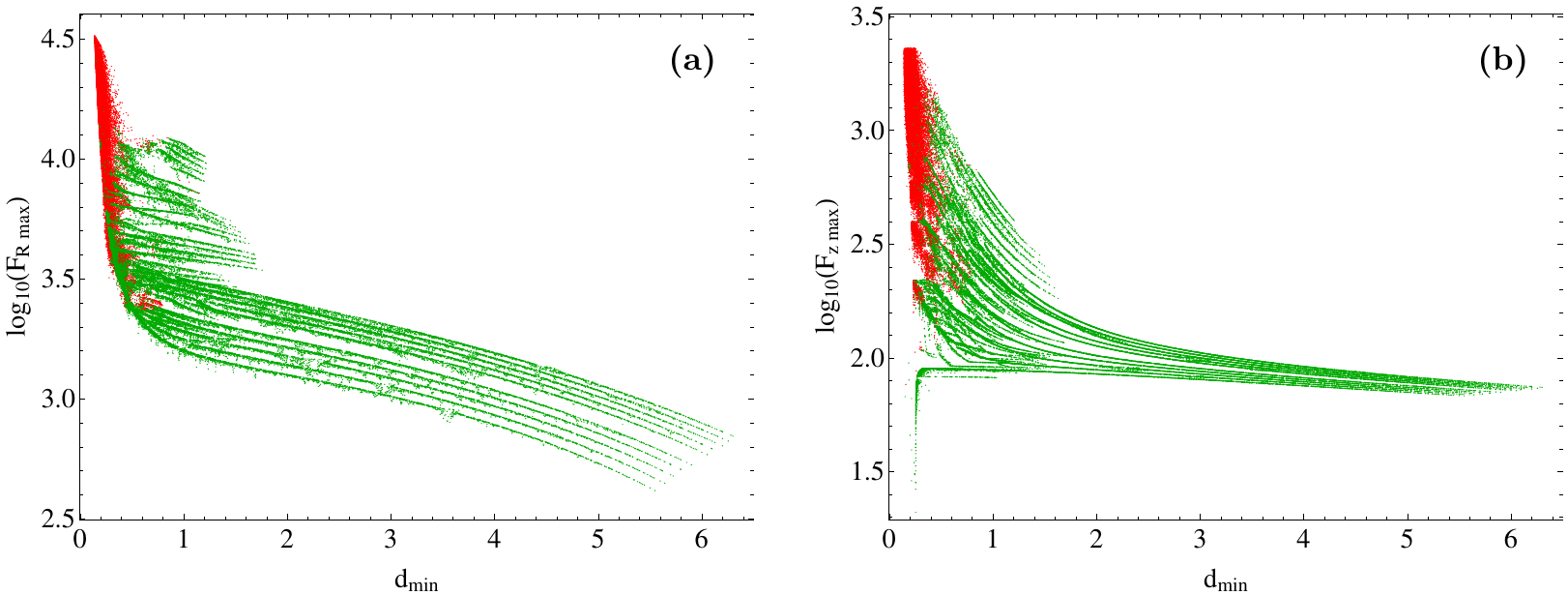}}
\caption{(a-left): Minimum distances of orbits to the origin versus (a-left): maximum force at the $R$ direction, (b-right): maximum force at the $z$ direction. Green color corresponds to regular motion, while red corresponds to chaotic.}
\label{RFmax}
\end{figure*}

The dynamical origin of the onset of chaos has proven very elusive so far. A promising line of investigation, namely the curvature of the phase space, although theoretically sound, came up against many experimental counterexamples \citep[e.g.,][]{S94}. Therefore, we will not attempt to explain which dynamical factors are responsible for the onset and growth of chaos, but we will try to isolate any behavior that may be correlated with that.

Almost 30 years ago, \citet{GB85} demonstrated that stars that pass near a density cusp, thus experiencing strong horizontal or vertical forces, may depopulate the family of box orbits that supports the triaxial figure of a galaxy. Thus, regions in which strong forces act, should be responsible for the onset of chaos. Since our potential is nowhere divergent, we do not have any cusps. Nevertheless, the centrifugal term $L_z^2/2R^2$ creates a cusp at the origin of the effective potential on the meridional plane\footnote{This statement is true only in our case when the value of the angular momentum is postulated to be constant and therefore, it should not be generalized}. Therefore, we seek to understand whether there is a relationship between chaos and a star passing near the origin. In Fig. \ref{FRz3D}(a-b), we present how the horizontal force $F_R$ and the vertical force $F_z$ act at every time step of the numerical integration on the path of two regulars orbits and one chaotic orbit. All three orbits were chosen randomly and integrated for a time interval of 500 time units, so that there is enough time for both forces to act on them. The regular 2:1 banana-type orbit shown in purple has initial conditions: $R_0 = 5.8$, $z_0 = \dot{R_0} = 0$, the 4:3 resonant orbit shown in red has initial conditions: $R_0 = 11.6$, $z_0 = \dot{R_0} = 0$, while the initial conditions of the chaotic orbit (blue) are: $R_0 = 0.18$, $z_0 = \dot{R_0} = 0$. The initial value of $\dot{z_0}$ was obtained from the energy integral (\ref{ham}) and all orbits belong to the $M_{\rm n} = 250$ OH galaxy model. At every time step of the numerical integration we recorded both forces, and their values are given in the vertical axis of the plots. It is evident that stars moving either in regular (i.e., the 4:3 resonant orbit) or chaotic orbits suffer from strong horizontal and vertical forces upon approaching the center of the galaxy. On the other hand, there are other types of regular orbits, such as the box, 2:1, and 8:5 resonances, which do not pass close to the center and therefore, the acting forces are immaterial along their entire orbital path. We found that the minimum distance of chaotic orbits to the origin $d_{min} = \sqrt{R^2 + z^2}$, is very small and in many cases it coincides with the minimum allowed $R$ value as it is defined by the corresponding ZVC. Moreover, we also observe that the maximum value of the horizontal force $F_R$ is about twice that of the vertical force $F_z$. Extensive numerical experiments indicate that the fundamental criterion that distinguishes ordered from chaotic orbits is how close to the center of the potential an orbit passes.

So far, we discussed three different quantities (the minimum distance to the origin $d_{min}$, the $F_R$, and the $F_z$ force) that somehow act differently on regular and chaotic orbits. Therefore, a question of great importance that arises is the following: can these quantities be used to safely distinguish between ordered and chaotic motion? The answer to this question is given in Fig. \ref{miscGR}(a-d). To test the efficiency of these quantities, we chose the $M_{\rm n} = 250$ OH galaxy model as a test field in our investigation. In Fig. \ref{miscGR}b, we reconstructed the grid of the initial conditions $(R_0,\dot{R_0})$ in which each point is colored according to the minimum to the origin distance $d_{min}$ of the orbits. In this plot, the reddish colors correspond to large values of $d_{min}$, while the blue/purple colors adhere to low values of $d_{min}$. The same philosophy is used in Figs. \ref{miscGR}(c-d) where each point is colored according to the maximum value of the forces $F_R$ and $F_z$ that act on the orbits. We observe that although these plots reveal the rough characteristics of the phase plane and we can indeed distinguish the different sets of islands corresponding to several resonant families. However, it is very difficult, or even impossible, to define numerical threshold values, thus separating safely between regular and chaotic motion. This is true because there will always be regular types of orbits, such as the 1:1 and 4:3 in our case, that approach very close to the center of the galaxy experiencing strong forces. On the other hand, we observe in Fig. \ref{miscGR}a the distinction between regular and chaotic motion is absolutely clear and beyond any doubt when using the SALI value as a criterion. Here we must note that in Fig. \ref{miscGR}(a-d) we actually reproduced the OH grid shown in Fig. \ref{Clas}b with four different methods.

In order to have a more complete view of the situation, we let $M_{\rm n}$ vary, and them we computed the following, for each star in our OH models: the minimum distance $d_{min}$ (in the meridional plane) to the origin of coordinates; the minimum distance to the minimum of the effective potential $Rg_{min}$; and the maximum values of the forces $F_R$ and $F_z$. Then, we tried to see if there is a correlation between these quantities. Fig. \ref{RminRS}a shows these minimum distances for all the orbits used to study the influence of the mass of the nucleus $M_{\rm n}$ of all the OH models, versus the value of their respective SALIs, while the horizontal, blue line indicates the threshold between ordered and chaotic orbits. In the same diagram, there is a dashed orange line at $\rm SALI = 10^{-5}$. We have strong numerical evidence that all the orbits between the two horizontal lines (orange and blue) are probably sticky orbits that require larger integration time so as to expose their true chaotic character. We show that \emph{all} the chaotic orbits pass near the center, thus suffering at one time or another some sudden acceleration due to the strong centrifugal force. However, this is not a sufficient condition to be chaotic: regular orbits can also pass near the center. Exactly the same behavior was found when using orbits from all the other PH and OH models analyzed in Section \ref{NumRes}. Therefore, we may draw the following conclusion: in the meridional plane of our galactic model, \emph{a necessary condition for an orbit to be chaotic is to pass near the center of the potential; a sufficient condition for an orbit to be regular is not to pass near the center of the potential}.

The analysis we conducted in the previous section revealed that almost all the studied parameters significantly influence the percentage of chaos in the meridional plane. For instance, the chaotic percentage grows with the increment of the mass of the nucleus (see Fig. \ref{percMn}), the decrement of the scale length of it (see Fig. \ref{perccn}), the increment of the scale length of the disk (see Fig. \ref{perca}), the increment of the scale length of the halo (see Fig. \ref{percch}), the decrement of the angular momentum (see Fig. \ref{percLz}), etc. Our numerical calculations indicate that in all the cases the position $R_g$ of the minimum of the effective potential, which is always located on the $R$ axis \citep[e.g.,][]{BT08}, nears the origin of coordinates whenever the percentage of chaos rises. Fig. \ref{RminRS}b shows the minimum distance to the origin for all the orbits shown in Fig. \ref{RminRS}a, versus their minimum distances to $R_g$. We can see a consistent correlation between those quantities, hinting that the position of the minimum of the effective potential might influence the degree of chaos, although we were not be able to find an analytic proof of this. Green dots correspond to regular orbits, while red dots correspond to chaotic orbits.

In Figs. \ref{RFmax}(a-b), we present diagrams connecting the maximum value of $F_R$ and $F_z$ forces respectively with the corresponding minimum distance of the orbits. In all cases, we observe a common behavior: the value of $d_{min}$ of the chaotic orbits is \emph{always} low ($\lesssim$ 1 kpc) exhibiting at the same time high maximum values of the forces. The picture described above is consistent with the rising of the percentage of the chaotic orbits with $M_{\rm n}$ (see Fig. \ref{percMn}b), since both forces grow with the mass of the nucleus. Also, it is consistent with the behavior of the chaotic percentage seen in Fig. \ref{perccn}b, considering that the more concentrated the nucleus, the more acceleration it causes near the center. It also explains why the percentage of chaotic motion diminishes when the angular momentum increases (Fig. \ref{percLz}b), given that a low angular momentum allows the star to approach the center of the potential. On the other hand, Fig. \ref{percMn}b shows that when the nucleus is absent, there is no chaotic motion at all. Whereas this proves that the onset of chaos is driven mainly by the presence of dark matter, it also poses a question about the above-mentioned role of the centrifugal force, since this force is at work even in this completely regular case.

\section{Discussion}
\label{Disc}

In the present work, we used an analytic, axially symmetric galactic gravitational model that embraces the general features of a disk galaxy with a dense, massive nucleus and a biaxial prolate or oblate dark matter halo component. To simplify our study, we chose to work in the meridional plane $(R,z)$, thus reducing three-dimensional to two-dimensional motion. Varying the values of all the involved parameters of the dynamical system, as well as the two global isolating integrals of the orbits, namely the angular momentum and the energy, we found that the level of chaos and the distribution in regular families is indeed very dependent on all of these parameters. Here we must point out that the present article belongs to a series of papers \citep{ZC13,CZ13,ZCar13,ZCar14} that have as their main objective the orbit classification (not only regular versus chaotic, but also separating regular orbits into different regular families) in different galactic gravitational potentials. Thus, we decided to follow a similar structure and the same numerical approach in all of them.

We found that in our PH and OH galaxy models several types of regular orbits exist, while there is also a unified chaotic domain separating the areas of regularity. In particular, most types of regular orbits, such as the box, 1:1, 2:1, 4:3, higher resonant, and chaotic orbits are common in both prolate and oblate dark halo models. Here we must clarify that by the term ``higher resonant orbits" we refer to resonant orbits with a rational quotient of frequencies made from integers $> 5$, which of course do not belong to the main families. However, each model type (PH or OH) has its own private families of regular orbits. Specifically, in prolate halo models we encountered subfamilies of the basic 2:1 family (i.e., the 6:3 and the 10:5 family) and resonant orbits of higher multiplicity, such as the 12:7 family. In the oblate halo models, on the other hand, the special families of orbits were the 3:2 and the 8:5. Our numerical calculations indicate that when the dark matter halo has a prolate shape with the vast majority of stars move in regular orbits, the 2:1 resonant family being the most populated one, while in the case of oblate dark halo chaos prevails, thus chaotic orbits are the all-dominant type.

Our investigation reveals that all the parameters of the dynamical system affect, more or less, the overall orbital structure of our PH and OH models. It was observed that the mass of the nucleus, the halo flattening parameter, the scale length of the halo, the angular momentum and the orbital energy are the most influential quantities, while the effect of all the other parameters is much weaker. In fact, we should point out that all the parameters corresponding to the disk potential have a minor influence on the nature of orbits, therefore we can conclude that of the three components of the model, the disk is the one that causes the least disturbance to the percentages of the orbits. In more precise terms, the influence of all the parameters can be summarized as follows:
\begin{enumerate}
 \item In PH galaxy models, the mass of the nucleus $M_{\rm n}$ mainly influences the box, the 2:1 banana-type, and the chaotic orbits. However, in OH models the mass of the nucleus although spherically symmetric and, therefore, maintaining the axial symmetry of the whole galaxy, generates chaos in the meridional plane as soon as it is above zero. As the mass increases, this chaotic motion grows in percentage at the expense of the box orbits. The chaotic percentage approached $\simeq 55\%$ once $M_{\rm n}$ has reached some $\simeq 7\%$ of the $M_{\rm d}$.
 \item The box and the chaotic orbits are the types of orbits mostly influenced by the concentration of the nucleus $c_{\rm n}$ in PH models. Moreover, in OH models, the percentage of chaotic motion depends almost linearly on this parameter. Once more, box orbits and high resonant orbits are the ones that give way to the chaotic orbits, while those with low resonances are less affected. The amount of chaos decreases as we proceed to OH models with less concentrated nuclei, corresponding between about $30\% - 50\%$ of the total orbits.
 \item The mass of the disk $M_{\rm d}$ mostly affects the percentages of the chaotic, box, and the 2:1 banana-type orbits in PH models, while in OH models only the chaotic and 2:1 resonant orbits are influenced by $M_{\rm d}$. In fact, the rate of chaotic orbits decreases linearly, while that of 2:1 resonant orbits increases linearly with increasing $M_{\rm d}$. In OH models, $M_{\rm d}$ acts as chaos regulator; the more massive is the disk, the less chaos is observed.
 \item The core radius of the disk-halo $b$ plays a similar role in the case of a prolate dark matter halo: the percentage of 2:1 resonant orbits depends almost linearly on this parameter. On the other hand, increasing the core radius of the disk-halo in OH galaxy models turns the majority of different types of low resonant orbits into chaotic orbits, while higher resonant orbits are considerably less affected. In all cases, chaotic orbits occupy between around one and two thirds of the total orbits.
 \item In galaxy models with prolate dark matter haloes, the scale length of the disk $\alpha$ affects mainly the chaotic, box, and 2:1 banana-type orbits. In contrast, increasing the scale length of the disk in OH galaxy models turns different kinds of low resonant orbits either into chaotic orbits or 1:1 resonant orbits, while high resonant families are practically unaffected. In the case of an oblate dark halo, the percentage of chaotic orbits seems to saturate at $\simeq 55\%$ of the orbits, when $\alpha > 4.5$.
 \item The scale height of disk $h$ is the least influential parameter of all that were studied. Nevertheless, in PH galaxy models only the chaotic and box orbits are affected by the variation of the value of $h$ (sharing about $20\%$ of the total orbits when $h = 1$), while the percentages of all the other families of orbits present a monotone evolution as $h$ varies. Moreover, chaotic and 1:1 resonant orbits are the orbits mostly influenced by $h$ in OH models.
 \item One of the most influential quantities is the halo flattening parameter $\beta$. In both PH and OH models the percentages of box, 2:1 resonant, and chaotic orbits are those mostly affected. The maximum percentage of regular orbits (around 65\%) was observed when $\beta = 0.4$, while that of the chaotic orbits (around 50\%) occurs at the extreme opposite values of $\beta$, that is when $\beta = 0.1$ (highly prolate halo), and $\beta = 1.9$ (highly oblate halo). Interestingly enough, the minimum amount of chaos (around 20\%) takes place at $\beta = 0.7$, thus rejecting our initial assumption according to which a spherical $(\beta = 1)$ dark matter halo should be the least chaotic.
 \item The scale length of the dark halo $c_{\rm h}$ is another parameter with a vital role on the orbital structure. We found that the percentage of chaos quickly grows as $c_{\rm h}$ increases, by collapsing the rates of regular orbits. In particular, the box, the 2:1 banana-type and 1:1 resonant orbits suffer the most in both PH and OH models. It was also observed, that at high values of $c_{\rm h}$, when the dark halo is much less concentrated, chaotic orbits is the most populated family holding high rates (around 60\%) and the overall orbital structure of the phase plane is the same, regardless of the particular type (prolate or oblate) of the dark halo.
 \item Our experiments suggest that the angular momentum of the orbits $L_{\rm z}$ greatly influences the level of chaos: orbits with low angular momenta have higher chances of being chaotic, especially in OH models, than those with high values of angular momentum. The relationship between chaos and angular momentum is close to linear in both cases and again, box and 2:1 banana-type orbits are the most affected by the percentage of chaos. At extreme high values of the angular momentum $L_{\rm z} \simeq 50$, box orbits are the all-dominant type of orbits possessing rates around 45\% and 65\% in PH and OH models, respectively.
 \item In models with very low negative values of the orbital energy $E$ the motion of stars is almost entirely regular. However, the percentage of chaotic orbits grows rapidly with increasing energy showing the maximum value when $E = 0$, while for all positive energies it decreases following an almost linear trend. In prolate dark halo models, the energy mostly affects the chaotic, box, 1:1, 2:1, and higher resonant orbits. When $E > 350$, 2:1 banana-type orbits take the field heavily increasing their rates covering about the two thirds of the phase plane. In contrast, in OH models, varying the value of the energy mainly shuffles the orbital content among the families of regular orbits and only chaotic orbits suffer the most, however, being the most populated family.
\end{enumerate}

When we take the fact that the effective potential in the meridional plane has a cusp caused by the centrifugal acceleration into account, all these behaviors turn out to be consistent with the analysis we made in Section \ref{Anal}, where we arrived at the conclusion that a necessary condition for an orbit to be chaotic is to pass near the center of the potential, while a sufficient condition for an orbit to be regular is \textbf{not} to pass near the center of the potential. In the same vein, we tested some dynamical quantities such as the minimum distance to the origin and the horizontal and vertical forces acting on stars, to determine if they can be used as chaos detectors. Unfortunately, our numerical experiments indicate that for all these quantities, even though they provide some general results, we cannot establish numerical threshold values that allow us to distinguish safely between ordered and chaotic motion. However, if we combine in pairs the outcomes of these quantities, we can extract useful information regarding the dynamical origin of the onset and the degree of chaos in our galaxy models.

We consider the results of the present research as an initial effort and also a promising step in the task of exploring the orbital structure of disk galaxies with dark matter haloes. Taking our encouraging outcomes into account, it is in our future plans to modify our dynamical model properly to expand our investigation into three dimensions. This will allow us to unveil how the parameters of the system influence the nature of three-dimensional orbits. Also, we would be particularly interested in obtaining the entire network of periodic orbits, revealing the evolution of the periodic points as well as their stability when varying the different parameters of our model.

\section*{Acknowledgments}

I would like to express my warmest thanks to Dr. D.D. Carpintero for all the illuminating and creative discussions during this research and also for his substantial contribution to our efforts to refine and improve further the orbit classification code. My thanks also go to the anonymous referee for the careful reading of the manuscript and for all the apt suggestions and comments that allowed us to improve both the quality and the clarity of the paper.

\Online

\begin{appendix}
\section{Orbital structure}
\label{os}

To show how the dynamical parameters of our galactic model influence the orbital structure of the system, we present for each case, color-coded grids of initial conditions $(R_0,\dot{R_0})$, equivalent to surfaces of section, which allow us to visualize what types of orbits occupy specific areas in the phase-space.

\begin{figure*}
\centering
\resizebox{\hsize}{!}{\includegraphics{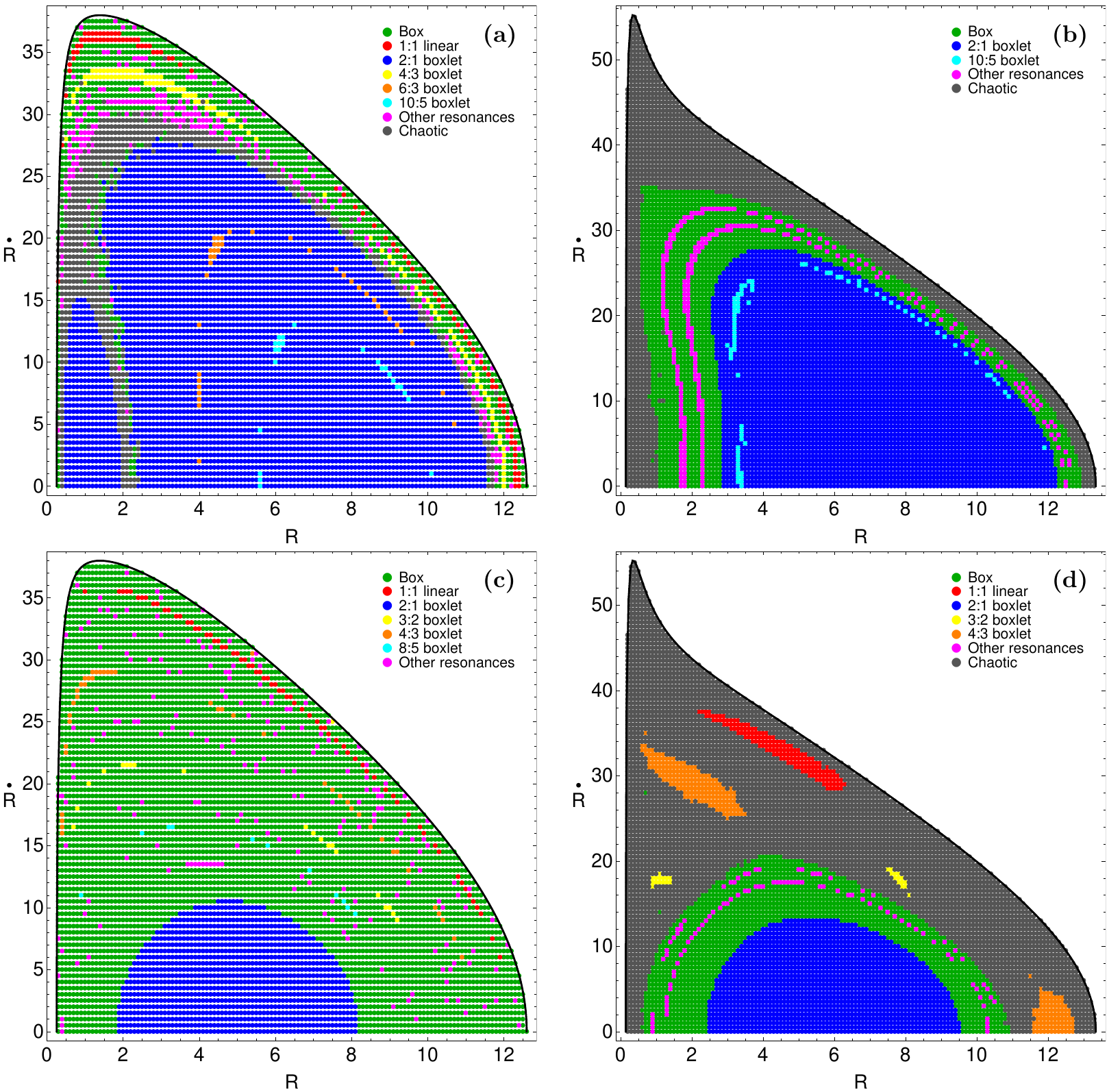}}
\caption{Orbital structure of the $(R,\dot{R})$ phase plane for the PH model when (a-upper left): $M_{\rm n} = 0$ and (b-upper right): $M_{\rm n} = 500$ and for the OH model when (c-lower left): $M_{\rm n} = 0$ and (b-lower right): $M_{\rm n} = 500$.}
\label{GridsMn}
\end{figure*}

Fig. \ref{GridsMn}a depicts the phase plane of the PH model when $M_{\rm n} = 0$. One can observe that most of the phase space is covered by 2:1 resonant orbits, while there is also a weak chaotic layer that separates the areas of regularity. The outermost thick curve is the ZVC. In Fig. \ref{GridsMn}b, we present a grid on the phase plane when $M_{\rm n} = 500$, i.e., a model with a more massive central nucleus. It is evident that there are many differences with respect to Fig. \ref{GridsMn}a, being the most visible: (i) the growth of the region occupied by chaotic orbits, (ii) an increase in the allowed radial velocity $\dot{R}$ of the stars near the center of the galaxy, and (iii) the absence of several families of resonant orbits (i.e., 1:1, 4:3, and 6:3 resonant orbits). In Fig. \ref{GridsMn}c, we can see the structure of the phase plane of the OH model when $M_{\rm n} = 0$. In this case, we observe that the phase plane is flooded with box orbits due to the absence of the central nucleus. On the other hand, in Fig. \ref{GridsMn}d, where we have an OH model with a massive nucleus ($M_{\rm n} = 500$), the portion of box orbits is confined considerably as a vast unified chaotic sea emerges surrounding several islands of secondary resonances.

\begin{figure*}
\centering
\resizebox{\hsize}{!}{\includegraphics{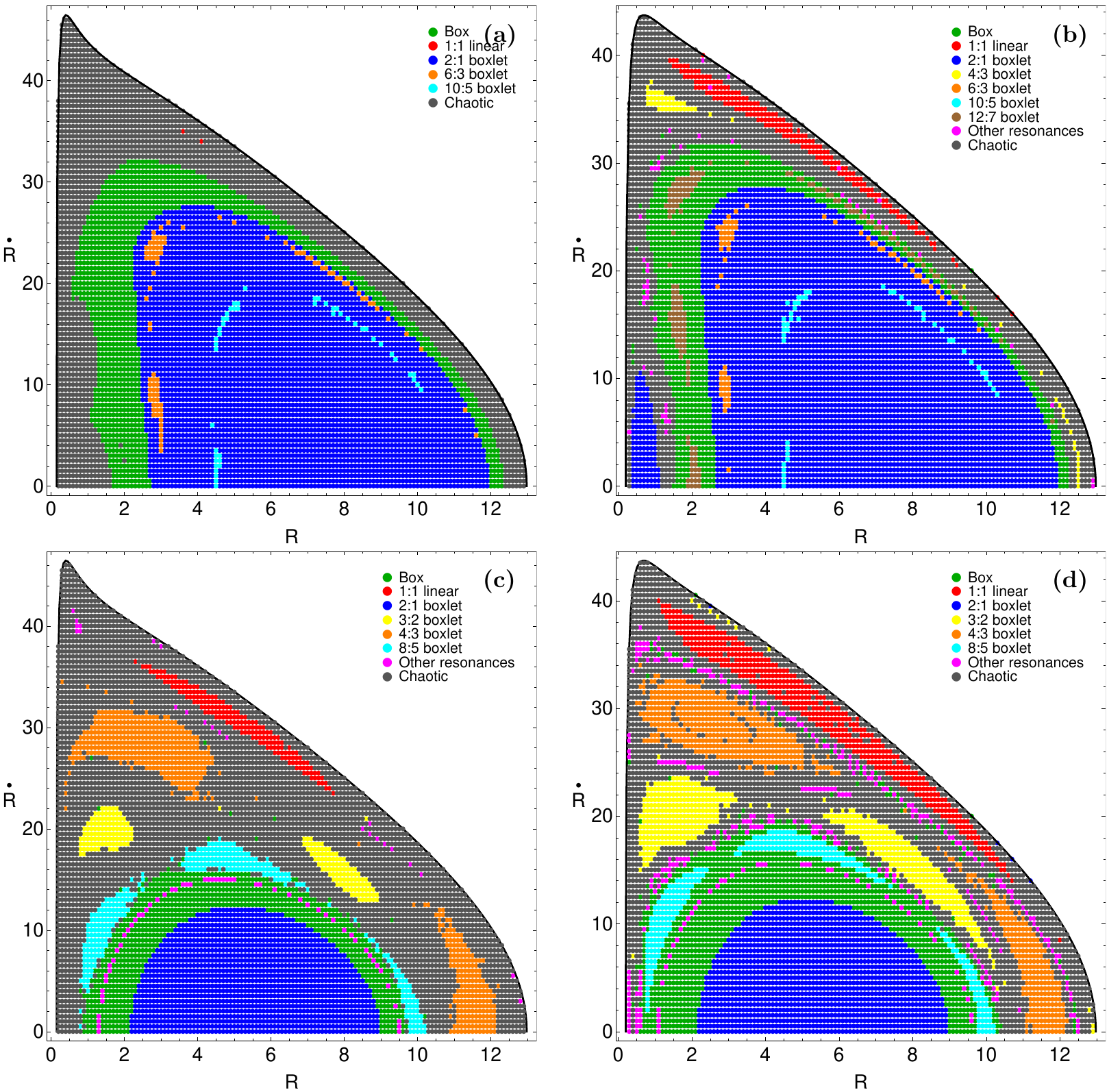}}
\caption{Orbital structure of the $(R,\dot{R})$ phase plane for the PH model when (a-upper left): $c_{\rm n} = 0.05$ and (b-upper right): $c_{\rm n} = 0.50$ and for the OH model when (c-lower left): $c_{\rm n} = 0.05$ and (b-lower right): $c_{\rm n} = 0.50$.}
\label{Gridscn}
\end{figure*}

The grid structure of the $(R,\dot{R})$ phase plane for the PH model when $c_{\rm n} = 0.05$ is presented in Fig. \ref{Gridscn}a. We see that when the concentration of the spherical nucleus is very high, a solid chaotic sea exists at the outer parts of the phase plane, resulting in the complete absence of secondary resonances. On the other hand, when $c_{\rm n} = 0.50$, we observe in Fig. \ref{Gridscn}b that the area on the phase plane occupied by chaotic orbits has shrunk, thus leaving space for several resonant families (i.e., 1:1, 4:3 and other resonances) to increase their rates. Furthermore, the 12:7 chain of islands emerges inside the box domain. A similar comparison between lower and higher concentrated nucleus for the OH models is made in Figs. \ref{Gridscn}c and \ref{Gridscn}d. We show that when $c_{\rm n} = 0.05$ (Fig. \ref{Gridscn}c) all the different resonant families are present and are surrounded by a unified chaotic sea. In the case where $c_{\rm n} = 0.50$ (Fig. \ref{Gridscn}d), the structure of the phase plane remains almost the same and the most prominent difference lies in the increasing rates of all of the regular families, which of course entails a reduction of the chaotic region.

\begin{figure*}
\centering
\resizebox{\hsize}{!}{\includegraphics{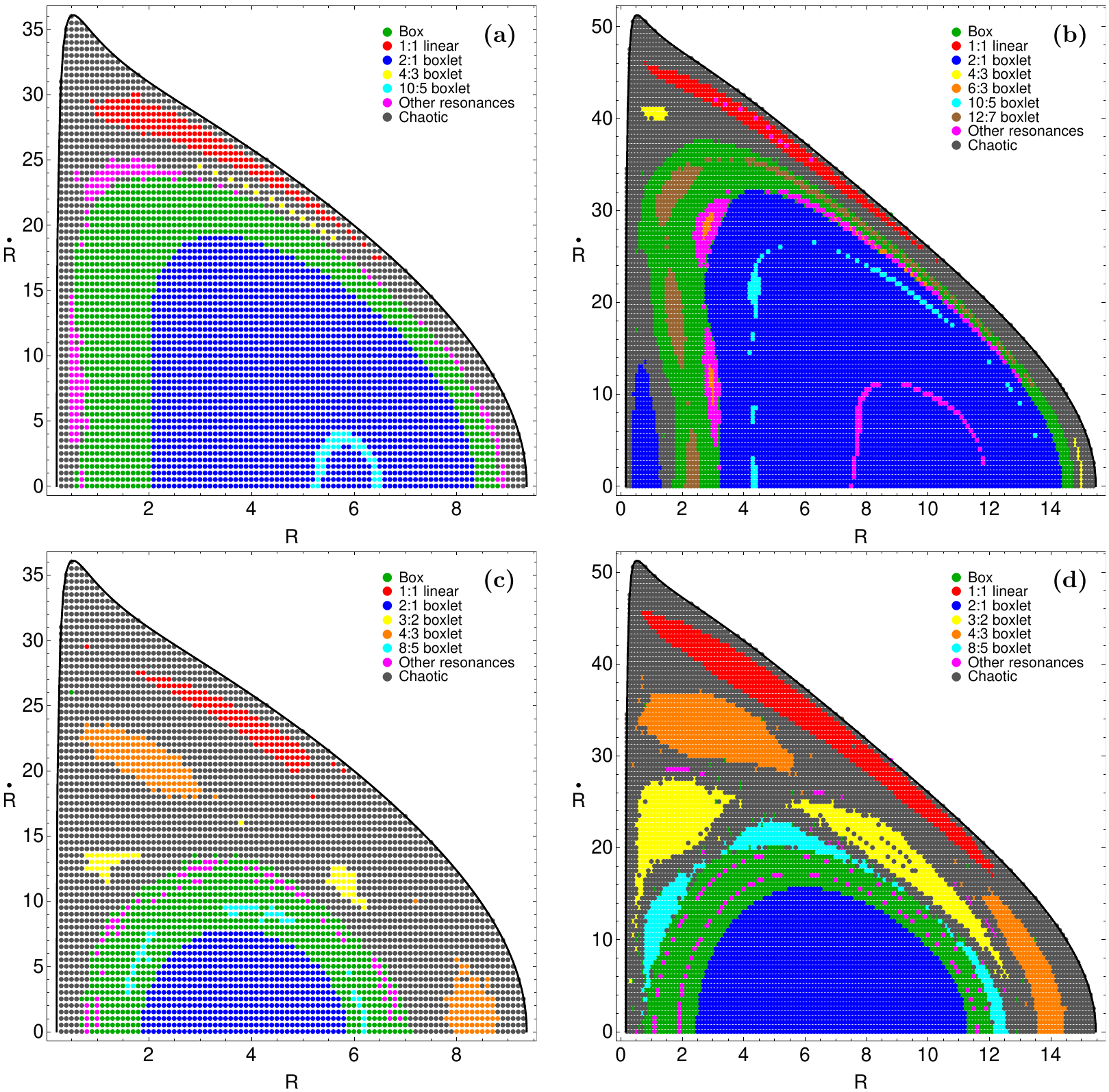}}
\caption{Orbital structure of the $(R,\dot{R})$ phase plane for the PH model when (a-upper left): $M_{\rm d} = 4500$ and (b-upper right): $M_{\rm d} = 9000$ and for the OH model when (c-lower left): $M_{\rm d} = 4500$ and (b-lower right): $M_{\rm d} = 9000$.}
\label{GridsMd}
\end{figure*}

A grid of initial conditions $(R_0,\dot{R_0})$ for the PH model when the disk has the minimum possible value ($M_{\rm d} = 4500$) is given in Fig. \ref{GridsMd}a. We observe that almost all of the resonant families are present forming well-defined sets of islands. It is interesting to note, the presence of a set of three islands corresponding to the so-called other resonances. In fact, this is the 5:3 resonance which, however, appears itself only in some isolated cases, so we do not feel it is necessary to include it in the list containing all the main resonant families under investigation. Fig. \ref{GridsMd}b shows a similar grid when $M_{\rm d} = 9000$ (i.e., the maximum possible value of the mass of the disk). It is evident that the structure of the phase plane has several differences with respect to the Fig. \ref{GridsMd}a which are: (i) the reduction of the region occupied by chaotic orbits; (ii) the approximately 40\% increase in the allowed radial velocity $\dot{R}$ of the stars near the central region of the galaxy; and (iii) the appearance of a second smaller area near the center occupied by 2:1 banana-type orbit. In Figs. \ref{GridsMd}c and \ref{GridsMd}d, we present two similar grids of initial conditions for the same values of $M_{\rm d}$ as in Figs. \ref{GridsMd}a and \ref{GridsMd}b, respectively, but applied to the OH models this time. Once more, as the disk becomes more massive the extent of the chaotic sea decreases, thus amplifying the rates of all the regular families.

\begin{figure*}
\centering
\resizebox{\hsize}{!}{\includegraphics{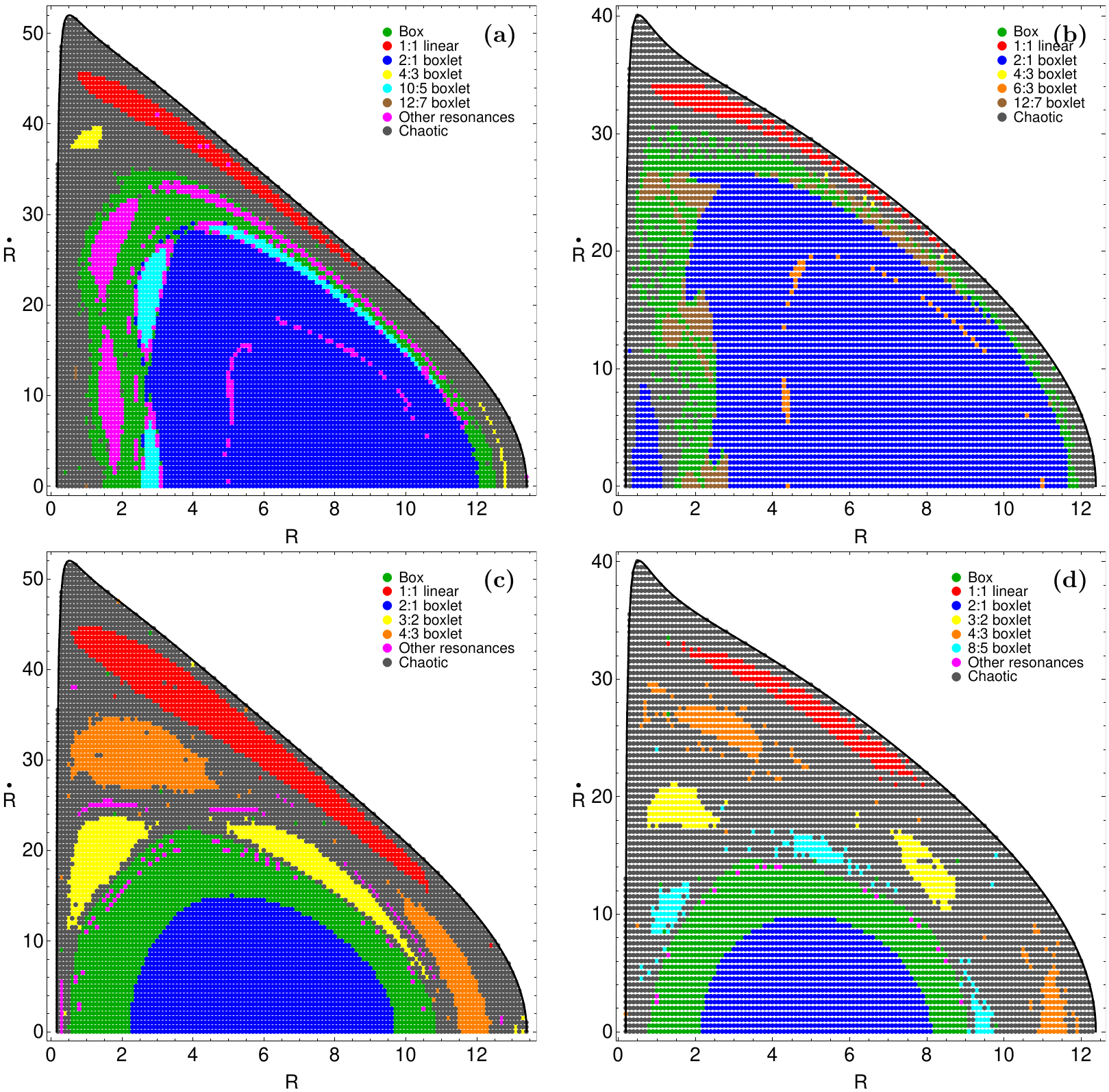}}
\caption{Orbital structure of the $(R,\dot{R})$ phase plane for the PH model when (a-upper left): $b = 4$ and (b-upper right): $b = 8$ and for the OH model when (c-lower left): $b = 4$ and (b-lower right): $b = 8$.}
\label{Gridsb}
\end{figure*}

In Fig. \ref{Gridsb}a, we present the orbital structure of a grid of initial conditions $(R_0,\dot{R_0})$ for the PH model when $b = 4$. We show that almost all resonant families are present, forming different sets of islands. Here we have to point out the existence of an additional resonant family, that of the 5:3 resonant orbits, which correspond to the so-called ``other resonances" and produce the set of the well-defined purple triple islands at the phase plane. Fig. \ref{Gridsb}b shows a similar grid of initial conditions when the core radius of the disk-halo possess its maximum possible value ($b = 8$). One may distinguish several differences in the structure of the phase plane with respect to that shown in Fig. \ref{Gridsb}a. The main differences are the following: (i) the area occupied by chaotic orbits has been reduced significantly and it is confined to the very outer parts of the phase plane; (ii) the bifurcated 10:5 family has been adsorbed by the main 2:1 family; (iii) the presence of higher resonances such as the 12:7 family is much stronger; and (iv) the 4:3 resonance is depopulated and the corresponding islands are so tiny that they appear as isolated points in the grid. Similar grids of initial conditions for the same values of $b$ as in Figs. \ref{Gridsb}a and \ref{Gridsb}b, respectively, but for the OH models, are shown in Figs. \ref{Gridsb}c and \ref{Gridsb}d. We observe that at highest value of $b$ (Fig. \ref{Gridsb}d) the amount of initial conditions corresponding to chaotic orbits is significantly greater with respect to Fig. \ref{Gridsb}c, thus limiting the extent of all the different areas of stability. Furthermore, higher resonant orbits (i.e., the 8:5 family) appear only in models with large values of $b$.

\begin{figure*}
\centering
\resizebox{\hsize}{!}{\includegraphics{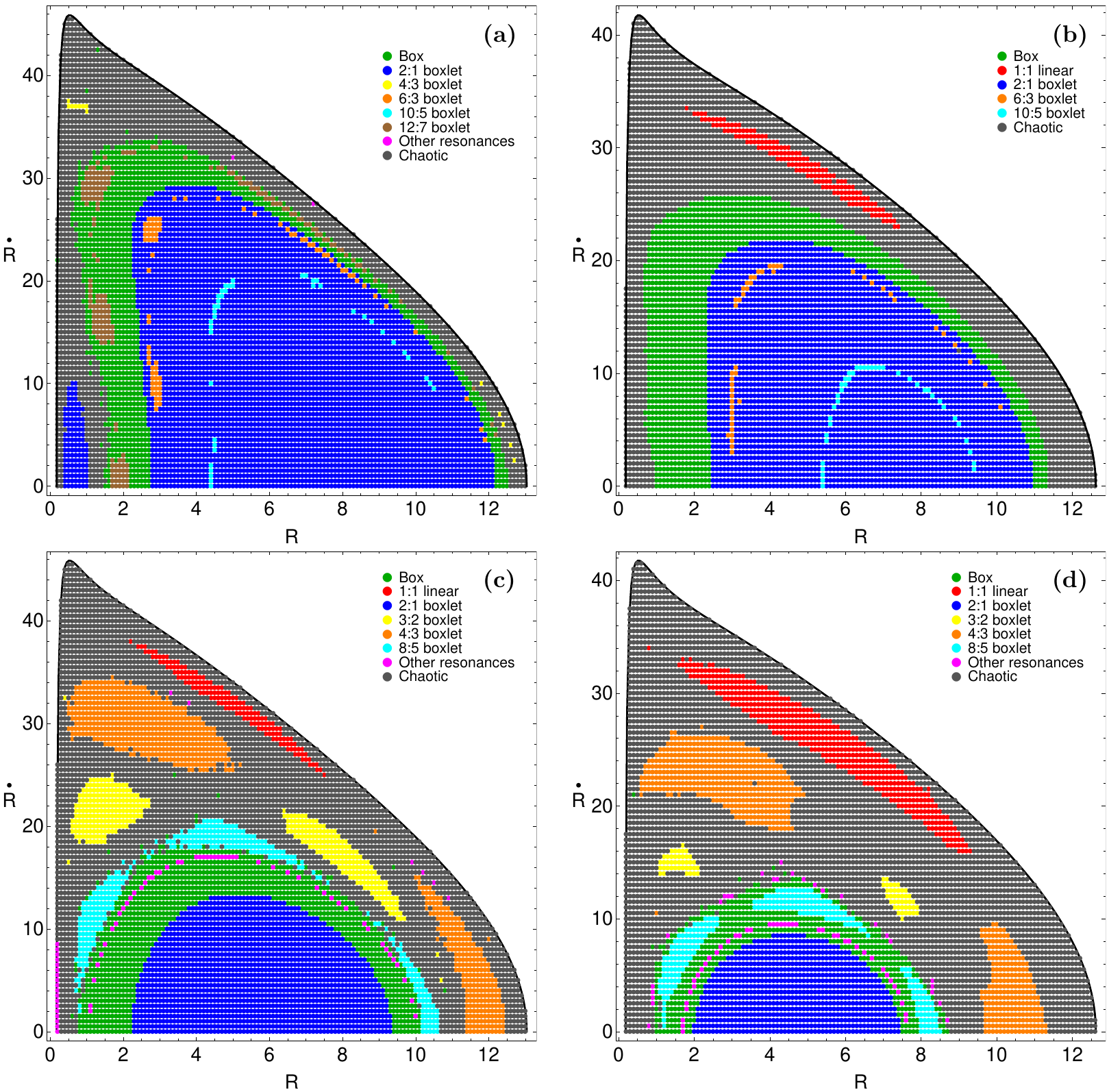}}
\caption{Orbital structure of the $(R,\dot{R})$ phase plane for the PH model when (a-upper left): $\alpha = 2.5$ and (b-upper right): $\alpha = 5$ and for the OH model when (c-lower left): $\alpha = 2.5$ and (b-lower right): $\alpha = 5$.}
\label{Gridsa}
\end{figure*}

The orbital structure of a grid of initial conditions $(R_0,\dot{R_0})$ for the PH model when $\alpha = 2.5$ is presented in Fig. \ref{Gridsa}a. We observe that the vast majority of the phase plane is covered by resonant 2:1 banana-type orbits. Specifically, there are two regions corresponding to 2:1 resonant orbits. It should be emphasized that the 1:1 resonant family, which is a basic family, is absent in PH galaxy models with sufficient small values of the scale length of the disk. A similar grid of initial conditions when $\alpha = 5$ is given in Fig. \ref{Gridsa}b. This grid has many similarities with respect to that shown in Fig. \ref{Gridsa}a, but it also has several important differences. The most visible differences are the growth of the area corresponding to chaotic orbits, the appearance of the main 1:1 family, the absence of the second small region occupied by 2:1 resonant orbits, and the disappearance of secondary resonances such as the 4:3 and the 12:7. In Figs. \ref{Gridsa}c and \ref{Gridsa}d we present grids of initial conditions for the same values of $\alpha$ as in Figs. \ref{Gridsa}a and \ref{Gridsa}b, respectively, but for the OH galaxy models. It is evident, that in the case of the highest value of $\alpha$, that is in Fig. \ref{Gridsa}d, the portion of the chaotic and the 1:1 resonant orbits is considerably larger, while at the same time, all the other regions of stability have been reduced (i.e., regions corresponding to box, 2:1, and 3:2 resonant orbits). Moreover, we should point out that in both cases, several higher, secondary resonant orbits (i.e., 11:7, 13:8, 14:9) emerge mainly inside the area of the box orbits.

\begin{figure*}
\centering
\resizebox{\hsize}{!}{\includegraphics{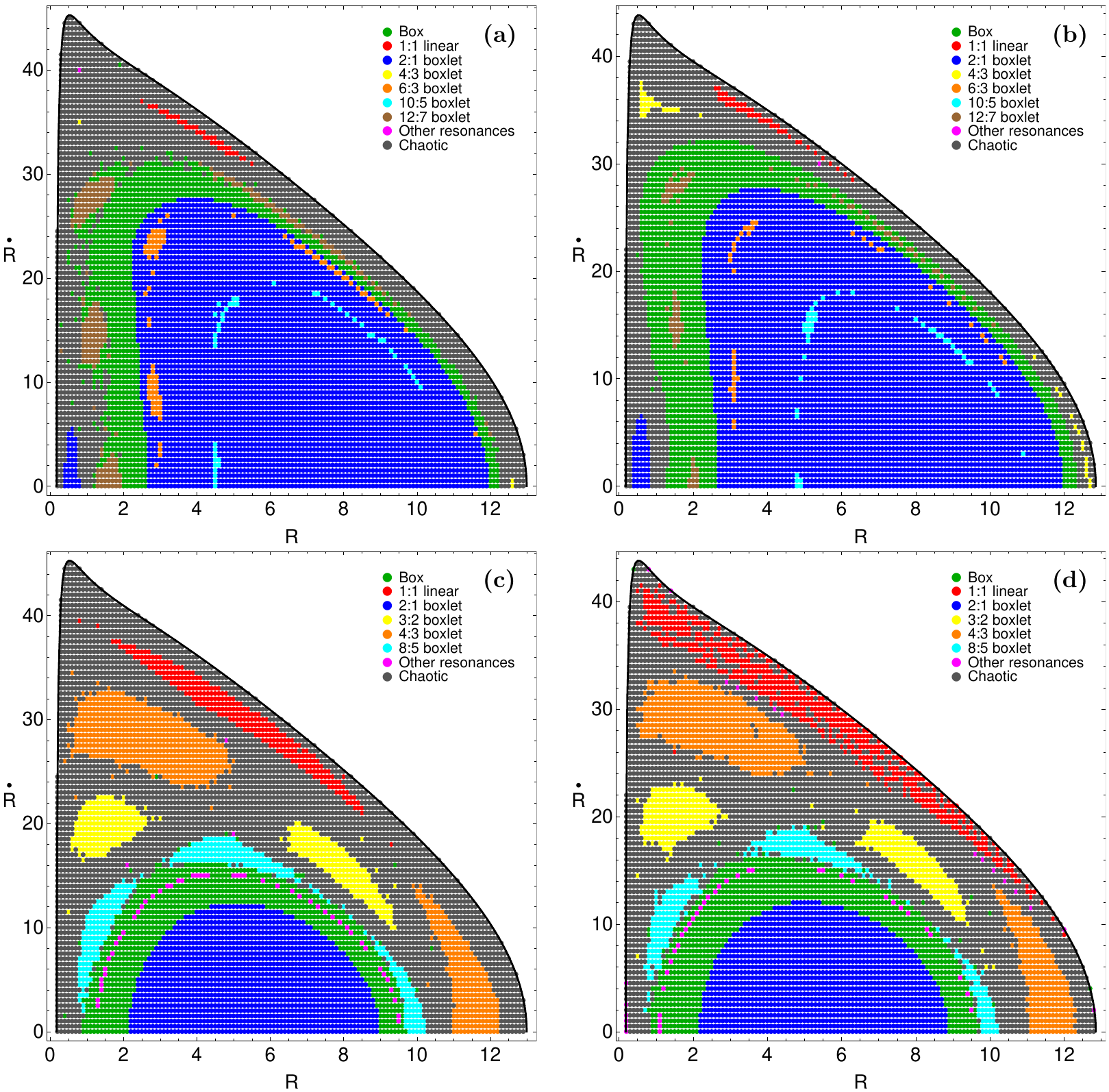}}
\caption{Orbital structure of the $(R,\dot{R})$ phase plane for the PH model when (a-upper left): $h = 0.1$ and (b-upper right): $h = 1$ and for the OH model when (c-lower left): $h = 0.1$ and (b-lower right): $h = 1$.}
\label{Gridsh}
\end{figure*}

Fig. \ref{Gridsh}a depicts the orbital structure of a grid of initial conditions $(R_0,\dot{R_0})$ for the prolate dark matter halo (PH) model when $h = 0.1$. As in all previous cases, more than half the phase plane is covered by 2:1 banana-type orbits. A unified chaotic layer exists at the outer parts of the phase plane and surrounds most of the different stability islands. However, the islands produced by the 4:3 resonant orbits are so small that they appear as lonely points in the gird. Things are quite similar in Fig. \ref{Gridsh}b where $h = 1$. We see that the overall grid structure is maintained and the observed differences are minor. In fact, the most noticeable differences are the following: (i) the extent of the chaotic layer is smaller, giving space to box orbits; (ii) the portion of the 4:3 resonant orbits looks more prominent; and (iii) the extent of the 12:7 resonant orbits has been reduced. Similar grids of initial conditions for the same values of $h$ exist, as in Figs. \ref{Gridsh}a and \ref{Gridsh}b, but for the OH galaxy models are shown in Figs. \ref{Gridsh}c and \ref{Gridsh}d, respectively. Once more, we note that the change of the value of the scale height of the disk does not cause significant influence on the structure of the phase plane. All it does is to shrink the area occupied by chaotic orbits thus, allowing mainly the 1:1 resonant orbits to enlarge their rate.

\begin{figure*}
\centering
\resizebox{\hsize}{!}{\includegraphics{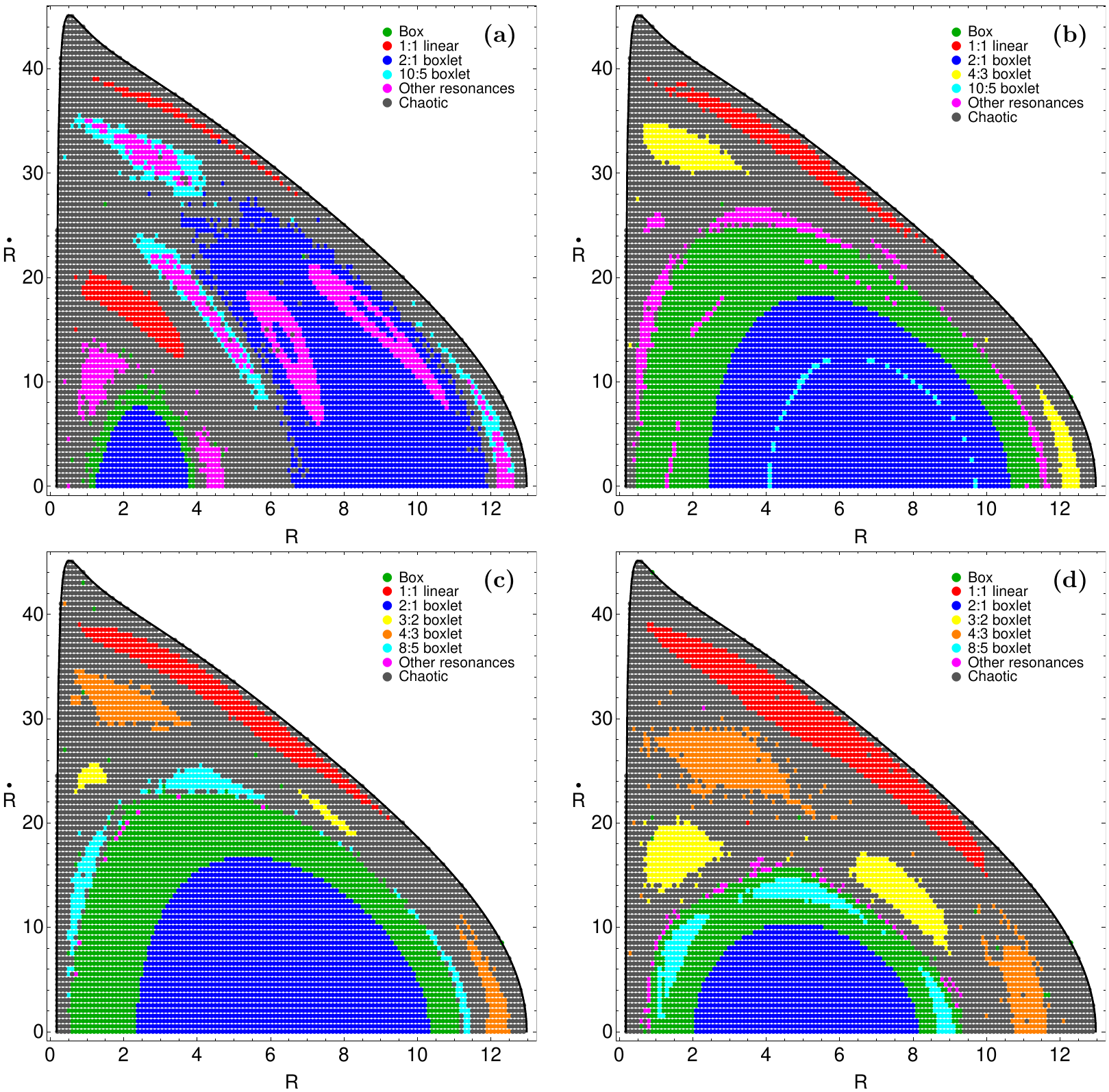}}
\caption{Orbital structure of the $(R,\dot{R})$ phase plane for the PH model when (a-upper left): $\beta = 0.1$ and (b-upper right): $\beta = 0.9$ and for the OH model when (c-lower left): $\beta = 1$ and (b-lower right): $\beta = 1.9$.}
\label{Gridsbeta}
\end{figure*}

In Fig. \ref{Gridsbeta}a, we present the orbital structure of a grid of initial conditions $(R_0,\dot{R_0})$ for the prolate dark matter halo (PH) model when $\beta = 0.1$. Undoubtedly, this is a very interesting phase plane that is very different from what we have seen so far. We observe the existence of a vast chaotic sea that embraces many islands of stability formed by different regular families of orbits. The meridional 2:1 banana-type and the linear 1:1 resonant orbits form two different islands on the grid, while box orbits have an anemic presence. Moreover, we distinguish several purple regions corresponding to other types of resonances. Here we have to point out that is the first time we have encountered such an intense presence of that type of orbits. Our numerical calculations suggest that these regions are produced by two different types of orbits. The set of the triple islands inside the region of the 10:5 orbits correspond to the 10:3 resonance, while the double set of islands inside the large 2:1 region and also above the box orbits is produced by the 8:4 resonance. The grid shown in Fig. \ref{Gridsbeta}b corresponds to the case where $\beta = 0.9$. We see that everything is now back to normal and the previous complicated structure has vanished. Again, we can distinguish many sets of islands formed by miscellaneous resonances. The small islands inside the region of box orbits correspond to the 13:8 resonance, the set right above the box orbits corresponds to the 8:5 resonance, while the set of the double islands embedded in the chaotic sea corresponds to the 3:2 resonant family. The orbital structure of the grid in the case of the spherical dark halo $(\beta = 1)$ is presented in Fig. \ref{Gridsbeta}c. Considering the previous analysis of the ``other resonances" shows that the orbital structure has remained almost unperturbed. Finally, Fig. \ref{Gridsbeta}d, shows the case of a highly flattened halo where $\beta = 1.9$. Here, the increase of several types of orbits (i.e., 1:1, 3:2, 4:3, chaotic) spurred both box and 2:1 orbits to reduce their rates and to be limited to the center of the phase plane.

\begin{figure*}
\centering
\resizebox{\hsize}{!}{\includegraphics{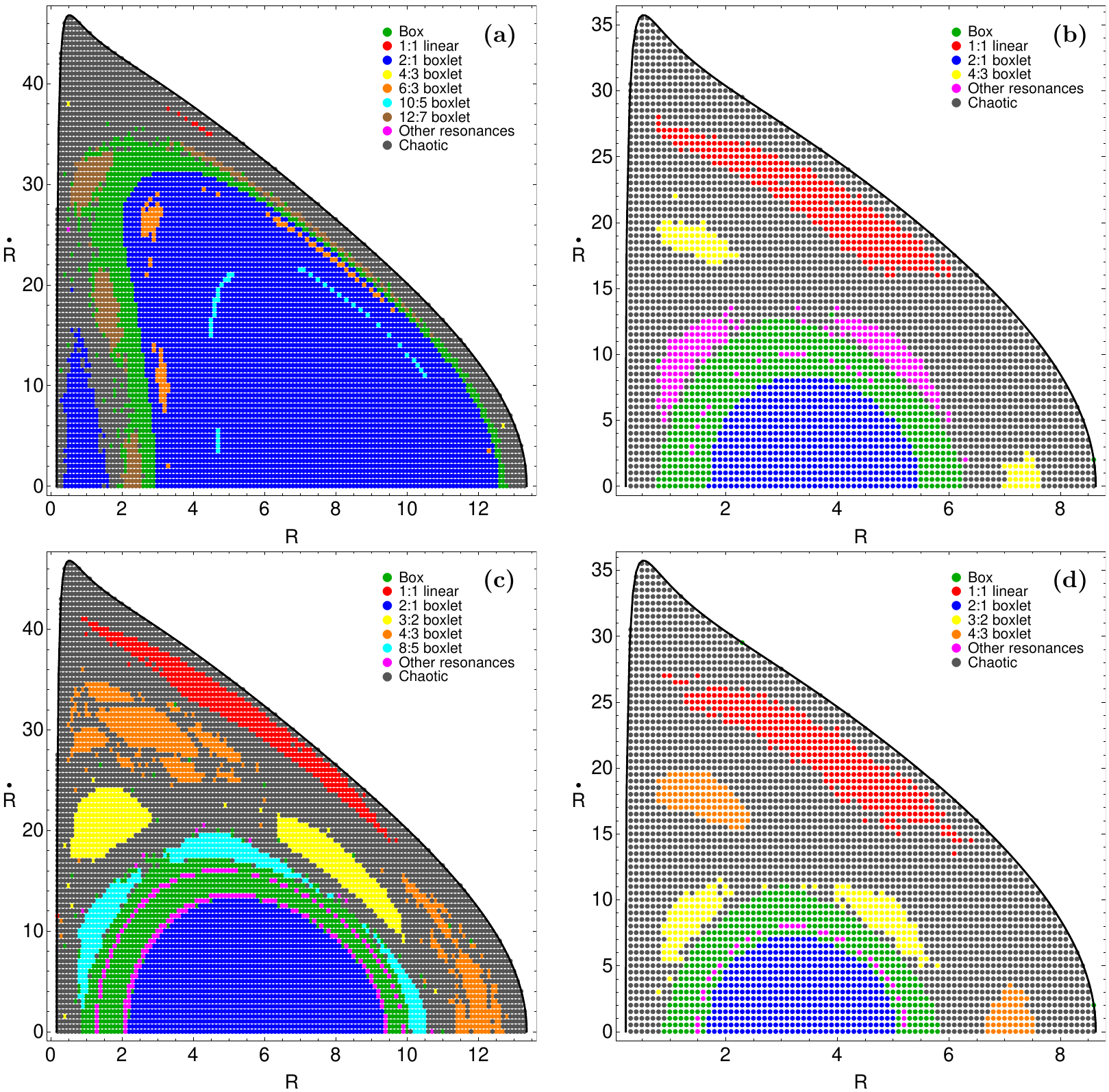}}
\caption{Orbital structure of the $(R,\dot{R})$ phase plane for the PH model when (a-upper left): $c_{\rm h} = 7$ and (b-upper right): $c_{\rm h} = 22$ and for the OH model when (c-lower left): $c_{\rm h} = 7$ and (b-lower right): $c_{\rm h} = 22$.}
\label{Gridsch}
\end{figure*}

The orbital structure of a grid of initial conditions $(R_0,\dot{R_0})$ for the prolate dark matter halo (PH) model when $c_{\rm h} = 7$ is shown in Fig. \ref{Gridsch}a. One may observe that the vast majority of the phase plane is covered by initial conditions corresponding to 2:1 banana-type orbits. In fact, there are two distinct regions formed by these orbits. At the outer parts of the phase plane, a chaotic layer is present that contains small stability islands. In particular, we can distinguish a tiny region formed by the 1:1 resonant orbits, while the points on the grid corresponding to the 4:3 resonant orbits are hardly visible. Fig. \ref{Gridsch}b shows a similar grid of initial conditions for a prolate galaxy model when $c_{\rm h} = 22$. It is evident that the orbital structure has many significant differences with respect to that presented in Fig. \ref{Gridsch}a. The most noticeable differences are: (i) the amount of chaos has increased considerably leading to a vast chaotic sea; (ii) box orbits have been greatly depopulated and now are confined only to the center of the grid; (iii) all the bifurcated resonances (i.e., 6:3 and 12:5) have disappeared, while new resonant families such as the 3:2 family appear; and (iv) the total area of the phase plane is reduced. In Fig. \ref{Gridsch}c, we present another grid of initial conditions for an oblate dark halo galaxy model when $c_{\rm h} = 7$. In this case, all the expected types of orbits are present forming well-defined regions in the phase plane. The purple dots correspond to resonant orbits of higher multiplicity (i.e., 11:7 and 13:8 resonant orbits). A similar oblate dark halo grid when $c_{\rm h} = 22$ is shown in Fig. \ref{Gridsch}d. We show that the extent of chaos is significantly larger than that observed in Fig. \ref{Gridsch}c. Moreover, the higher resonant 8:5 family is completely absent. With a closer look at the overall orbital structure, we realize that the grid of Fig. \ref{Gridsch}d is very similar to that shown in Fig. \ref{Gridsch}b. Those two grids correspond to the same value of scale length of the halo $c_{\rm h} = 22$, but to different shapes of the halo (prolate and oblate, respectively). Therefore, we may conclude that in high values of $c_{\rm h}$, or in other words, when the halo is much less concentrated, the orbital structure is the same, regardless of the particular shape of the halo (prolate or oblate).

\begin{figure*}
\centering
\resizebox{\hsize}{!}{\includegraphics{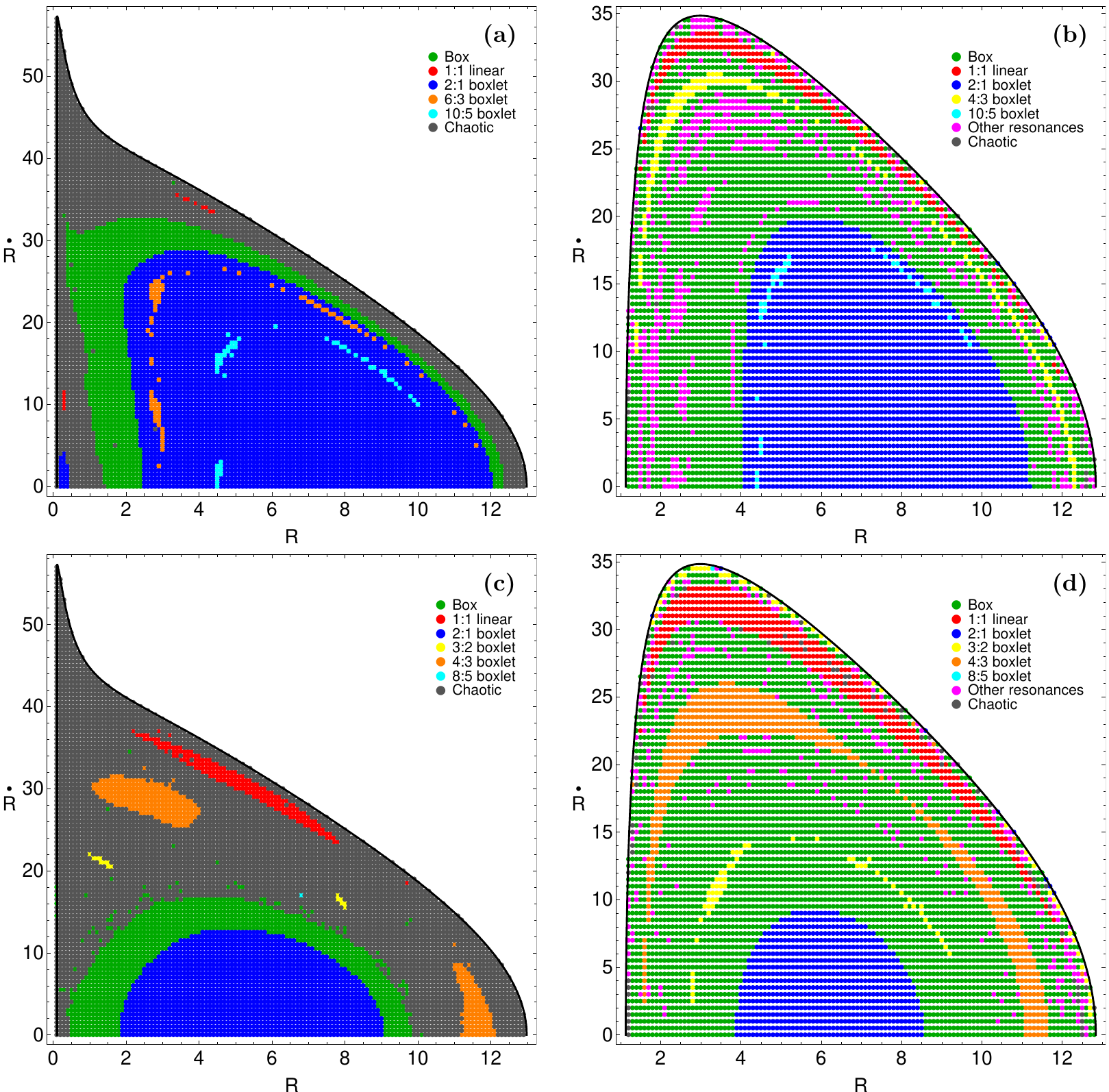}}
\caption{Orbital structure of the $(R,\dot{R})$ phase plane for the PH model when (a-upper left): $L_{\rm z} = 1$ and (b-upper right): $L_{\rm z} = 50$ and for the OH model when (c-lower left): $L_{\rm z} = 1$ and (b-lower right): $L_{\rm z} = 50$.}
\label{GridsLz}
\end{figure*}

A grid of initial conditions $(R_0,\dot{R_0})$ showing the orbital structure of the prolate dark matter halo model when $L_{\rm z} = 1$ is presented in Fig. \ref{GridsLz}a. We observe the existence of a dense and unified chaotic sea, while the majority of the regular domains are located near the central region of the phase plane, although there are important stability islands surrounding them. Moreover, we should notice the complete absence of 4:3 or higher resonant orbits. Fig. \ref{GridsLz}b shows a similar grid of initial conditions corresponding to the $L_{\rm z} = 50$ prolate halo model. It is evident that the structure of this phase plane differs greatly from the previous one. The most significant differences are: (i) the entire phase plane is covered by regular orbits therefore chaotic motion, if any, is negligible; (ii) numerous types of miscellaneous resonant orbits belonging to the ``other resonances" class are spread all over the phase plane (3:2, 5:3, 4:5, 8:5, 5:7, 6:7, 10:7, 11:7, 13:8, 14:9 resonant orbits, mentioning most of them); (iii) the allowed radial velocity $\dot{R}$ of stars passing near the center of the galaxy is almost decreased by half; and (iv) the permissible area on the $(R,\dot{R})$ plane is reduced. In Fig. \ref{GridsLz}c, we present another grid of initial conditions for the $L_{\rm z} = 1$ oblate halo model. A vast chaotic sea is observed surrounding all the different stability islands. Furthermore, we should note, a lack of higher resonant orbits, while the 8:5 resonant orbits appear as extreme isolated points on the grid. Things are very different in Fig. \ref{GridsLz}d where the grid of the $L_{\rm z} = 50$ oblate halo model is depicted. The main differences with respect to the structure of the grid shown in Fig. \ref{GridsLz}c are very similar to those described earlier in the prolate dark halo case. Again, the 8:5 resonant orbits are hardly visible. However, we should point out that in this case, the number of the ``other resonances" orbits (4:5, 6:5, 7:5, 6:7, 8:7, 10:7, 8:9, 12:11, 13:11, mentioning the most important of them) is lower.

\begin{figure*}
\centering
\resizebox{\hsize}{!}{\includegraphics{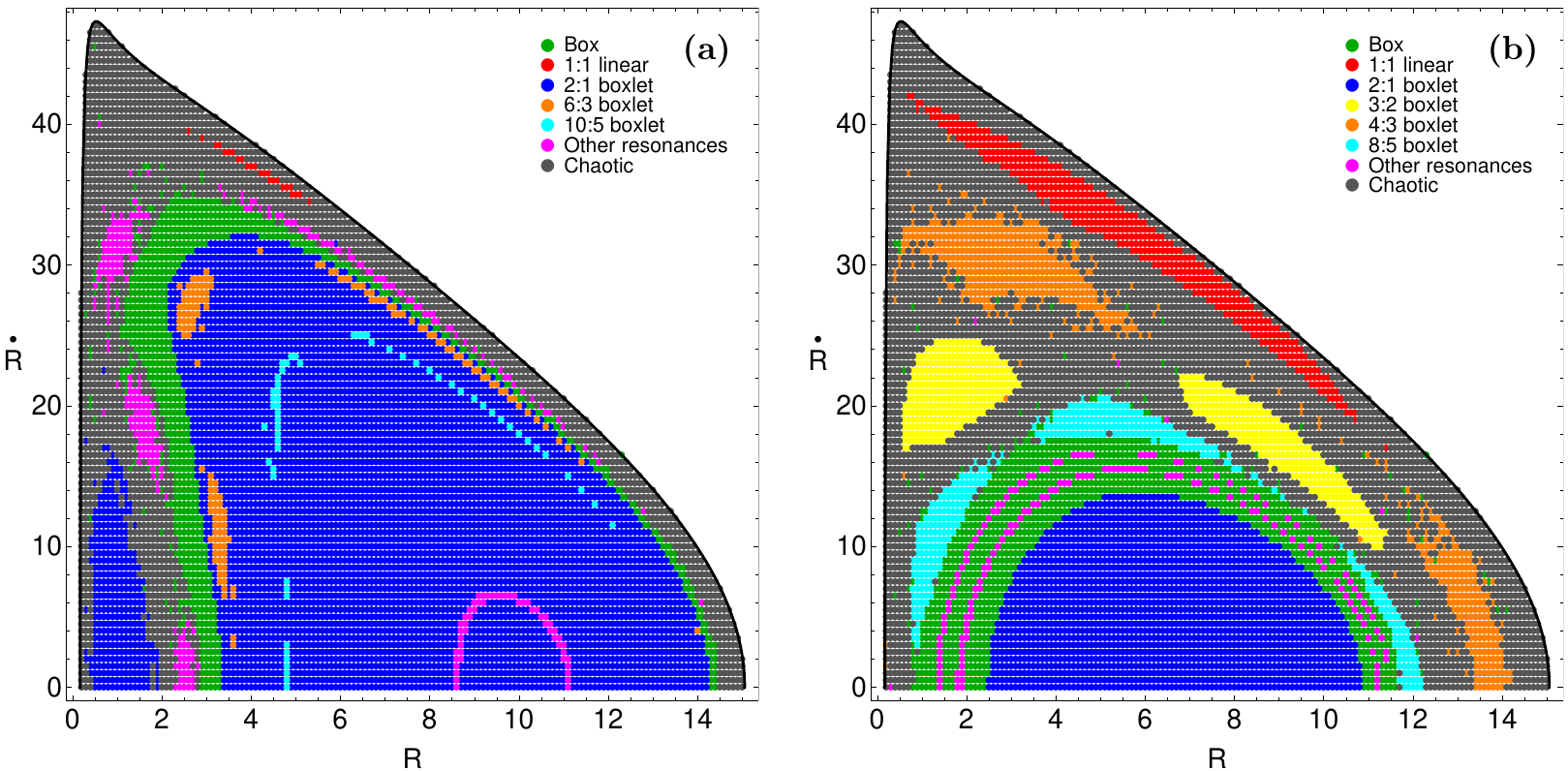}}
\caption{Orbital structure of the $(R,\dot{R})$ phase plane for (a-left): the PH model and (b-right): the OH model when $E = 700$.}
\label{GridsEn}
\end{figure*}

The orbital structure of a grid of initial conditions $(R_0,\dot{R_0})$ of the prolate dark matter halo model when $E = 700$ (the maximum energy level studied) is presented in Fig. \ref{GridsEn}a. It is clearly seen that the vast majority (about 60\%) of the phase plane is covered by initial conditions corresponding to the 2:1 banana-type orbits. As in many previous prolate halo grids, the 2:1 resonant orbits form two separate islands of stability on the phase plane. In Fig. \ref{GridsEn}a, we note a complete lack of 4:3 and 12:7 boxlet orbits, while we can identify the presence of ``other resonant" orbits. Inside the main 2:1 region, there is a thin ribbon of initial conditions produced by the 8:4 resonance, which is a bifurcated subfamily of the main 2:1 family (like the 6:3 and the 10:5 families of orbits). Moreover, just outside the box orbits we see a set of five islands of initial conditions corresponding to the 12:5 resonance. In Fig. \ref{GridsEn}b, a similar grid of initial conditions for the same value of the energy $(E = 700)$, but for the oblate case, is given. We observe that all the orbit families are present forming distinct well-defined stability islands, which are embraced by a unified chaotic sea. We should mention, that there is a strong presence of resonant orbits of higher multiplicity inside the area occupied by box orbits. Our numerical calculations indicate that these initial conditions correspond either to 10:7 or 13:8 resonant orbits.

\end{appendix}


\begin{thebibliography}{}

\bibitem[\protect\citeauthoryear{Allen \& Santill\'an}{1991}]{AS91} Allen, C., Santill\'an, A. 1991, Rev. Mex. Astron. Astrof., 22, 255

\bibitem[\protect\citeauthoryear{Allgood et al.}{2006}]{AFP06} Allgood, B., Flores, R.A., Primack, J.R., Kravtsov, A.V., Wechsler, R.H., Faltenbacher, A., Bullock, J.S. 2006, MNRAS, 367, 1781

\bibitem[\protect\citeauthoryear{Binney \& Spergel}{1982}]{BS82} Binney, J., Spergel, D. 1982, ApJ, 252, 308

\bibitem[\protect\citeauthoryear{Binney \& Spergel}{1984}]{BS84} Binney, J., Spergel, D. 1984, MNRAS, 206, 159

\bibitem[\protect\citeauthoryear{Binney \& Tremaine}{2008}]{BT08} Binney, J., Tremaine, S. 2008, Galactic Dynamics, Princeton Univ. Press, Princeton, USA

\bibitem[\protect\citeauthoryear{Bosma}{1981}]{B81} Bosma, A. 1981, AJ, 86, 1825

\bibitem[\protect\citeauthoryear{Bountis et al.}{2012}]{BMA12} Bountis, T., Manos, T., Antonopoulos, Ch. 2012, CeMDA, 113, 63

\bibitem[\protect\citeauthoryear{Capuzzo-Dolcetta et al.}{2007}]{CLMV07} Capuzzo-Dolcetta, R., Leccese, L., Merritt, D., Vicari, A. 2007, ApJ, 666, 165

\bibitem[\protect\citeauthoryear{Caranicolas}{1997}]{C97} Caranicolas, N.D. 1997, Ap\&SS, 246, 15

\bibitem[\protect\citeauthoryear{Caranicolas \& Innanen}{1991}]{CI91} Caranicolas, N.D., Innanen, K.A. 1991, AJ, 102, 1343

\bibitem[\protect\citeauthoryear{Caranicolas \& Zotos}{2009}]{CZ09} Caranicolas, N.D., Zotos, E.E. 2009, Baltic Astronomy, 18, 205

\bibitem[\protect\citeauthoryear{Caranicolas \& Zotos}{2010}]{CZ10} Caranicolas, N.D., Zotos, E.E. 2010, Astron. Nachr., 331, 330 (Paper I)

\bibitem[\protect\citeauthoryear{Caranicolas \& Zotos}{2011}]{CZ11} Caranicolas, N.D., Zotos, E.E. 2011, Research in Astron. Astrophys., 11, 811

\bibitem[\protect\citeauthoryear{Caranicolas \& Zotos}{2013}]{CZ13} Caranicolas, N.D., Zotos, E.E. 2013, PASA, 30, 49

\bibitem[\protect\citeauthoryear{Carlberg \& Innanen}{1987}]{CI87} Carlberg, R.G., Innanen, K.A. 1987, AJ, 94, 666

\bibitem[\protect\citeauthoryear{Carpintero \& Aguilar}{1998}]{CA98} Carpintero, D.D., Aguilar, L.A. 1998, MNRAS, 298, 1

\bibitem[\protect\citeauthoryear{Clemens}{1985}]{C85} Clemens, D.P. 1985, ApJ, 295, 422

\bibitem[\protect\citeauthoryear{Contopoulos \& Grosb{\o}l}{1989}]{CG89} Contopoulos, G., Grosb{\o}ol, P. 1989, A\&AR, 1, 261

\bibitem[\protect\citeauthoryear{Cooray}{2000}]{C00} Cooray, A.R. 2000, MNRAS, 313, 783

\bibitem[\protect\citeauthoryear{Dubinski \& Carlberg}{1991}]{DC91} Dubinski, J., Carlberg, R.G. 1991, ApJ, 378, 496

\bibitem[\protect\citeauthoryear{Evans \& Bridle}{2009}]{EB09} Evans, A.K.D., Bridle, S. 2009, ApJ, 695, 1446

\bibitem[\protect\citeauthoryear{Frenk et al.}{1988}]{FWDE88} Frenk, C.S., White, S.D.M., Davis, M., Efstathiou, G. 1988, ApJ, 327, 507

\bibitem[\protect\citeauthoryear{Gerhard \& Binney}{1985}]{GB85} Gerhard, O., Binney, J. 1985, MNRAS, 216, 467

\bibitem[\protect\citeauthoryear{G\'omez et al.}{2010}]{GHBL10} G\'omez, F., Helmi, A., Brown, A.G.A., Li, Y.S. 2010, MNRAS, 408, 935

\bibitem[\protect\citeauthoryear{Hasan \& Norman}{1990}]{HN90} Hasan, H., Norman, C.A. 1990, ApJ, 361, 69

\bibitem[\protect\citeauthoryear{Hasan et al.}{1993}]{HPN93} Hasan, H., Pfenniger, D., Norman, C. 1993, ApJ, 409, 91

\bibitem[\protect\citeauthoryear{Helmi}{2004}]{H04} Helmi, A. 2004, PASA, 21, 212

\bibitem[\protect\citeauthoryear{Honma \& Sofue}{1997}]{HS97} Honma, M., Sofue, Y. 1997, PASJ, 49, 453

\bibitem[\protect\citeauthoryear{Irrgang et al.}{2013}]{IWTS13} Irrgang, A., Wilcox, B., Tucker, E., Schiefelbein, L. 2013, A\&A, 549, A137

\bibitem[\protect\citeauthoryear{Jing \& Suto}{2002}]{JS02} Jing, Y.P., Suto, Y. 2002, ApJ, 574, 538

\bibitem[\protect\citeauthoryear{Kasun \& Evrard}{2005}]{KE05} Kasun, S.F., Evrard, A.E. 2005, ApJ, 629, 781

\bibitem[\protect\citeauthoryear{Kaufmann \& Patsis}{2005}]{KP05} Kaufmann, D., Patsis, P. 2005, ApJ, 624, 693

\bibitem[\protect\citeauthoryear{Kunihito et al.}{2000}]{KTT00} Kunihito, I., Takahiro, T., Takashi, N. 2000, ApJ, 528, 51

\bibitem[\protect\citeauthoryear{Manos \& Athanassoula}{2011}]{MA11} Manos, T., Athansssoula, E. 2011, MNRAS, 415, 629

\bibitem[\protect\citeauthoryear{Manos et al.}{2013}]{MBS13} Manos, T., Bountis, T., Skokos, Ch. 2013, J. Phys. A: Math. Theor. 46, 254017

\bibitem[\protect\citeauthoryear{McLaughlin}{1999}]{ML99} McLaughlin, D.E. 1999, ApJ, 512, L9

\bibitem[\protect\citeauthoryear{Merritt}{1999}]{M99} Merritt, D. PASP, 111, 129

\bibitem[\protect\citeauthoryear{Merritt \& Fridman}{1996}]{MF96} Merritt, D., Fridman, T. 1996, ApJ, 460, 136

\bibitem[\protect\citeauthoryear{Miyamoto \& Nagai}{1975}]{MN75} Miyamoto, W., Nagai, R. 1975, PASJ, 27, 533

\bibitem[\protect\citeauthoryear{Muzzio et al.}{2005}]{MCW05} Muzzio, J.C., Carpintero, D.D., Wachlin, F.C. 2005, CeMDA, 91, 173

\bibitem[\protect\citeauthoryear{Navarro et al.}{1996}]{NFW96} Navarro, J.F., Frenk, C.S., White, S.D.M. 1996, ApJ, 462, 563

\bibitem[\protect\citeauthoryear{Navarro et al.}{1997}]{NFW97} Navarro, J.F., Frenk, C.S., White, S.D.M. 1997, ApJ, 490, 493

\bibitem[\protect\citeauthoryear{Oll\'{e} \& Pfenniger}{1998}]{OP98} Oll\'{e}, M., Pfenniger, D. 1998, A\&A, 334, 829

\bibitem[\protect\citeauthoryear{Olling \& Merrifield}{2000}]{OM00} Olling, R.P., Merrifield, M.R. 2000, MNRAS, 311, 361

\bibitem[\protect\citeauthoryear{Ollongren}{1962}]{O62} Ollongren, A. 1962, Bulletin of the Astronomical Institutes of the Netherlands, 16, 241

\bibitem[\protect\citeauthoryear{Ostriker et al.}{1974}]{OPY74} Ostriker, J.P., Peebles, P.J.E., Yahil, A. 1974, ApJ, 193, 2, L1

\bibitem[\protect\citeauthoryear{Papadopoulos \& Caranicolas}{2006}]{PC06} Papadopoulos, N.J., Caranicolas, N.D. 2006, New Astronomy, 12, 11

\bibitem[\protect\citeauthoryear{Persic \& Salucci}{1995}]{PS95} Persic, M., Salucci, P. 1995, ApJS, 99, 501

\bibitem[\protect\citeauthoryear{Pfenniger}{1984}]{P84} Pfenniger, D. 1984, A\&A, 134, 373

\bibitem[\protect\citeauthoryear{Pfenniger}{1996}]{P96} Pfenniger, D., 1996, in Buta R., Crocker D. A., Elmegreen B. G. eds, ASP Conf. Ser. Vol. 91, Barred Galaxies. Astron. Soc. Pac., San Francisco, p. 273

\bibitem[\protect\citeauthoryear{Pichardo et al.}{2004}]{PMM04} Pichardo, B., Martos, M., Moreno, E. 2004, ApJ, 609, 144

\bibitem[\protect\citeauthoryear{Press et al.}{1992}]{PTVF92} Press, H.P., Teukolsky, S.A, Vetterling, W.T., Flannery, B.P. 1992, Numerical Recipes in FORTRAN 77, 2nd Ed., Cambridge Univ. Press, Cambridge, USA

\bibitem[\protect\citeauthoryear{Rubin et al.}{1980}]{RFT80} Rubin, V.C., Ford, W.K., Thonnard, N. 1980, ApJ, 238, 471

\bibitem[\protect\citeauthoryear{Rubin \& Burstein}{1988}]{RB85} Rubin, V.C., Burstein, D. 1985, ApJ, 297, 423

\bibitem[\protect\citeauthoryear{Ru\v{z}i\v{c}ka et al.}{2007}]{RPT07} Ru\v{z}i\v{c}ka, A., Palou\v{s}, J., Theis, C. 2007, A\&A, 461, 155

\bibitem[\protect\citeauthoryear{Satoh}{1980}]{S80} Satoh, C. 1980, PASJ, 32, 41

\bibitem[\protect\citeauthoryear{Sellwood \& Wilkinson}{1993}]{SW93} Sellwood, J., Wilkinson, A. 1993, Rep. Prog. Phys., 56, 173

\bibitem[\protect\citeauthoryear{\v Sidlichovsk\'y and Nesvorn\'y}{1996}]{SN96} \v Sidlichovsk\'y, M., Nesvorn\'y, D. 1996, CeMDA, 65, 137

\bibitem[\protect\citeauthoryear{Skokos}{2001}]{S01} Skokos, C. 2001, J. Phys. A: Math. Gen., 34, 10029

\bibitem[\protect\citeauthoryear{Skokos et al.}{2002a}]{SPA02a} Skokos, Ch., Patsis, P.A., Athanassoula, E. 2002 MNRAS, 333 847

\bibitem[\protect\citeauthoryear{Skokos et al.}{2002b}]{SPA02b} Skokos, Ch., Patsis, P.A., Athanassoula, E. 2002, MNRAS, 333, 861

\bibitem[\protect\citeauthoryear{Steidel et al.}{2002}]{SKS02} Steidel, C.C., Kollmeier, J.A., Shapley, A.E., Churchill, C.W., Dickinson, M., Pettini, M. 2002, ApJ, 570, 526

\bibitem[\protect\citeauthoryear{Szydlowski}{1994}]{S94} Szydlowski, M. 1994, J. Math. Phys., 35, 1850

\bibitem[\protect\citeauthoryear{Wang et al.}{2009}]{WMJ09} Wang, H., Mo, H.J., Jing, Y.P., Guo, Y., van den Bosch, F.C., Yang, X. 2009, MNRAS, 394, 398

\bibitem[\protect\citeauthoryear{Wechsler et al.}{2002}]{WBPKD02} Wechsler, R.H., Bullock, J.S., Primack, J.R., Kravtsov, A.V., Dekel, A. 2002, ApJ, 568, 52

\bibitem[\protect\citeauthoryear{Zotos}{2011}]{Z11} Zotos, E.E. 2011, New Astronomy, 16, 391

\bibitem[\protect\citeauthoryear{Zotos}{2012a}]{Z12a} Zotos, E.E. 2012a, New Astronomy, 17, 576

\bibitem[\protect\citeauthoryear{Zotos \& Carpintero}{2013}]{ZC13} Zotos, E.E., Carpintero, D.D. 2013, CeMDA, 116, 417

\bibitem[\protect\citeauthoryear{Zotos \& Caranicolas}{2013a}]{ZCar13} Zotos, E.E., Caranicolas, N.D. 2013, A\&A, 560, A110

\bibitem[\protect\citeauthoryear{Zotos \& Caranicolas}{2013b}]{ZCar14} Zotos, E.E., Caranicolas, N.D. 2014, Nonlinear Dynamics, 76, 323 

\bibitem[\protect\citeauthoryear{Zwicky}{1933}]{Z33} Zwicky, F. 1933, Helvetica Physica Acta, 6, 110

\end{thebibliography}
\end{document}